\documentclass[11pt]{amsart}
\usepackage{pictexwd,
graphicx}
\usepackage{amsmath,amsfonts,amscd,amssymb,epsfig,euscript, eufrak}
\usepackage{amsxtra,mathrsfs}
\newtheorem{theorem}{Theorem}[section]

\newtheorem{lemma}[theorem]{Lemma}
\newtheorem{corollary}[theorem]{Corollary}
\theoremstyle{definition}
\newtheorem{definition}[theorem]{Definition}

\theoremstyle{remark} \newtheorem{remark}[theorem]{Remark}
\numberwithin{equation}{section}
\newcommand{\field}[1]{\ensuremath{\mathbb{#1}}}
\newcommand{\CC}{\field{C}}

\newcommand{\PP}{\field{P}}
\newcommand{\RR}{\field{R}}

\newcommand{\Up}{\Upsilon}
\newcommand{\Ka}{K\"{a}hler\:}
\newcommand{\Te}{Teichm\"{u}ller\:}                    
\DeclareMathOperator{\id}{id}

\DeclareMathOperator{\im}{Im}
 \DeclareMathOperator{\PSL}{PSL}
\DeclareMathOperator{\PSU}{PSU} 

\newcommand{\del}{\partial}

\newcommand{\vp}{\varphi}

\newcommand{\Sch}{\mathfrak{S}}
\newcommand{\D}{\mathfrak{D}}

\newcommand{\up}{{\mathbb{U}}}
\newcommand{\lo}{{\mathbb{L}}}

\newcommand{\U}{{\mathbb{U}}}
\newcommand{\C}{{\mathbb{C}}}
\newcommand{\R}{{\mathbb{R}}}
\newcommand{\curly}[1]{\mathscr{#1}}

\newcommand{\cC}{\curly{C}}
\newcommand{\cD}{\curly{D}}
\newcommand{\cE}{\curly{E}}

\newcommand{\cM}{\curly{M}}

\newcommand{\cT}{\curly{T}}

\newcommand{\al}{\alpha}
\newcommand{\be}{\beta}

\newcommand{\s}{\gamma}

\newcommand{\bk}{\backslash}
\newcommand{\Ga}{\Gamma}

\newcommand{\pa}{\partial}
\newcommand{\la}{\langle}
\newcommand{\ra}{\rangle}

\newcommand{\ov}{\overline}
\newcommand{\ep}{\epsilon}
\newcommand{\vep}{\varepsilon}

\newcommand{\z}{\bar{z}}
\newcommand{\w}{\bar{w}}

\newcommand{\ma}[4]{(\begin{smallmatrix}
              #1 & #2 \\ #3 & #4
             \end{smallmatrix})}

\newcommand{\wpm}{\scriptscriptstyle{WP}}

\begin{document}
\title[Quantum Liouville Theory]{Quantum Liouville theory in the background field formalism I. Compact Riemann surfaces}
\author{Leon A. Takhtajan} \address{Department of Mathematics \\
Stony Brook University\\ Stony Brook, NY 11794-3651 \\ USA}
\email{leontak@math.sunysb.edu}
\author{ Lee-Peng Teo}\address{Faculty of Information Technology \\
Multimedia University \\ Jalan Multimedia, Cyberjaya\\
63100, Selangor, Malaysia} 
\email{lpteo@mmu.edu.my}
\begin{abstract} 
Using Polyakov's functional integral approach and the Liouville action functional defined in \cite{ZT2} and \cite{LTT}, we formulate quantum Liouville theory on a compact Riemann surface $X$ of genus $g>1$.  For the partition function $\la X\ra$ and correlation functions with the stress-energy tensor components $\la \prod_{i=1}^{n}T(z_{i})\prod_{k=1}^{l}\bar{T}(\w_{k})X\ra$, we describe Feynman rules in the background field formalism by expanding corresponding functional integrals around a classical solution, the hyperbolic metric on $X$.
Extending analysis in \cite{LT1,LT2,LT-Varenna,LT3}, we define the  regularization scheme  for any choice of the global coordinate on $X$. For the Schottky and quasi-Fuchsian global coordinates, we rigorously prove that one- and two-point correlation functions satisfy conformal Ward identities in all orders of the perturbation theory. Obtained results are interpreted in terms of complex geometry of the projective line bundle $\cE_{c}=\lambda_{H}^{c/2}$ over the moduli space $\mathfrak{M}_{g}$, where $c$ is the central charge and $\lambda_{H}$ is the Hodge line bundle, and provide the Friedan-Shenker \cite{FS} complex geometry approach to CFT with the first non-trivial example besides rational models.
\end{abstract}
 \maketitle

\tableofcontents

\section{Introduction}\label{introduction}
Classical Liouville theory is a 
Euclidean field theory associated with hyperbolic Riemann surfaces.   
Complete conformal metrics $ds^{2}$ on a Riemann surface $X$ are classical fields of the theory, and the so-called Liouville equation --- the equation $K(ds^{2})=-1$, where $K(ds^{2})$ is a Gaussian curvature, is the corresponding Euler-Lagrange equation. According to the uniformization theorem, it has a unique solution --- the hyperbolic metric on $X$. The quantized Liouville theory describes ``quantum corrections'' to hyperbolic geometry of $X$ by taking into account  fluctuations around the hyperbolic metric. In 1981, Polyakov formulated a functional integral approach to bosonic string theory, and made a fundamental discovery that quantum Liouville theory is a conformal anomaly for non-critical strings \cite{Pol1}. 
Thus in order to find correlation functions of vertex operators of the bosonic string in any dimension $D$ (and not only for $D=26$), one needs to know correlation functions of the Liouville vertex operators $V_{\al}(z)=e^{\alpha\vp(z)}$, where $ds^{2}=e^{\vp(z)}|dz|^{2}$ is the Liouville field --- a conformal metric on $X$. 
The fundamental property that classical fields and equation of motion of the Liouville theory are conformally invariant, led 
Belavin, Polyakov and Zamolodchikov to their formulation of the two-dimensional Conformal Field Theory (CFT) \cite{BPZ}. Though the problem of computing correlation functions of the Liouville vertex operators, needed for non-critical string theory, is still outstanding, in the works of Dorn and Otto \cite{DornOtto94;structureconstants}, and of Zamolodchikov and Zamolodchikov \cite{ZamZam96;Structureconstants} the quantum Liouville theory was formulated as a non-rational model of CFT with a continuous spectrum of conformal dimensions (see the review \cite{Teschner} for a complete account and references). 

In \cite{Pol2}, Polyakov proposed a functional integral representation for correlation functions of the Liouville vertex operators in the form needed for the non-critical string theory. This so-called geometric approach to the quantum Liouville theory was formalized and developed in \cite{LT1,LT2,LT-Varenna,LT3}. In this formulation, correlation functions of Liouville vertex operators on the Riemann sphere $\PP^{1}$ are defined by
\begin{equation} \label{X-A}
\la V_{\al_{1}}(z_{1})\dots V_{\al_{n}}(z_{n})\ra =\underset{\cC\cM_{\al}(\PP^{1})}{\pmb{\int}}\!
e^{-\frac{1}{2\pi\hbar}S_{\al}(\vp)}\;\cD\vp,
\end{equation}  
where $\hbar>0$ plays the role of Planck's constant, $\cC\cM_{\mathbf{\al}}(\PP^{1})$ is the space of all smooth conformal metrics $e^{\vp(z)}|dz|^{2}$ on $\PP^{1}\setminus\{z_{1},\dots,z_{n}\}$ which have conical singularities at the insertion points
\begin{equation} \label{conical}
e^{\vp(z)}\simeq \frac{1}{|z-z_{i}|^{2\hbar\al_{i}}}\quad\text{as}\quad z\rightarrow z_{i},\;\; i=1,\dots,n,
\end{equation}
and $S_{\mathbf{\al}}(\vp)$ is the Liouville action functional defined in \cite{LT3}. Here $\hbar\al_{i}\leq 1$ and $\sum_{i=1}^{n}\hbar\al_{i}>2$.
When $\hbar\al_{i}=1$, which corresponds to the puncture vertex operator, asymptotic \eqref{conical} is replaced by
$e^{\vp(z)}\simeq |z-z_{i}|^{-2}(\log|z-z_{i}|)^{-2}$. According to \cite{BPZ}, conformal symmetry of the theory manifests itself through conformal Ward identities for correlation functions with insertions of components of the stress-energy tensor. The Ward identity for  the $(2,0)$ component $T(\vp)=\frac{1}{\hbar}(\vp_{zz}-\frac{1}{2}\vp_{z}^{2})$ has the form
\begin{equation} \label{WTX}
\la T(z)X_{\mathbf{\al}}\ra =\sum_{i=1}^{n}\left(\frac{\Delta_{\al_{i}}}{(z-z_{i})^{2}}+\frac{\pa_{z_{i}}}{z-z_{i}}\right)\la X_{\mathbf{\al}}\ra,
\end{equation}
where $X_{\al}= V_{\al_{1}}(z_{1})\dots V_{\al_{n}}(z_{n})$,
\begin{equation} \label{TX}
\la T(z)X_{\mathbf{\al}}\ra =\underset{\cC\cM_{\al}(\PP^{1})}{\pmb{\int}}\!T(\vp)(z)
e^{-\frac{1}{2\pi\hbar}S_{\al}(\vp)}\;\cD\vp,
\end{equation}
and 
$\displaystyle{\Delta_{\al_{i}}=\al_{i}(2-\hbar\al_{i})}/2$ are conformal dimensions of the vertex operators $V_{\al_{i}}$. Note that since $\la X_{\al}\ra$ and $\la T(z)X_{\al}\ra$ have been already defined by functional integrals \eqref{X-A} and \eqref{TX}, the Ward identity \eqref{WTX} actually requires a proof\footnote{This should be compared with the standard CFT approach to quantum Liouville theory, where Ward identities are built into the construction of the Hilbert space of states which carries a representation of the Virasoro algebra.}. BPZ conformal Ward identities were generalized to higher genus Riemann surfaces in \cite{EO}.

At the classical level equation \eqref{WTX} (and a similar equation for compact Riemann surfaces)
represents a non-trivial relation between the accessory parameters of the Fuchsian uniformization of the Riemann surface $X=\PP^{1}\setminus\{z_{1},\dots,z_{n}\}$ and the classical Liouville action --- the critical value of the Liouville action functional. It was proved in \cite{ZT1} (and in \cite{ZT2} for the compact case; see also the discussion in \cite{T-PSPM,T-Cargese}).  

The background field formalism for puncture operators --- a perturbative expansion in $\hbar$ around the classical solution for the partition function and correlation functions with insertions of the stress-energy tensor, was developed in \cite{LT1,LT2}. The results,
summarized in \cite{LT-Varenna}, are the following.
\begin{itemize}
\item Rigorous definition of $\la X_{\al}\ra$ and 
$\la \prod_{i=1}^{n}T(z_{i})\prod_{k=1}^{l}\bar{T}(\w_{k})X_{\al}\ra$ in all orders of the perturbation theory, and the proof of the conformal Ward identity \eqref{WTX} at the one-loop level.
The latter follows from the formula for the first variation of the Selberg zeta function $Z(s)$ at $s=2$ in \cite{TZ}. 
\item The proof of conformal Ward identities with two insertions of the stress-energy tensor at the classical level, based on results in \cite{ZT1,ZT2}.
Equivalence between the Ward identity for $\la T(z)\bar{T}(\w)X_{\al}\ra$ at the one-loop level, and the local index theorem for families of $\bar{\pa}$-operators on punctured Riemann surfaces, proved in \cite{TZ}.  
\item The asymptotic
$$\la T(z)T(w)X_{\al}\ra =\frac{c/2}{(z-w)^{4}} + O(|z-w|^{-2})\quad\text{as}\quad w\rightarrow z,$$
valid in all orders of the perturbation theory, where
\begin{equation*} 
c=\frac{12}{\hbar} + 1
\end{equation*}
is the central charge of quantum Liouville theory, given by 
the sum of classical contribution $\frac{12}{\hbar}$ and one-loop correction $1$. 
\end{itemize}

The present paper is a long overdue sequel to \cite{LT-Varenna}. We extend the results in \cite{LT-Varenna} to all orders of the perturbation theory, with precise formulations and complete proofs. To emphasize the invariant geometric meaning of our results, and avoid any additional analytic ramifications due to non-compactness, we concentrate on the case of compact Riemann surfaces.  The important case of the Riemann surfaces with punctures will be considered elsewhere. 

Namely, let $X$ be a compact Riemann surface of genus $g>1$. We define the partition function $\la X\ra$ as the following functional integral
\begin{equation} \label{X-compact}
\la X \ra =\underset{\cC\cM(X)}{\pmb{\int}}\!
e^{-\frac{1}{2\pi\hbar}S(\vp)}\;\cD\vp,
\end{equation}  
where $\cC\cM(X)$ is the set of all smooth conformal metrics on $X$, and $S$ is the Liouville action functional. It is known \cite{LTT} that the definition of $S$ depends on the choice of a global coordinate on $X$ --- a representation $X\simeq\Ga\bk\Omega$, where $\Ga$ is a Kleinian group with an invariant component $\Omega$. For our purposes it is sufficient to consider the case when  $\Ga$ is either a Schottky group, or a quasi-Fuchsian group. Corresponding action functionals were defined in \cite{ZT2} and \cite{LTT} respectively.

A comparison with \eqref{X-A} shows that $\la X\ra$ can be interpreted as a ``correlation function of handle operators''. Ultimately, for every $\hbar>0$ we would like to define $\la X\ra$ as a function on the corresponding Schottky space $\mathfrak{S}_{g}$, or Teichm\"{u}ller space $\mathfrak{T}_{g}$, which parameterizes marked Riemann surfaces of genus $g>1$. However, we can only define $\la X\ra$  perturbatively as a ``formal function'' --- a formal power series in $\hbar$ with coefficients being smooth functions on $\mathfrak{S}_{g}$ or $\mathfrak{T}_{g}$. This is done in the background field formalism by considering the perturbation expansion of the functional integral \eqref{X-compact} around a classical solution.  Corresponding UV-divergencies, following \cite{LT1}, are regularized
by using a reparametrization-invariant definition of the propagator at coincident points\footnote{In the non-compact case one should also regularize IR-divergencies at the punctures.}. In this regularization scheme only classical contribution to the partition function $\la X\ra$  --- a term of order $\hbar^{-1}$ --- depends on the choice of a global coordinate on $X$. All other terms in the formal Taylor series expansion of $\la X\ra$ are well-defined functions on the moduli space $\mathfrak{M}_{g}$ of compact Riemann surfaces of genus $g>1$. 

Multi-point correlation functions $\la \prod_{i=1}^{n}T(z_{i})\prod_{k=1}^{l}\bar{T}(\w_{k})X\ra$ with insertions of the stress-energy tensor are defined in a similar way. The UV-divergence arising from a tadpole graph is regularized as in \cite{LT1,LT-Varenna}, whereas divergencies arising from graphs with self-loops
are regularized in a way similar to the regularization of $\la X\ra$. In this definition, only classical and one-loop contributions to one-point correlation functions depend non-trivially on the choice of a global coordinate on $X$. All other higher loop contributions to $\la T(z)X\ra$ and $\la \bar{T}(\z)X\ra$ are correspondingly $(2,0)$ and $(0,2)$ tensors on $X$. Similarly, all terms in the irreducible multi-point correlation functions $\la\la \prod_{i=1}^{n}T(z_{i})\prod_{k=1}^{l}\bar{T}(\w_{k})X\ra\ra$  with $n+l\geq 2$ are tensors of type $(2,0)$ and $(0,2)$ on $X$ in $z_{1},\dots,z_{n}$ and $w_{1},\dots,w_{l}$, which are symmetric with respect to these two groups of variables.

Our main results are given in Theorems \ref{one-point}, \ref{TT-equation} and \ref{T-barT-equation}. Succinctly, partition and correlation functions of the quantum Liouville theory, defined in Section \ref{QLT}, satisfy the conformal Ward identities in all orders of the perturbation theory. As a corollary, the quantum Liouville theory in the background field formalism is a conformal field theory with the central charge $c=\frac{12}{\hbar}+1$. 

To present the first result (we refer to Theorem \ref{one-point} 
for the invariant geometric formulation), let $X\simeq\Ga\bk\Omega$ be a Riemann surface of genus $g>1$ with a Schottky global coordinate, $\mu$ be a harmonic Beltrami differential for $\Ga$, and let $X^{\vep\mu}\simeq\Ga^{\vep\mu}\bk\Omega^{\vep\mu}$ be the corresponding holomorphic family of Riemann surfaces (see Section \ref{TT-VF}). Then
\begin{equation} \label{T-Ward} 
\left.\frac{\pa}{\pa\vep}\right|_{\vep=0}\log\la X^{\vep\mu}\ra = -\frac{1}{\pi} \iint\limits_{F}\left(\la\la
T(z)X\ra\ra-\tfrac{1}{12}\,\mathcal{S}(J^{-1})(z)\right)\mu(z)d^{2}z,
\end{equation}
where $F$ is a fundamental domain for $\Ga$ in $\Omega$, 
$J:\up\rightarrow \Omega$ is the covering map of $\Omega$ by
the upper half-plane $\up$, and $\mathcal{S}(f)$ stands for the Schwarzian derivative of a holomorphic function $f$. Equation \eqref{T-Ward} is a precise analog of the BPZ conformal Ward identity \eqref{WTX} for a compact Riemann surface $X$. We emphasize that
both sides of \eqref{T-Ward} are defined by corresponding functional integrals, and the equation is valid in all orders of the perturbation expansion.

To state the second result (see Theorem \ref{TT-equation} for the invariant formulation), let $G(z,w)$ be the propagator of the quantum Liouville theory  --- the kernel of a resolvent operator $\frac{1}{2}(\Delta_{0}+\frac{1}{2})^{-1}$, where $\Delta_{0}$ is the Laplace operator of the hyperbolic metric  $ds^{2}=e^{\vp_{cl}(z)}|dz|^{2}$ on $X$, acting on functions, and let
$R(z,w)=4e^{-\vp_{cl}(z)}\pa_{\z}\mathcal{D}_{w}G(z,w)$,
where $\mathcal{D}_{w}=\pa_{w}^{2}-(\pa_{w}\vp_{cl})(w)\pa_{w}$ (see Section \ref{PPV}). Then in all orders of the perturbation expansion,
\begin{gather} 
\left.\frac{\pa^{2}}{\pa\vep_{1}\pa\vep_{2}}\right|_{\vep_{1}=\vep_{2}=0}\log\la X^{\vep_{1}\mu+\vep_{2}\nu}\ra  \label{TT-Ward} \\
=
\frac{1}{\pi^{2}}\iint\limits_{F}\iint\limits_{F}\bigl(\la\la
T(z)T(w) X\ra\ra  -\tfrac{6}{\hbar}K(z,w) 
 -\tfrac{\pi}{6}\mathcal{D}_{z}\mathcal{D}_{w}G(z,w) \nonumber \\
-\pi(2\pa_{z}R(z,w)+R(z,w)\pa_{z})(\la\la T(z)X\ra\ra-\tfrac{1}{12}\mathcal{S}(J^{-1})(z))\bigr)\mu(z)\nu(w)d^{2}zd^{2}w \nonumber \\
=
\frac{1}{\pi^{2}}\iint\limits_{F}\iint\limits_{F}\bigl(\la\la
T(z)T(w) X\ra\ra  -\tfrac{6}{\hbar}K(z,w) 
-\tfrac{\pi}{6}\mathcal{D}_{z}\mathcal{D}_{w}G(z,w))\mu(z)\nu(w)d^{2}zd^{2}w. \nonumber
\end{gather}
Here $\mu$, $\nu$ are harmonic Beltrami differentials for $\Ga$, and 
$$K(z,w)=\sum_{\gamma\in\Ga}\frac{\gamma'(w)^{2}}{(z-\gamma w)^{4}}.$$
It follows from \eqref{TT-Ward} that $\la\la T(z)T(w)X\ra\ra$ is a meromorphic quadratic differential for $\Ga$ in $z$ and $w$, with the only fourth order pole at $z=w$, and that for $w\rightarrow z$ 
\begin{align}
 \la\la T(z)T(w)X \ra\ra & =\frac{c/2}{(z-w)^4} + \left(\frac{2}{(z-w)^2}
- \frac{\pa_{z}}{z-w}\right)\la\la T(z)X \ra\ra  \label{TT-near} \\ 
&\quad+\;\text{regular terms as}\; w\rightarrow z,\nonumber
\end{align}
where $c=\frac{12}{\hbar}+1$. Equation \eqref{TT-Ward} is a precise analog of the BPZ conformal Ward identity with two insertions of the $(2,0)$ component of the stress-energy tensor \cite{BPZ}. This proves that the quantum Liouville theory in the background field formalism is  a CFT model with the central charge $c$.

Finally, the third result (see Theorem  \ref{T-barT-equation} and Corollary \ref{L-barL-X} for invariant formulations) can be stated as
\begin{gather} 
\left.\frac{\pa^{2}}{\pa\vep_{1}\pa\bar{\vep}_{2}}\right|_{\vep_{1}=\vep_{2}=0}\log\la X^{\vep_{1}\mu+\vep_{2}\nu}\ra  \label{T-barT-Ward} \\
=
\frac{1}{\pi^{2}}\iint\limits_{F}\iint\limits_{F}\bigl(\la\la
T(z)\bar{T}(\w) X\ra\ra  
 -\tfrac{\pi}{6}\mathcal{D}_{z}\mathcal{D}_{\w}G(z,w)\bigr)
 \mu(z)\ov{\nu(w)}d^{2}zd^{2}w.\nonumber
\end{gather}
Equation \eqref{T-barT-Ward}, which is valid in all orders of the perturbation expansion, is a conformal Ward identity with single insertions of $(2,0)$ and $(0,2)$ components of the stress-energy tensor (the case not considered in \cite{BPZ}). In particular, $\la\la T(z)\bar{T}(\w) X\ra\ra$ is a holomorphic $(2,0)$ tensor on $X$ in variable $z$ and anti-holomorphic $(0,2)$ tensor in variable $w$. 
The classical contribution to $\la\la T(z)\bar{T}(\w) X\ra\ra$ is a multiple of the Weil-Petersson metric on $\mathfrak{M}_{g}$, and \eqref{T-barT-Ward} at the classical level states that the classical Liouville
action is its K\"{a}hler potential \cite{ZT2}. At the one-loop level,  as we show in Appendix A, \eqref{T-barT-Ward} is another way of presenting the Belavin-Knizhnik theorem \cite{BK}. 
Finally, in Remark \ref{Potential} we explain the sense in which the two-point correlation function $\la\la T(z)\bar{T}(\w) X\ra\ra$
defines a family of K\"{a}hler metrics on $\mathfrak{M}_{g}$ with K\"{a}hler potential $\pi^{2}(\log\la X\ra +\frac{\hbar}{12}\log\la X\ra_{cl})$.  

For the reader's convenience, we make the paper relatively self-contained by presenting the background material necessary for the computations. To keep the length of the paper under control, we refer to \cite{ZT2,LTT}  for the construction of the Liouville action functional, and to \cite{T-Cargese,LT1,LT2,LT-Varenna,LT3} for the history of the geometric approach and discussion of conformal Ward identities at classical and one-loop levels.
 
 Here is a more detailed content of the paper. Section \ref{LAF} is devoted to the classical Liouville theory.  We briefly discuss Schottky, Fuchsian and quasi-Fuchsian uniformizations of the compact Riemann surfaces, introduce the Liouville action functional and the stress-energy tensor, and describe their main properties. In Section \ref{QLT}, we formulate the quantum Liouville theory in the background field formalism. Specifically, in Section \ref{QFT-FR} we describe the Feynman rules and the regularization scheme for the perturbative expansion of the partition function $\la X\ra$, and in Section \ref{EMT-OCF} we describe the Feynman rules and the regularization scheme for multi-point correlation functions with insertions of the stress-energy tensor. 
 
 Section \ref{TT-VF} is a ``crash course'' on deformation theory of compact Riemann surfaces. In Section \ref{TT-VF-TS} we define the deformation space $\D(\Ga)$, where $\Ga$ is either a Schottky or a quasi-Fuchsian group, and describe a complex manifold structure on $\D(\Ga)$. The  
 Schottky space $\mathfrak{S}_{g}$ is a deformation space $\D(\Ga)$ where $\Ga$ is a Schottky group, and the Teichm\"{u}ller space $\mathfrak{T}_{g}$ is a complex-analytic submanifold of $\D(\Ga)$ where $\Ga$ is a Fuchsian group. 
In Section \ref{formal-geo} we define the formal function on $\D(\Ga)$ as a formal power series in $\hbar$ whose coefficients are smooth functions on $\D(\Ga)$, and show that the partition function $\la X\ra$ and the free energy $\mathcal{F}_{X}=-\log\la X\ra$ give rise to formal functions on $\mathfrak{S}_{g}$ and $\mathfrak{T}_{g}$. In Section \ref{TT-VF-VF} we collect necessary variational formulas, from classical results of Ahlfors \cite{Ahl} and Wolpert \cite{Wol} to the formulas in \cite{ZT2} and \cite{LTT}. To the reader without prior knowledge of the deformation theory we recommend classical works \cite{A2,B1,Ahl}, briefly summarized in \cite{LTT}. 

Section \ref{PPV}, where we study the propagator $G(P,Q)$ of the quantum Liouville theory, is crucial for our approach. The propagator $G(P,Q)$ is defined as the integral kernel of the resolvent  operator $\tfrac{1}{2}(\Delta_{0}+\tfrac{1}{2})^{-1}$, where $\Delta_{0}$ is the Laplace operator of the hyperbolic metric on $X$ acting on functions. In Section \ref{diagonal}, using the Fuchsian global coordinate $z$ on $X\simeq\Ga\bk\up$, we represent $G(z,w)$ as the average over a Fuchsian group $\Ga$ of the propagator $\mathcal{G}(z,w)$ on the upper half-plane $\up$ (method of images), which is given by an explicit formula.  We determine the short-distance behavior of $G(z,w)$, 
and define a regularization of $G(z,w)$ and of $\pa_{z}\pa_{w}G(z,w)$ at $w=z$ by subtracting corresponding contributions of the identity element of $\Ga$.
Thus defined $G(z,z)$ gives rise to a smooth function on
$X$, whereas the corresponding $H(z)=\pa_{z}\pa_{z}G(z,z)$ is a smooth quadratic differential for $\Ga$, which behaves like  ``$1/12\pi$ of the projective connection'' under the changes of global coordinates. We present an explicit formula $P(z,w)=4\mathcal{D}_{z}\mathcal{D}_{\w}G(z,w)$ for the integral kernel of the projection operator $P$ onto the subspace of holomorphic quadratic differentials on $X$.  Though just being another form of the Ahlfors classical result, it plays a fundamental role in the computations in Sections \ref{VA-OPCF}--\ref{TPCF-TTbar}. In Section 
\ref{VF-propagator} we prove variational formulas for the propagator and its derivatives, collected in Lemmas \ref{var-propagator} and \ref{variation-H}. 

In Section \ref{VA-OPCF}, we prove the conformal Ward identity with single insertion of the stress-energy tensor --- Theorem \ref{one-point}, by computing $\pa \log\la X \ra$ in all orders of the perturbative expansion. As we have already mentioned, at the classical level the corresponding result was proved in \cite{ZT2} and \cite{LTT}. In Section \ref{VA-OPCF-OLL}, we compute $\pa \log\la X \ra$ at the one-loop level, and show that  the result coincides with the representation  
of $\la\la T(z) X\ra\ra_{1-\text{loop}}$ as a sum of Feynman graphs. The computation uses the formula for $\pa\log Z(2)$ from \cite[Section 3]{TZ}, the explicit form of the kernel $P(z,w)$, and the  Stokes' theorem. For higher loop terms, the statement of Theorem \ref{one-point} is valid ``graph by graph''. The actual computation splits into three cases, analyzed in Section \ref{VA-OPCF-HLL} by repeated use of the Stokes' theorem and careful analysis of boundary contributions. 

In Sections \ref{TPCF-TT} and \ref{TPCF-TTbar} we prove Theorems \ref{TT-equation} and \ref{T-barT-equation} --- conformal Ward identities with two insertions of the stress-energy tensor, which express the two-point correlation functions $\la\la T(z)T(w)X\ra\ra$ and $\la\la T(z)\bar{T}(\w)X\ra\ra$ in terms of the one-point correlation functions $\la\la T(z) X\ra\ra$ and $\la\la \bar{T}(\z) X\ra\ra$. By Theorem \ref{one-point}, the correlation function $\la\la T(z)X\ra\ra$ is a holomorphic quadratic differential for $\Ga$ which corresponds to an exact
$(1,0)$-form on the Schottky space $\mathfrak{S}_{g}$, 
and Theorem \ref{T-barT-equation} states that the $(1,1)$-form 
$\bar{\pa}\la\la T(z)X\ra\ra$ on $\mathfrak{S}_{g}$ corresponds to $-\frac{1}{\pi}\la\la T(z)\bar{T}(\w)X\ra\ra$, which is a holomorphic quadratic differential for $\Ga$ in variable $z$ and an anti-holomorphic quadratic differential for $\Ga$ in variable $w$.
On the other hand, the two-point correlation function $\la\la T(z)T(w)X\ra\ra$ is symmetric in $z$ and $w$, so it can not be represented by a $(2,0)$-form on  $\mathfrak{S}_{g}$.
Theorem \ref{TT-equation} expresses  $\la\la T(z)T(w)X\ra\ra$  as an application of a ``symmetrized $(1,0)$-differential'' $\pa_{s}$ to $\la\la T(z)X\ra\ra$.
It is defined in Section \ref{TT-VF-VF} as follows: if $\theta=\sum_{i=1}^{d}a_{i}dt_{i}$ is a $(1,0)$-form on $\mathfrak{S}_{g}$, then 
$\pa_{s}\theta =\sum_{i,j=1}^{d}\frac{\pa a_{j}}{\pa t_{i}} dt_{i}\otimes_{s} dt_{j}$, where $\otimes_{s}$ stands for the symmetrized tensor product. This explains why Theorem \ref{TT-equation}, which is a statement about second partial derivatives of a certain formal function on $\mathfrak{S}_{g}$ rather than a statement about differential forms, looks complicated when compared with Theorem \ref{T-barT-equation}.  The actual proof of  Theorems \ref{TT-equation} and \ref{T-barT-equation} is based on the computation of $\pa\la\la T(z)X\ra\ra$ and $\bar{\pa}\la\la T(z)X\ra\ra$ in all orders of the perturbative expansion. Again, at the classical level, the corresponding result was proved in \cite{ZT2} and \cite{LTT}, and the major computation is at the one-loop level. It is based on Theorem \ref{one-point}, variational formulas in Section \ref{VF-propagator}, the explicit formula for the kernel $P(z,w)$, and repeated application of the Stokes' theorem. 

In Section \ref{MG} we show that Theorem \ref{TT-equation} at $w\rightarrow z$ agrees precisely with the BPZ conformal Ward identity with two insertions of the $(2,0)$ component of the stress-energy tensor, where $c=\frac{12}{\hbar}+1$. This proves that the quantum Liouville theory in the background field formalism is conformal with the central charge $c$. Using one result of Zograf \cite{Zo}, we interpret Theorems \ref{one-point}, \ref{TT-equation} and \ref{T-barT-equation} in terms of complex geometry of the projective line bundle $\cE_{c}=\lambda_{H}^{c/2}$ over the moduli space $\mathfrak{M}_{g}$, where $\lambda_{H}$ is the Hodge line bundle. This agrees with (and clarifies) the Friedan-Shenker ``modular geometry'' approach to conformal field theory \cite{FS}.

We conclude the paper with two appendices. In Appendix A, we show that the one-loop contribution to Theorem \ref{T-barT-equation} gives the Belavin-Knizhnik theorem \cite{BK} for the case of Laplace operators acting on quadratic differentials on Riemann surfaces. In Appendix B we show how to obtain the stress-energy tensor from the Liouville action functional for the Schottky global coordinate. The corresponding result --- Lemma \ref{ste-action} --- follows from the proof of Theorem 1 in \cite{ZT2}.
\vspace{0.75mm}

\noindent
\textbf{Acknowledgments.} The first author was partially supported by the NSF grant DMS-0204628.

\section{Classical Liouville theory}\label{LAF} 
Let $X$ be a compact Riemann surface of genus $g>1$,  and let
$\{U_{\alpha}\}_{\alpha\in A}$ be a complex-analytic atlas  on $X$ with charts $U_{\al}$, local coordinates $z_\al:U_\al\rightarrow \CC$, and transition functions
$f_{\al\be}: z_{\be}(U_\al\cap U_\be)\rightarrow z_{\al}(U_\al\cap U_\be)$. Denote
by $\cC\cM(X)$ the space (actually a cone) of smooth conformal
metrics on $X$. Every metric $ds^2\in \cC\cM(X)$ is a collection
$\left\{e^{\varphi_\alpha}|dz_\alpha|^2\right\}_{\al\in A}$, where the functions
$\varphi_\alpha\in C^\infty(z_{\al}(U_\al),\RR)$ satisfy
\begin{equation} \label{glue}
\varphi_\alpha\circ f_{\alpha\beta} + \log |f_{\alpha\beta}'|^{2}
=\varphi_\beta\quad\text{on}\quad z_{\be}(U_\al\cap U_\be).
\end{equation}
According to the uniformization theorem, $X$ has a unique conformal metric of the constant Gaussian curvature $-1$, called hyperbolic metric. The corresponding functions $\varphi_\al$
on $z_{\al}(U_\al)$ satisfy the so-called Liouville equation,
\begin{equation} \label{Liouville}
\frac{\pa^2\varphi_\al}{\pa z_\al \pa\bar{z}_\al}
=\frac{1}{2}\,e^{\varphi_\al}.
\end{equation} 

The Lagrangian formulation of the classical Liouville field theory is based on the action functional $S: \cC\cM(X)\rightarrow\RR$, characterized by the property that its unique critical point is the hyperbolic metric on $X$, and the corresponding Euler-Lagrange equation is the Liouville equation. Classical Liouville field theory 
is \emph{conformally invariant}. This fundamental property is a manifestation of the fact that the ``Liouville field'' $e^{\vp}=\{e^{\vp_{\alpha}}\}_{\alpha\in A}$, as it follows from the transformation law \eqref{glue}, is a
$(1,1)$-tensor on $X$. 
For the two-dimensional classical field theory conformal invariance implies that the corresponding stress-energy tensor is traceless (see \cite{BPZ}).
In this section we recall the definition of the action functional for the Liouville theory, introduce the stress-energy tensor and describe its properties. In Appendix B we show how to derive the stress-energy tensor from the action functional.
\subsection{Liouville action functional}
It is well-known (see \cite{ZT2} and the discussion in \cite{LTT}) that a rigorous definition of the
Liouville action functional on a genus $g>1$ Riemann surface is a nontrivial issue.
This is due to the fact that the Liouville field $e^{\vp}$ is a conformal metric on $X$ rather than a function, so that a naive Dirichlet type functional is not well-defined as an integral of a $(1,1)$-form over $X$ when $g>1$. In \cite{ZT2}, this problem was solved by using a global coordinate on
$X$ given by the Schottky uniformization.  In \cite{LTT}, we were able to tackle this problem when a Riemann surface $X$ is equipped with a global coordinate provided by the uniformization of $X$ by a rather general class of Kleinian groups. 
Here by a global coordinate on a Riemann surface $X$ we understand the complex-analytic
covering $J:\Omega\rightarrow X$ of $X$ by a plane domain $\Omega\subset\hat{\C}=\CC\cup\{\infty\}$, such that the corresponding group of deck transformations $\Gamma$ is a Kleinian group with the invariant component $\Omega$.
For the purposes of this paper, it will be sufficient to consider
global coordinates on $X$ given by the Schottky and quasi-Fuchsian uniformizations.
\subsubsection{Schottky uniformization}\label{LAF-CRS-S} 
Marked Riemann surface is a compact Riemann surface $X$ of genus $g>1$ equipped with a canonical system of generators $a_{1},\dots,a_{g}, b_{1},\dots,b_{g}$ of the fundamental group $\pi_{1}(X,x_{0})$ (defined up to an inner automorphism). 
Schottky uniformization of a marked compact Riemann surface $X$ of genus $g$ is a complex-analytic isomorphism $X\simeq\Gamma\bk\Omega$, where
$\Gamma$ is a marked Schottky group --- a strictly loxodromic freely generated Kleinian group with a choice of free generators $\gamma_1, \ldots, \gamma_g\in \PSL(2,\C)$ and with the domain of discontinuity $\Omega$. As an abstract group, $\Gamma\simeq\pi_{1}(X,x_{0})/\mathcal{N}$, where $\mathcal{N}$ is the smallest normal subgroup in $\pi_{1}(X,x_{0})$ containing $a_{1},\dots,a_{g}$, and marked generators $\gamma_{1},\dots,\gamma_{g}$ correspond to the cosets $b_{1}\mathcal{N},\dots,b_{g}\mathcal{N}$.
The holomorphic covering map $J_{S} : \Omega\rightarrow X$ provides a marked Riemann surface $X$ with the Schottky global coordinate. 
It is always assumed that $\Gamma$ is normalized, i.e., the attracting and repelling fixed points of $\gamma_1$ are
$0$ and $\infty$, and the attracting fixed point of $\gamma_2$ is $1$.
The space $\cC\cM(X)$ is identified with the affine subspace of
$C^{\infty}(\Omega, \R)$ consisting of functions $\vp$ satisfying condition
\begin{align}\label{covariant2}
\varphi\circ \gamma + \log |\gamma'|^2 = \varphi,\quad 
\gamma\in \Gamma.
\end{align}
According to \cite{ZT2} (see also \cite{LTT} for the cohomological interpretation), the Liouville action functional $S: \cC\cM(X) \rightarrow \R$ is defined by the following formula,
\begin{equation} \label{action-Schottky}
S(\varphi) = \frac{i}{2}\iint\limits_F \omega[\varphi]+
\frac{i}{2}\sum_{k=2}^g \oint_{C_k} \theta_{\gamma_k^{-1}}[\varphi]+
4\pi\sum_{k=2}^g \log|c(\s_k)|^2.
\end{equation}
Here 
\begin{align*}
\omega[\varphi]&=\left(\left|\varphi_{z}\right|^2
+ e^{\varphi}\right)dz\wedge
d\bar{z},\\
\theta_{\gamma^{-1}}[\varphi] &= \left(\varphi -
\frac{1}{2}\log|\gamma'|^2\right) \left(\frac{\gamma''}{\gamma'}
dz - \frac{\ov{\gamma''}}{\ov{\gamma'}}d\z\right),
\end{align*}
where the subscript $z$ stands for the partial derivative,
$c(\gamma)=c$ for $\gamma= \ma{a}{b}{c}{d}$,
and $F \subset \Omega$ is a fundamental domain for the marked Schottky group $\Gamma$ --- a region bounded by $2g$ non-intersecting smooth Jordan
curves $C_1, C_1^{\prime}, \ldots, C_g,C_g^{\prime}$, satisfying $C_k^{\prime}=-\gamma_k(C_k),\, k=1,\dots,g$. 

The Liouville action functional satisfies the property 
\begin{equation} \label{difference}
S(\vp +\chi)-S(\vp)=\iint\limits_{X}(e^{-\vp}|\chi_{z}|^{2} +e^{\chi}+K\chi -1)e^{\vp}d^{2}z,
\end{equation}
for all $ds^{2}=e^{\vp}|dz|^{2}\in\cC\cM(X)$ and $\chi\in C^{\infty}(X,\R)$, where $K=-2e^{-\vp}\vp_{z\z}$ is the Gaussian curvature of the metric $ds^{2}$, and 
$e^{\vp}d^{2}z =e^{\vp}dx\wedge dy$, $z=x+iy,$ 
is the corresponding area form on $X$ (see \cite{ZT2} and \cite[Lemma 2.1]{LTT}).
It follows from \eqref{difference} that $S$ has a unique non-degenerate critical point given by the hyperbolic metric on $X$. We will denote the corresponding solution of the Liouville equation by $\vp_{cl}$ and, using the physics terminology, will call the corresponding critical value of $S$ the \emph{classical action} $S_{cl}$. We have for $\chi\in C^{\infty}(X,\R)$,
\begin{equation} \label{difference-qf}
S(\vp_{cl} +\chi) - S_{cl}=\iint\limits_{X}(e^{-\vp_{cl}}|\chi_{z}|^{2} +e^{\chi}-\chi -1)e^{\vp_{cl}}d^{2}z.
\end{equation} 

The classical action $S_{cl}$ for varying Riemann surfaces defines a function on the Schottky space $\mathfrak{S}_{g}$, and $-S_{cl}$ is a \Ka potential for the Weil-Petersson metric on $\mathfrak{S}_{g}$ \cite{ZT2}. 
\subsubsection{Fuchsian and quasi-Fuchsian uniformizations}\label{LAF-CRS-F} 
The Fuchsian uniformi\- zation of a compact Riemann
surface $X$ of genus $g>1$ is a complex-analytic
isomorphism $X\simeq \Gamma\bk\U$,
where $\Gamma$ is a torsion-free, strictly hyperbolic Fuchsian
group, and $\U$ is the upper half-plane. Equivalently, the Fuchsian uniformization is a
holomorphic covering $J_{F} : \U \rightarrow X$, with the group of deck transformations $\Gamma\simeq\pi_{1}(X,x_{0})$. It
equips the Riemann surface $X$ with the Fuchsian global coordinate,
and the space $\cC\cM(X)$ is identified with
the affine subspace of $C^{\infty}(\U, \R)$ consisting of
functions $\varphi$ satisfying condition \eqref{covariant2}.

The Liouville action functional $S: \cC\cM(X) \rightarrow
\R$ is defined explicitly by the formula similar to \eqref{action-Schottky}. It is based on the homological algebra machinery associated 
with the $\Gamma$-action on $\up$, developed in \cite{AT}, and we refer to \cite{LTT} for the details. 
As in the Schottky case, the action functional $S$ has a unique non-degenerate critical point given by the hyperbolic metric on $X$, and satisfies property \eqref{difference-qf}. It is an easy computation (see \cite[Corollary 2.1]{LTT}) that $S_{cl}=4\pi(2g-2)$ --- twice the hyperbolic area of $X$.

To describe the quasi-Fuchsian uniformization of $X$, fix a Riemann surface $Y$ of the same genus as $X$ but with the opposite orientation. According to the Bers' simultaneous uniformization theorem, there exists a quasi-Fuchsian group $\Gamma$ with the
domain of discontinuity $\Omega\subset\hat{\C}=\C\cup\{\infty\}$, such that $X\sqcup Y\simeq\Gamma\bk\Omega$.  
The group $\Gamma$ is unique up to a conjugation in $\PSL(2,\C)$ if $X$ and $Y$ are marked Riemann surfaces, and domain $\Omega$ consists of two disjoint components $\Omega_1$ and $\Omega_2$, which cover the Riemann surfaces $X$ and $Y$ respectively. The covering $J_{QF}: \Omega_{1}\rightarrow X$ defines a quasi-Fuchsian global coordinate on $X$ (which depends on $Y$). 

The definition of the Liouville action functional on the space $\cC\cM(X\sqcup Y)$ of conformal metrics on $X\sqcup Y$ is a generalization of the Fuchsian case. We refer to  \cite{LTT} for the explicit representation and details. Here we just emphasize that the action functional on $\cC\cM(X\sqcup Y)$ satisfies property \eqref{difference} and has a unique non-degenerate critical point, given by the hyperbolic metric on $X\sqcup Y$. Moreover, the choice of the hyperbolic metric on $Y$ defines the embedding $\cC\cM(X)\hookrightarrow\cC\cM(X\sqcup Y)$, and the restriction of the action functional to $\cC\cM(X)$ is the Liouiville action functional $S$ for the quasi-Fuchsian global coordinate on $X$,
which satisfies property \eqref{difference-qf}. Corresponding classical action $S_{cl}$ depends non-trivially on $X$, and for varying $X$ (and fixed $Y$) defines a function on the \Te space $\mathfrak{T}_{g}$ of marked compact Riemann surfaces of genus $g>1$. It is proved in \cite{LTT} that the function $-S_{cl}$ is a \Ka potential for the 
Weil-Petersson metric on $\mathfrak{T}_{g}$.  

\subsection{The stress-energy tensor}
The stress-energy tensor is associated with local deformations of classical fields --- conformal metrics on $X$, and is defined by corresponding variational derivatives of the action functional (see
Appendix B for details). For the classical Liouville theory,  its $(2,0)$ and $(0,2)$ components are given by
\begin{align*}
T(\varphi)  =  \varphi_{zz} - \frac{1}{2}
\varphi_z^2\quad\text{and}\quad
\bar{T}(\varphi)  =\ov{T(\vp)}=
\varphi_{\z\z} - \frac{1}{2} \varphi_{\z}^2.
\end{align*}
Here $\vp$ is a Liouville field --- a function $\vp\in C^{\infty}(\Omega,\R)$ satisfying transformation law \eqref{covariant2}.
It follows from \eqref{covariant2} that the function $T(z)=T(\vp)(z)$ on $\Omega$ satisfies
$$T\circ\gamma\,(\gamma')^{2} = T, 
\quad\gamma\in\Gamma,$$
i.e., is a quadratic differential for $\Gamma$. Corresponding \emph{classical stress-energy tensor} $T_{cl}=T(\vp_{cl})$ satisfies the ``conservation law''
$$\partial_{\z} T_{cl}=0,$$
i.e., is a holomorphic quadratic differential for $\Gamma$. This property expresses the fact that the stress-energy tensor for the classical Liouville theory is traceless. The same result holds for Fuchsian and quasi-Fuchsian global coordinates as well and,  in particular, for the Fuchsian case $T_{cl}=0$. 
In this form the stress-energy tensor $T$ for the Liouville theory was introduced by Poincar\'{e}  \cite{Poin} more than a hundred years ago in his proof of the uniformization theorem for Riemann surfaces which uses the Liouville equation. 

The stress-energy tensor $T$ has the following geometric interpretation. For every $ds^{2}=e^{\vp}|dz|^{2}=\{e^{\vp_{\alpha}}|dz_{\alpha}|^{2}\}_{\alpha\in A}\in\cC\cM(X)$ define the following functions on $z_{\al}(U_{\alpha})$,
\begin{equation} \label{STE}
T_{\alpha}(\varphi)  =  \pa^{2}_{z_{\al}}\varphi_{\al} -\tfrac{1}{2}(\pa_{z_{\al}}\varphi_{\al})^{2}
\quad\text{and}\quad\,
\bar{T}_{\alpha}(\varphi)  =\bar{\pa}^{2}_{z_{\al}}\varphi_{\al} -\tfrac{1}{2}(\bar\pa_{z_{\al}}\varphi_{\al})^{2}.
\end{equation}
It follows from \eqref{glue} that on every $z_{\be}(U_{\alpha}\cap U_{\beta})$,
\begin{equation} \label{proj-conn}
T_{\beta}=T_{\alpha}\circ f_{\alpha\beta}\,(f_{\alpha\beta}^{\prime})^{2} +\mathcal{S}(f_{\alpha\beta}),
\end{equation}
where
\begin{align*}
\mathcal{S}(f) =\frac{f'''}{f'} -
\frac{3}{2}\left(\frac{f''}{f'}\right)^2
\end{align*}
is the Schwarzian derivative of a holomorphic function $f$. By definition, collection
$T(\vp)=\{T_{\alpha}(\vp)\}_{\alpha\in A}$ satisfying \eqref{proj-conn} is a non-holomorphic projective connection on $X$, and it follows from the Liouville equation that $T_{cl}$ is a holomorphic projective connection. Since the hyperbolic metric $e^{\vp_{cl}}|dz|^{2}$ is a push-forward of the Poincar\'{e} metric on $\up$ by the covering map $J_{F}:\up\rightarrow X$, a simple computation gives
$T_{cl}=\{\mathcal{S}_{z_{\alpha}}(J_{F}^{-1})\}_{\alpha\in A}$. 
Using the property $\mathcal{S}(\s) = 0$ for all $\s\in\PSL(2,\CC)$, and the Caley identity
\begin{align*}
\mathcal{S}(f \circ g) & = \mathcal{S}(f) \circ g\;(g')^2 + \mathcal{S}(g),
\end{align*}
it is easy to verify directly
that $\mathcal{S}_{z_{\alpha}}(J_{F}^{-1})$  are well-defined functions
on $z_{\al}(U_{\al})$, which satisfy \eqref{proj-conn}. Slightly abusing notations, we will write $T_{cl}=\mathcal{S}(J_{F}^{-1})$.
 
Let $z_{K}$ be a global coordinate on $X$ given by the covering
$J_{K}:\Omega_{K}\rightarrow X$, and
let $e^{\vp_{K}}|dz_{K}|^{2}=J_{K}^{\ast}(ds^{2})$ be the pull-back of $ds^{2}=e^{\vp}|dz|^{2}\in\cC\cM(X)$ by $J_{K}$. 
From \eqref{STE} we obtain
\begin{equation} \label{push-ste}
T(\vp) =T(\vp_{K})\circ J_{K}^{-1}\,(J_{K}^{-1})_{z}^{2} +\mathcal{S}(J_{K}^{-1}),
\end{equation}
where $z$ is a local coordinate on $X$.
Thus the push-forward to $X$ of the quadratic differential $T(\vp_{K})$ on $\Omega_{K}$ is not a quadratic differential on $X$, but  a projective connection. The stress-energy tensor also behaves like a projective connection under changes of global coordinates. Namely, consider the following commutative diagram
$$
\begin{CD}
\up@>J>>\Omega\\
@VJ_{F}VV @VVJ_{S}V\\
X@>\mathrm{id}>>X
\end{CD}
$$
where $J=J_{S}^{-1}\circ J_{F}$ describes the relation between Fuchsian and Schottky global coordinates.  
Denoting by $e^{\vp_{F}}|dz_{F}|^{2}$ and $e^{\vp_{S}}|dz_{S}|^{2}$ the pull-backs of $ds^{2}=e^{\vp}|dz|^{2}\in\cC\cM(X)$ by the mappings $J_{F}$ and $J_{S}$ respectively, we obtain
\begin{align} \label{push-ste-FS}
T(\vp_{S}) & =T(\vp_{F})\circ J^{-1}\,(J_{z}^{-1})^{2} +\mathcal{S}(J^{-1}). 
\end{align}
In particular, $T_{cl}=\mathcal{S}(J^{-1})$.
The same formula  \eqref{push-ste-FS} holds if we replace $J_{S}$ by a covering $J_{K}:\Omega_{K}\rightarrow X$ associated with any global coordinate $z_{K}$ on $X$, e.g., by $J_{K}=J_{QF}$. 

For every $\chi\in C^{\infty}(X)$ set $\chi_{\al}=\chi\circ z^{-1}_{\al}$ and let $q_{\al}=(\pa_{z_{\al}}\chi_{\al})^{2}\in C^{\infty}(z_{\al}(U_{\al}))$. On every $z_{\be}(U_{\alpha}\cap U_{\beta})$ these functions satisfy
$$q_{\beta}=q_{\al}\circ f_{\al\beta}\,(f_{\al\beta}')^{2},$$
so that the collection $q=\{q_{\al}\}_{\al\in A}$ is a quadratic differential on $X$. If $z$ is a local coordinate on $X$, then $q=\chi_{z}^{2}$. Now let $\mathcal{D}_{\al}$ be the following second order differential operator acting on functions on $z_{\al}(U_{\al})$,
$$\mathcal{D}_{\al}  =
e^{\vp_{cl}}\circ \pa_{z_{\al}}\circ e^{-\vp_{cl}}\circ \pa_{z_{\al}}
= \pa^{2}_{z_{\al}}-(\pa_{z_{\al}}\vp_{cl})\pa_{z_{\al}}.$$
It follows from \eqref{glue} that for every $\chi\in C^{\infty}(X)$ the collection $\mathcal{D}\chi =\{\mathcal{D}_{\al}\chi_{\al}\}_{\al\in A}$ is also a quadratic differential\footnote{This is true for every $\vp$ satisfying \eqref{glue}.} on $X$. If $z$ is a local coordinate on $X$, then
$$\mathcal{D}_{z}\chi=e^{\vp_{cl}}(e^{-\vp_{cl}}\chi_{z})_{z}=\chi_{zz}-(\pa_{z}\vp_{cl})\chi_{z}.$$
For every $e^{\vp}|dz|^{2}\in\cC\cM(X)$ setting $\vp=\vp_{cl} +\chi$, where $\chi\in C^{\infty}(X,\RR)$, we get
\begin{equation} \label{ste-difference}
T(\vp)
=T_{cl} +\mathcal{D}_{z}\chi -\tfrac{1}{2}\chi_{z}^{2}.
\end{equation}
Quadratic differentials $\mathcal{D}_{z}\chi -\tfrac{1}{2}\chi_{z}^{2}$ for
$\chi\in C^{\infty}(X,\RR)$ describe ``fluctuations'' around the classical stress-energy tensor $T_{cl}$.

\section{Quantum Liouville theory}\label{QLT}
Here we formulate quantum Liouville theory using the Feynman functional integral formalism. The space  $\cC\!\cM(X)$ of conformal metrics on $X$ is an infinite-dimensional Fr\'{e}chet manifold with a natural Riemannian metric defined by
\begin{align*} \left\Vert \delta
\varphi\right\Vert^2 = \iint\limits_X |\delta \varphi|^2 e^{\varphi} d^2z, \quad \delta\vp\in T_{\vp}\cC\!\cM(X)\simeq C^{\infty}(X,\RR).
\end{align*}
Assuming that the metric $\Vert\cdot\Vert^{2}$ gives rise to the ``volume element'' $\cD\vp$ and choosing a global coordinate on $X$ (Schottky, or quasi-Fuchsian), we define the partition function $\la X\ra$ ---  ``expectation value'' of the Riemann surface $X$, by the following functional integral
\begin{align} \label{X}
\la X\ra = \underset{\cC\cM(X)}{\pmb{\int}}\!
e^{-\frac{1}{2\pi\hbar}S(\vp)}\;\cD\vp.
\end{align}
Here the dimensionless parameter $\hbar>0$ plays the role of the Planck constant. For varying $X$ the partition function $\la X\ra$ gives rise to a real-valued function on the corresponding deformation space, Schottky space $\mathfrak{S}_{g}$, or \Te space $\mathfrak{T}_{g}$, defined in
Section \ref{TT-VF-TS}. 
The correlation functions of multi-local fields $\mathcal{O}$ --- functionals on $\cC\!\cM(X)$ which depend on the values of $\vp$ at finitely many points on $X$, are defined by  
\begin{align} \label{O}
\la \mathcal{O}X\ra = \underset{\cC\cM(X)}{\pmb{\int}}\!\mathcal{O}(\vp)
e^{-\frac{1}{2\pi\hbar}S(\vp)}\;\cD\vp.
\end{align}
For $\mathcal{O}=\prod_{i=1}^{k} T(\vp)(z_i) \prod_{j=1}^{l}
\bar{T}(\vp)(\bar{w}_j)$, where $z$ is a global coordinate on $X$, we get multi-point correlation functions with insertions of the stress-energy tensor. Correlation functions $\la \prod_{i=1}^{k} T(\vp)(z_i) \prod_{j=1}^{l}\bar{T}(\vp)(\bar{w}_j)X\ra$ are tensors of type $(2,0)$ in $z_{1},\dots,z_{k}$, and tensors
of type $(0,2)$ in $w_{1},\dots,w_{l}$, and are symmetric with respect to these two groups of variables.  

Here we do not attempt to give a rigorous mathematical definition of functional integrals \eqref{X} and \eqref{O}\footnote{This would require rigorous definition of the probability measure on the space of distributions on $\cC\cM(X)$, similar to what has been done in constructive quantum field theory in two dimensions \cite{simon, glimm-jaffe}.}. 
Instead, we define  \eqref{X} and \eqref{O} perturbatively using \emph{background field formalism} --- the expansion around the critical point of the action, i.e., around the classical solution $\vp_{cl}$. The result is a formal power series in $\hbar$ with coefficients given by the \emph{Feynman rules}. The combinatorics of the perturbative expansion in QFT is well-established (see, e.g., 
\cite{ramon, witten} and \cite{kazhdan} for mathematically oriented exposition). Here we describe the formal power series in $\hbar$ for partition and correlation functions, and give a rigorous regularization scheme for 
the coefficients of these series.
\subsection{Feynman rules for the partition function}\label{QFT-FR}
Let $\Delta_{0}$ be the Laplace operator of the hyperbolic metric acting on functions on $X$,
$$\Delta_{0}=-e^{-\vp_{cl}}\frac{\del^{2}}{\del z_{\al}\del\z_{\al}}\quad\text{on}\quad z_{\al}(U_{\al}).$$
The Laplacian $\Delta_{0}$ on a compact Riemann surface $X$ is a positive, elliptic operator. Let 
$$G=\tfrac{1}{2}(\Delta_{0} +\tfrac{1}{2})^{-1}$$
be one-half of the resolvent of $\Delta_{0}$ at the point $-\tfrac{1}{2}$. It is well-known that $G$ is an integral operator with a point-wise positive  
kernel $G(P,Q)$, which is a smooth function on $X\times X$, except  for the diagonal $P=Q$, where it has a logarithmic singularity. The function 
$G(P,Q)$ is called the \emph{propagator} of the quantum Liouville theory. We describe its properties in Section 5.

The perturbative expansion of Feynman integrals is based on Wick's theorem --- the following formula for Gaussian integrals,
\begin{gather} \label{Gauss}
\underset{C^{\infty}(X,\RR)}{\pmb{\int}}\chi(P_{1})\dots\chi(P_{n})e^{-\tfrac{1}{2}\iint\limits_{X}\chi(\Delta_{0}+\tfrac{1}{2})\chi\,e^{\vp_{cl}} d^{2}z}\cD\chi \\
=\begin{cases}  0, & \text{if $n$ is odd},\\
\frac{D}{\left(\tfrac{n}{2}\right)!}\sum_{\sigma \in S_n} G(P_{\sigma(1)},
P_{\sigma(2)}) \ldots G(P_{\sigma(n-1)},
P_{\sigma(n)}),& \text{if $n$ is even}.
\end{cases} \nonumber
\end{gather}
Here $P_{1},\dots,P_{n}$ are distinct points on $X$, $D^{-2}=\det(\Delta_{0}+\tfrac{1}{2})$ is the zeta-function regularized determinant of $\Delta_{0} +\tfrac{1}{2}$, and $S_{n}$ is the permutation group on $n$ elements. The integration measure $\cD\chi$ is defined as the volume form of the Riemannian metric
$$\Vert\chi\Vert^{2}=\iint\limits_{X}|\chi|^{2}e^{\vp_{cl}}d^{2}z$$
on $C^{\infty}(X,\R)$.
Effectively, formula \eqref{Gauss} is a definition of the Gaussian measure $\cD\chi$ (see, e.g., \cite{glimm-jaffe}).

To use \eqref{Gauss}, we expand $\varphi$ around
the critical point $\varphi_{cl}$,
\begin{align*}
\varphi= \varphi_{cl} +\sqrt{\pi \hbar}\, \chi,
\end{align*}
where $\chi\in C^{\infty}(X,\RR)$. By \eqref{difference-qf} we have
\begin{align*}
S(\varphi_{cl}+\chi) =S_{cl} + \pi\hbar\iint\limits_X \chi(\Delta_{0}+\tfrac{1}{2})\chi e^{\vp_{cl}}d^{2}z 
 +\sum_{n=3}^{\infty} \frac{(\sqrt{\pi\hbar})^{n}}{n!}\iint\limits_X  \chi^n e^{\vp_{cl}}d^2z.
\end{align*}
Substituting this expansion into \eqref{X} and using $\cD\vp=\cD\chi$ (which may be considered as a perturbative definition of $\cD\vp$), we obtain
\begin{align} \label{pre-graph}
\la X\ra & = e^{-\frac{1}{2\pi\hbar}S_{cl}}\;\sum_{\mathbf{m}}c_{\mathbf{m}} \!\!\!\underset{C^{\infty}(X,\RR)}{\pmb{\int}}\,
\prod_{n=1}^{\infty}\left(
\iint\limits_X  \chi^n e^{\vp_{cl}}d^{2}z\right)^{m_n}\!\! e^{-\Vert\chi\Vert^{2}_{2}} 
\cD\chi,
\end{align}
where 
$$\Vert\chi\Vert^{2}_{2}=\frac{1}{2}\iint\limits_{X}\chi(\Delta_{0}+\tfrac{1}{2})\chi\,e^{\vp_{cl}} d^{2}z$$
is essentially a Sobolev norm square of a function $\chi$. The
summation in \eqref{pre-graph} goes over all multi-indices $\mathbf{m}=(m_{1},m_{2},\dots)$, such that $m_{n}\geq 0,\,m_{1}=m_{2}=0$, $m_{n}=0$ for $n>N$ for some $N>0$,
and
\begin{align*}
c_{\mathbf{m}}=\frac{(\pi\hbar)^{\frac{|\tilde{\mathbf{m}}|}{2}}}{(-2)^{|\mathbf{m}|}\mathbf{m}! \mathbf{n}^{\mathbf{m}}!},
\end{align*}
where 
\begin{alignat*}{2}
|\mathbf{m}| & =\sum_{n=1}^{\infty}m_{n},&\qquad 
|\tilde{\mathbf{m}}| & =\sum_{n=1}^{\infty}(n-2)m_{n},\\
\mathbf{m}!& =\prod_{n=1}^{\infty}m_{n}!, &\qquad\mathbf{n}^{\mathbf{m}}! & =\prod_{n=1}^{\infty}(n!)^{m_{n}}.
\end{alignat*}

Using  \eqref{Gauss} and standard combinatorics of functional integration (see, e.g, \cite{kazhdan, witten}), it is easy to convert \eqref{pre-graph} into the following sum over Feynman diagrams,
\begin{gather} \label{X-graph-1}
\langle X\rangle = De^{-\frac{1}{2\pi\hbar}S_{cl}}\left(1
+\sum_{\Upsilon\in\mathcal{G}_{\geq 3}}
 (-1)^{|V(\Upsilon)|} (2\pi
\hbar)^{-\chi(\Upsilon)}\frac{W_{\Upsilon}(X)}{|\mathrm{Aut}\Upsilon|}\right).
\end{gather}
Here $\mathcal{G}_{\geq 3}$ is a set of graphs $\Upsilon$ with all vertices of valency $\geq 3$. For $\Upsilon\in\mathcal{G}_{\geq 3}$ 
$V(\Upsilon)$ and $E(\Upsilon)$ are, respectively, the set of vertices\footnote{By definition, the set $V(\Upsilon)$ is not empty.} and the set of edges of a graph $\Upsilon$, and $|V(\Up)|=\#V(\Up), \,|E(\Up)|=\#E(\Up)$. Also,
$|\mathrm{Aut}\Upsilon|$ is the order of the group of automorphisms
of $\Upsilon$, and   
$$\chi(\Up)=|V(\Up)|-|E(\Up)|=m-|L(\Up)|$$
is the Euler characteristic of $\Upsilon$, where $|L(\Up)|$ is the number of loops of $\Up$, and $m$ is the number of connected components of
$\Up$.
The weights $W_{\Up}(X)$ are given by the following formula,
\begin{equation} \label{W-integral}
W_{\Up}(X)=\idotsint\limits_{X^{V}}H(P_{1},\dots, P_{V})\prod_{k=1}^{V}dP_{k}.
\end{equation}
Here $V=|V(\Up)|$, $dP_{k}=e^{\varphi_{cl}(z_{k})}d^{2}z_{k}$ is the area form of the hyperbolic metric on the $k$-th factor in $X^{V}=\underbrace{X\times\dots\times X}_{V \mathrm{times}}$, and
\begin{equation} \label{W-integrand}
H(P_{1},\dots, P_{V})=\prod_{e\in E(\Up)}G\left(P_{v_{0}(e)},P_{v_{1}(e)}\right),
\end{equation}
where $\del e=\{v_{0}(e), v_{1}(e)\}\subset V(\Up)$ are the endpoints of the edge $e\in E(\Up)$, and $G(P,Q)$ is the propagator. 
 
Formulas \eqref{X-graph-1}-\eqref{W-integrand} give a formal definition of the partition function $\la X \ra$. However, for graphs with self-loops, i.e., graphs having edges that start and end at the same vertex, corresponding weights are infinite, since the propagator $G(P,Q)$ diverges at $Q=P$. To make sense of the formal power series expansion \eqref{X-graph-1}, one needs to redefine the propagator at coincident points. It follows from the short-distance behavior of the resolvent kernel in Section \ref{diagonal}, that the following expression 
\begin{equation} \label{G-reg}
G(P,P):=\lim_{Q\rightarrow P}\left(G(P,Q) +\frac{1}{2\pi}\left(\log|z(P)-z(Q)|^{2} +\varphi_{cl}(z(P))\right)\right),
\end{equation}
where $z$ is a local coordinate in the neighborhood $U\subset X$ containing $P$ and $Q$, defines a smooth real-valued function on $X$. It is this ``regularization at the coincident points'' (see \cite{LT1, LT-Varenna}) that we use in \eqref{W-integrand}. 

As it is customary in quantum field theory, we introduce the \emph{free energy}
$$\mathcal{F}_{X} =-\log\la X\ra.$$
It is well-known (see, e.g., \cite{ramon, kazhdan, witten}) that passing from the partition function to free energy results in replacing the sum over all graphs in the expansion \eqref{X-graph-1} by the sum over connected graphs only. 
\begin{definition} \label{free-energy}
The free energy $\mathcal{F}_{X}=-\log\la X\ra$ of the quantum Liouville theory on a compact Riemann surface $X$ in the background field formalism is given by the following formal power series in $\hbar$,
\begin{gather*}
\mathcal{F}_{X} = \frac{1}{2\pi \hbar}S_{cl} + \frac{1}{2} \log\det(\Delta_{0} +\tfrac{1}{2}) 
- \sum_{\Up\in\mathcal{G}^{(c)}_{\geq 3}}
 (-1)^{|V(\Up)|} (2\pi
\hbar)^{-\chi(\Up)}\frac{W_{\Up}(X)}{|\mathrm{Aut}\,\Up|},
\end{gather*}
 where $\mathcal{G}_{\geq 3}^{(c)}$ is a subset of all connected graphs $\Up\in \mathcal{G}_{\geq 3}$.
\end{definition}
\begin{remark} The term of order $\hbar^{-1}$ in $\mathcal{F}_{X}$ represents classical contribution to the free energy. 
The constant in $\hbar$ term is a $1$-loop contribution associated with the circle diagram. By definition, it is equal to one-half of the logarithm of the regularized determinant of the elliptic operator $\Delta_{0}+\tfrac{1}{2}$.  The higher order terms correspond to graphs with loops: the $n$-loop term --- the coefficient in front of $\hbar^{n-1}$ --- is the contribution of all connected graphs with $n$ loops in $\mathcal{G}_{\geq 3}$.  
\end{remark}
\begin{remark} \label{canonical}
It follows from Definition \ref{free-energy} that different choices of global coordinates on $X$ affect only classical contribution to the
free energy. All other terms in the perturbative expansion of $\mathcal{F}_{X}$ are canonical in the sense that they only depend on the hyperbolic metric on $X$ through the resolvent kernel $G(P,Q)$. In what follows it will be convenient, though not really necessary, to consider Schottky, quasi-Fuchsian and Fuchsian global coordinates on $X$.
In Section \ref{MG} we will interpret the free energy in terms of complex geometry of the moduli space $\mathfrak{M}_{g}$.
\end{remark}
\begin{remark} According to \cite{dhoker-phong} and \cite{Sa},  the $1$-loop contribution, up to an additive constant $c_{g}$ depending only on the genus $g$, can be expressed solely in terms of the hyperbolic geometry of $X$ as follows,
$$\log \det(\Delta_0 +\tfrac{1}{2})=\log Z_{X}(2) +c_{g}.$$
Here $Z_{X}(s)$ is the Selberg zeta function of a Riemann surface
$X$, defined for $\mathrm{Re}\,s>1$ by the following absolutely convergent product:
\begin{align*}
Z_{X}(s) = \prod_{\{\ell\}} \prod_{n=0}^{\infty} ( 1 - e^{-(s+n)|\ell|}),
\end{align*}
where $\ell $ runs through the set of all simple closed oriented geodesics on $X$ with respect to the hyperbolic metric, and $|\ell|$ is the length of $\ell$.
\end{remark}
\subsection{Feynman rules for correlation functions}\label{EMT-OCF} 
Let $z$ be a global coordinate on $X$. Here we define the multi-point correlation functions
\begin{align*}
\langle \prod_{i=1}^{k} T(z_i) \prod_{j=1}^{l}
\bar{T}(\bar{w}_j)X\rangle =
 \underset{\cC\cM(X)}{\pmb{\int}}\;\prod_{i=1}^{k} T(\vp)(z_i)
\prod_{j=1}^{l} \bar{T}(\vp)(\bar{w}_j)
e^{-\frac{1}{2\pi\hbar}S(\vp)}\;\cD\vp
\end{align*}
as formal power series in $\hbar$. 
It will be convenient to replace
$T(\vp)$ and $\bar{T}(\vp)$ by $\tfrac{1}{\hbar}T(\vp)$ and
$\tfrac{1}{\hbar}\bar{T}(\vp)$ respectively, and in what follows we will always use this normalization.
As in the definition
of the partition function $\la X\ra$, we use the
substitution $\varphi=\varphi_{cl} +\sqrt{\pi \hbar}\, \chi$. It follows from
\eqref{ste-difference} that 
\begin{align*}
T(\varphi) =T_{cl}
+ {\sqrt{\frac{\pi}{\hbar}}}\,\mathcal{D}_{z}\chi-\frac{\pi}{2}\chi_z^2.
\end{align*}
Using \eqref{Gauss} we get that Feynman diagrams for
$\langle \prod_{i=1}^{k} T(z_i) \prod_{j=1}^{l}
\bar{T}(\bar{w}_j)X\rangle $ are labeled graphs with $k+l$ vertices with valencies $1$ and $2$ carrying the labels $z_1,\ldots, z_k$, $\bar{w}_1, \ldots, \bar{w}_l$, and with all other vertices of valency $\geq 3$. In order to sum only over connected graphs, we introduce irreducible correlation functions,
\begin{align*}
\langle\langle T(z)X\rangle\rangle &=\frac{\langle
T(z)X\rangle}{\langle X\rangle},\hspace{1cm}\langle\langle
\bar{T}(\bar{z})X\rangle\rangle =\frac{\langle
\bar{T}(\bar{z})X\rangle}{\langle X\rangle},\\
\langle\la T(z)T(w)X\ra\ra&=\frac{\la T(z)T(w)X\ra}{\la X\ra}-
\la\la T(z)X\ra\ra \la\la T(w)X\ra\ra,\\
\langle\la T(z)\bar{T}(\bar{w})X\ra\ra&=\frac{\la
T(z)\bar{T}(\bar{w})X\ra}{\la X\ra}- \la\la T(z)X\ra\ra \la\la
\bar{T}(\bar{w})X\ra\ra.
\end{align*}
In general, denoting $I=\prod_{i=1}^{k} T(z_i)
\prod_{j=1}^{l} \bar{T}(\bar{w}_j)$, we have inductively (see, e.g., \cite{LT-Varenna})
\begin{align*}
\langle\la IX\ra\ra = \frac{\la IX\ra}{\la X \ra} -\sum_{r}
\sum_{I=I_1 \ldots I_r} \la\la I_1X\ra\ra\ldots\la\la
I_r X\ra\ra,
\end{align*}
where the sum goes over all representations of $I$ as a product of
$I_1, \dots, I_r$ corresponding to the partition of the set $\{z_{1},\dots,z_{k}, \bar{w}_{1},\dots,\bar{w}_{l}\}$ into $r$ non-empty subsets.  

The perturbative expansion for the one-point irreducible correlation function with the $(2,0)$ component of the stress-energy tensor is given by
\begin{align} \label{ste-1-point}
\la \la T(z)X\ra\ra =T_{cl}(z) +
2\pi\!\!\sum_{\Upsilon\in\mathcal{G}^{(c)}_{\{z\}}}
  (-1)^{|V(\Up)|+\vep_{1}(\Up)}(2\pi \hbar)^{-\chi(\Up)}\frac{W_{\Upsilon}(X; z)}{|\mathrm{Aut}\Upsilon|}.
\end{align}
Here $\mathcal{G}^{(c)}_{\{z\}}$ is the set of all connected graphs $\Upsilon$ with a single vertex  of valency $1$ or $2$ with the label 
$z$ and all other vertices of valency $\geq 3$, and $\vep_{1}(\Up)=1$ or $0$ depending on whether the labeled vertex has valency $1$ or $2$.  Also $\mathrm{Aut}\Upsilon$ is the group of automorphisms of $\Upsilon$ which preserves the labeling, and   
$\chi(\Up)=|V(\Up)|-|E(\Up)|=1-|L(\Up)|$
is the Euler characteristic of $\Upsilon$, where $|L(\Up)|$ is the number of loops of $\Up$.
The weights $W_{\Up}(X; z)$ are given by the following formula,
\begin{equation} \label{W-integral-ste-1}
W_{\Up}(X; z)=\idotsint\limits_{X^{\tilde{V}}}H(P_{1},\dots, P_{V})\prod_{k=1}^{\tilde{V}}dP_{k},
\end{equation}
where $\tilde{V}=|V(\Up)|-1$, $P_{V}=z$ corresponds to vertex $v$ labeled by $z$, and
\begin{equation} \label{W-integrand-ste-1}
H(P_{1},\dots, P_{V})=\prod_{e\in E(\Up)}\mathcal{D}_{v_{0}(e)}\mathcal{D}_{v_{1}(e)}G\left(P_{v_{0}(e)},P_{v_{1}(e)}\right).
\end{equation}
Here 
$$
\begin{cases}
\mathcal{D}_{v}=\id & \text{if vertex $v$ has valency $\geq 3$}, \\ 
\mathcal{D}_{v}=\mathcal{D}_{z}  & \text{if vertex $v$ has valency $1$},\\ 
\mathcal{D}_{v}=\del_{z} & \text{if vertex $v$ has valency $2$}.
\end{cases}
$$
For self-loops we use the same regularization \eqref{G-reg} at coincident points, except
for the case when vertex $v$ has $n$ self-loops and is connected by an edge to a labeled vertex of valency $1$. In this case we replace one of the factors in $G(P_{v},P_{v})^{n}$ by $G(P_{v},P_{v})+\tfrac{n}{2\pi}$. The contribution of the vertex of valency $1$ with label $z$ to \eqref{W-integrand-ste-1} is the factor $\mathcal{D}_zG(z,w)$, which has a singularity of the form $1/(z-w)^2$ as $w\rightarrow z$. The same singularity arises when the labeled vertex of valency $2$ is attached to a self-loop.
The corresponding integrals in \eqref{W-integral-ste-1} are understood in the principal value sense.

To complete this definition, we need to assign the weight to a tadpole graph (see Fig. 1)

\begin{center}
\epsfig{file=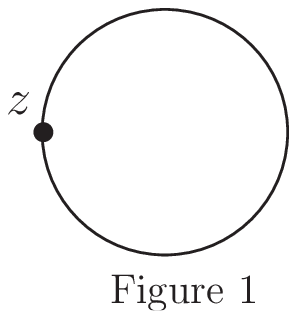, %
height=3cm}
\end{center}

\noindent
According to \cite{LT-Varenna}, we define
\begin{align} \label{tadpole} H(z)& =\pa_z\pa_zG(z,z)\\
&:=\lim_{w\rightarrow z} \left( \pa_z\pa_{w}
G(z,w)
+\frac{1}{2\pi}\left(\frac{1}{(z-w)^2}-\frac{1}{2}e^{\varphi_{cl}(z)}\frac{\z-\bar{w}}{z-w}\right)\right). \nonumber
\end{align}
We will show in Section 5.1 that $H(z)$ is a quadratic differential for a group $\Gamma$ corresponding to a global coordinate $z$, which behaves like a projective connection under changes of global coordinates. Analyzing formula \eqref{ste-1-point} it is easy to see that
$\la\la T(z)X\ra\ra$ is a formal power series in $\hbar$ whose coefficients are quadratic differentials for $\Gamma$. Except the classical and $1$-loop terms, their push-forwards to $X$ are quadratic differentials. In other words, higher loop terms in $\la\la T(z)X\ra\ra$ do not depend on the choice of global coordinate on $X$.

The correlation function $\la \la
\bar{T}(\bar{z})X\ra\ra$ is defined similarly, with $z$ replaced
by $\bar{z}$. The perturbative definition of multi-point correlation functions is the following.  
\begin{definition} \label{multi-point} Irreducible multi-point correlation functions with insertions of the stress-energy tensor are given by the following formal power series in $\hbar$,
\begin{gather*} \la\la\prod_{i=1}^{k} T(z_i) \prod_{j=1}^{l}
 \bar{T}(\bar{w}_j)X\ra\ra
 =(2\pi)^{k+l}\!\!\sum_{\Upsilon\in\mathcal{G}^{(c)}_{\mathcal{I}}}
  (-1)^{|V(\Up)|+\vep_{1}(\Up)}(2\pi \hbar)^{-\chi(\Up)}\frac{W_{\Up}(X;\mathcal{I})}{|\mathrm{Aut}\Upsilon|}.
\end{gather*}
Here $\mathcal{G}^{(c)}_{\mathcal{I}}$ is the set of all connected graphs $\Up$ with $k+l$ vertices of valencies $1$ or $2$ labeled by the set
$\mathcal{I}=\{z_{1},\dots,z_{k},\bar{w}_{1},\dots,\bar{w}_{l}\}$ and with all other vertices of valency $\geq 3$, $\vep_{1}(\Up)$ is the number of vertices of valency $1$, and $\mathrm{Aut}\Up$ is the group of automorphisms of $\Up$ which preserve the labeling.
The weights $W_{\Up}(X;\mathcal{I})$ are given by
\begin{equation*} 
W_{\Up}(X;\mathcal{I})=\idotsint\limits_{X^{\tilde{V}}}H(P_{1},\dots, P_{V})\prod_{k=1}^{\tilde{V}}dP_{k},
\end{equation*}
where $\tilde{V}=|V(\Up)|-k-l$. Here for a labeled vertex $v$ the point $P_{v}$ is  from the labels set $\mathcal{I}$, and
\begin{equation*} 
H(P_{1},\dots, P_{\tilde{V}};\mathcal{I})=\prod_{e\in E(\Up)}\mathcal{D}_{v_{0}(e)}\mathcal{D}_{v_{1}(e)}G\left(P_{v_{0}(e)},P_{v_{1}(e)}\right),
\end{equation*}
where
$$
\begin{cases}
\mathcal{D}_{v}=\id & \text{if $v$ has valency $\geq 3$}, \\ 
\mathcal{D}_{v}=\mathcal{D}_{z_{i}} & \text{if $v$ has valency $1$ and label $z_{i}$},\\ 
\mathcal{D}_{v}=\del_{z_{i}} & \text{if $v$ has valency $2$ and label $z_{i}$}, \\
\mathcal{D}_{v}=\mathcal{D}_{\bar{w}_{j}} & \text{if $v$ has valency $1$ and label $\bar{w}_{j}$},\\ 
\mathcal{D}_{v}=\del_{\bar{w}_{j}} & \text{if $v$ has valency $2$ and label $\bar{w}_{j}$}. 
\end{cases}
$$
For graphs with self-loops the weights are regularized by \eqref{G-reg}, except for the case when a vertex $v$ with $n$ self-loops is connected by an edge to a vertex of valency $1$, in which case one of the factors in $G(P_{v},P_{v})^{n}$ is replaced by $G(P_{v},P_{v}) +\tfrac{n}{2\pi}$. For the cases $k=1, l=0$ and $k=0, l=1$ one should add, correspondingly, the tree-level terms 
$$T_{cl}(z)=\tfrac{1}{\hbar}(\del_{z}^{2}\vp_{cl}-\tfrac{1}{2}(\del_{z}\vp_{cl})^{2})
\quad\text{and}\quad T_{cl}(\bar{w})=\tfrac{1}{\hbar}(\del_{\bar{w}}^{2}\vp_{cl}-\tfrac{1}{2}(\del_{\bar{w}}\vp_{cl})^{2}).$$
The weights for the tadpole graphs --- graphs with a single vertex of valency $2$ labeled by $z$ or $\bar{w}$, are given, correspondingly, by $H(z)$ and $\bar{H}(\w)=\overline{H(w)}$, where $H(z)$ is defined in \eqref{tadpole}.
\end{definition}

It follows from Definition \ref{multi-point} that
$\la\la\prod_{i=1}^{k} T(z_i) \prod_{j=1}^{l} \bar{T}(\bar{w}_j)X\ra\ra$ are symmetric with respect to the variables 
 $z_1,\ldots, z_k$ and $\bar{w}_1, \ldots, \bar{w}_l$ respectively. When $k+l\geq 2$, all coefficients in these formal power series are quadratic differentials for $\Gamma$ in variables $z_{i}$ and $\w_{j}$, whose push-forward to $X$ are quadratic differentials\footnote{Here by a quadratic differential in $\w$ we understand a complex-conjugate of a quadratic differential in $w$.}. In other words, for 
$k+l\geq 2$  correlation functions
$\la\la\prod_{i=1}^{k} T(z_i) \prod_{j=1}^{l} \bar{T}(\bar{w}_j)X\ra\ra$ do not depend on the choice of a global coordinate on $X$. 
We will prove in Sections \ref{VA-OPCF}-\ref{TPCF-TTbar} that these correlation functions are quadratic differentials in $z_{i}$ and $\w_{j}$, which are meromorphic in $z_{i}$ and anti-meromorphic in $w_{j}$, and with only poles at coincident points.

\section{Deformation theory}\label{TT-VF}
For the convenience of the reader, here we present necessary basic facts from deformation theory (see e.g., \cite{A2,B1,Ahl} or our discussion in \cite{LTT}).
\subsection{ Schottky and Teichm\"uller spaces}\label{TT-VF-TS}
Let $\Gamma$ be either a Schottky group, or a quasi-Fuchsian group\footnote{In fact, $\Gamma$ could be any non-elementary, finitely generated Kleinian group.}, with the domain of discontinuity $\Omega$.
Let $\mathcal{A}^{-1,1}(\Gamma)$ be
a space of bounded Beltrami differentials for $\Gamma$ --- the Banach space of $\mu \in L^{\infty}(\C)$ satisfying
\begin{align*}
\mu( \gamma z)
\frac{\ov{\gamma^{\prime}(z)}}{\gamma^{\prime}(z)}= \mu(z)
\quad \text{for}\; \;z\in\Omega, \; \gamma\in \Gamma,
\end{align*}
and let $\mathcal{B}^{-1,1}(\Ga)$ be the open unit ball in
$\mathcal{A}^{-1,1}(\Ga)$ with respect to $\parallel
\cdot\parallel_{\infty}$ norm, 
\begin{equation*}
\parallel \mu \parallel_{\infty} = \sup_{z\in \C} |\mu(z)| <1.
\end{equation*}
For every $\mu\in\mathcal{B}^{-1,1}(\Ga)$
there exists a unique quasiconformal
homeomorphism $f^{\mu}: \hat{\C} \rightarrow \hat{\C}$ satisfying
the Beltrami equation
\begin{equation*}
f^{\mu}_{\z} = \mu f^{\mu}_z
\end{equation*}
and fixing the points $0,1~\text{and}~\infty$. Set $\Ga^{\mu} =
f^{\mu}\circ\Ga \circ(f^{\mu})^{-1}$ and define the deformation
space of $\Gamma$ to be
\begin{equation*}
\D(\Gamma)=\mathcal{B}^{-1,1}(\Ga)/ \thicksim\,,
\end{equation*}
where $\mu \thicksim \nu$ if and only if $f^\mu=f^\nu$ on
$\hat{\CC}\setminus\Omega$, which is equivalent to the condition $f^\mu\circ\s\circ
(f^\mu)^{-1}=f^\nu\circ\s\circ (f^\nu)^{-1}$ for all $\s\in\Ga$.

The deformation space $\mathfrak{D}(\Gamma)$ has a natural structure of a complex manifold, explicitly described as follows (see, e.g., \cite{A2}). Let
$\mathcal{H}^{-1,1}(\Ga)$ be the Hilbert space of Beltrami
differentials for $\Ga$ with the inner product
\begin{equation}\label{pairing}
(\mu_1, \mu_2) 
= \iint\limits_{F} \mu_1(z)
\ov{\mu_2(z)} \rho(z)\,d^2z,
\end{equation}
where $\mu_1, \mu_2 \in \mathcal{H}^{-1,1}(\Ga)$, $F$ is a fundamental domain for $\Ga$ in $\Omega$, and
$\rho=e^{\varphi_{cl}}$ is density of the hyperbolic metric on
$\Omega$. Denote by $\Omega^{-1,1}(\Ga)$ the finite-dimensional
subspace of harmonic Beltrami differentials with respect to the
hyperbolic metric. It consists of $\mu\in\mathcal{H}^{-1,1}(\Ga)$
satisfying
\begin{equation} \label{mu-harm}
\pa_z(\rho \mu) = 0.
\end{equation}
The complex vector space $\Omega^{-1,1}(\Ga)$ is identified with
the holomorphic tangent space to $\D(\Ga)$ at the origin. Choose a
basis $\mu_1,\dots, \mu_d$ for $\Omega^{-1,1}(\Ga)$, set $\mu =
t_1 \mu_1 + \dots + t_d \mu_d$, where $t_{1},\dots, t_{d}$ are such that $\Vert\mu\Vert_{\infty}<1$, and let $f^\mu$ be the
normalized solution of the Beltrami equation. Then the
correspondence $(t_1, \dots, t_d) \mapsto \Ga^{\mu}= f^{\mu}
\circ \Ga \circ (f^{\mu})^{-1}$ defines complex coordinates in a
neighborhood of the origin in $\D(\Ga)$, called the Bers coordinates.
The holomorphic cotangent space to $\D(\Ga)$ at the origin can be
naturally identified with the vector space $\Omega^{2,0}(\Ga)$ of
holomorphic quadratic differentials
--- holomorphic functions $q$ on $\Omega$ satisfying
\begin{equation*}
q(\gamma z)\gamma'(z)^2=q(z),\quad \gamma\in\Ga.
\end{equation*}
The pairing between holomorphic cotangent and tangent spaces to
$\D(\Ga)$ at the origin is given by
\begin{equation*}
(q, \mu)  = \iint\limits_{F} q(z)\mu(z)\,d^2z.
\end{equation*}
Corresponding anti-holomorphic cotangent space to $\D(\Ga)$ at $\Ga$ is identified with the vector space $\Omega^{0,2}(\Ga)=\ov{\Omega^{2,0}(\Ga)}$ of anti-holomorphic quadratic differentials. 

There is a natural isomorphism $\Phi^\mu$ between the deformation
spaces $\D(\Ga)$ and $\D(\Ga^{\mu})$, which maps $\Ga^{\nu} \in
\D(\Ga)$ to $(\Ga^{\mu})^{\lambda} \in \D(\Ga^{\mu})$, where, in
accordance with $f^\nu=f^\lambda \circ f^\mu$,
\[
\lambda = \left(\frac{\nu - \mu}{ 1- \nu \bar{\mu}}
\frac{f_z^{\mu}}{\bar{f}_{\z}^{\mu}} \right)\circ (f^{\mu})^{-1}.
\]
Isomorphism $\Phi^\mu$ allows us to identify the holomorphic
tangent space to $\D(\Ga)$ at $\Ga^{\mu}$ with the complex vector
space $\Omega^{-1,1}(\Ga^{\mu})$, and holomorphic cotangent space
to $\D(\Ga)$ at $\Ga^{\mu}$ with the complex vector space
$\Omega^{2,0}(\Ga^{\mu})$. It also allows us to introduce Bers
coordinates in the neighborhood of $\Ga^\mu$ in $\D(\Ga)$, and to
show directly that these coordinates transform
complex-analytically. For the de Rham differential $d$ on
$\D(\Ga)$ we denote by $d=\pa + \bar\pa$ decomposition into
$(1,0)$ and $(0,1)$ components.

The differential of isomorphism $\Phi^\mu: \D(\Ga)\simeq
\D(\Ga^{\mu})$ at $\nu=\mu$ is given by the linear map $D^{\mu} :
\Omega^{-1,1}(\Ga) \rightarrow \Omega^{-1,1} (\Ga^{\mu})$,
\begin{equation*}
\nu \mapsto D^{\mu} \nu = P_{-1,1}^{\mu}\left[ \left(\frac{\nu}{
1- |\mu|^2} \frac{f_z^{\mu}}{\bar{f}_{\z}^{\mu}} \right)\circ
(f^{\mu})^{-1} \right],
\end{equation*}
where $P_{-1,1}^{\mu}$ is orthogonal projection from
$\mathcal{H}^{-1,1}(\Ga^{\mu})$ to $\Omega^{-1,1}(\Ga^{\mu})$. The
map $D^\mu$ allows us to extend a tangent vector $\nu$ at the origin
of $\D(\Ga)$ to vector field $\frac{\pa}{\pa t_{\nu}}$ defined on the
coordinate neighborhood of the origin,
\begin{equation*}
\left.\frac{\pa}{\pa t_{\nu}}\right|_{\Ga^{\mu}}  = D^{\mu}
\nu\in\Omega^{-1,1}(\Ga^{\mu}).
\end{equation*}

The scalar product \eqref{pairing} in $\Omega^{-1,1}(\Ga^{\mu})$
defines a Hermitian metric on the deformation space $\D(\Ga)$.
This metric is called the Weil-Petersson metric and it is
K\"ahler. We denote its symplectic form by $\omega_{\wpm}$,
\begin{equation*}
\omega_{\wpm} \left.\left(\frac{\pa}{\pa t_\mu}, \frac{\pa}{\pa
\bar{t}_\nu}\right) \right|_{\Ga^{\lambda}} = \frac{i}{2}
(D^{\lambda} \mu, D^{\lambda} \nu),\quad \mu, \nu
\in \Omega^{-1,1}(\Ga).
\end{equation*}

When $\Gamma$ is a Schottky group of marked compact Riemann surface $X$ of genus $g>1$, the deformation space $\D(\Gamma)$ is the Schottky space $\mathfrak{S}_g$ of $X$.
When $\Gamma$ is a Fuchsian group, such that components $\U$
and $\lo$ cover, respectively, the marked compact Riemann surface $X$ of genus $g>1$ and its mirror image $\bar{X}$, the
deformation space $\D(\Gamma)$ contains the Teichm\"uller space
$\mathfrak{T}_{g}$ of $X$ as a submanifold. Namely, the choice of the subspace of $\mathcal{B}^{-1,1}(\Gamma)$ consisting of $\mu$ with the reflection symmetry:
\[
\mu(\z)=\ov{\mu(z)},\quad z\in \C ,\] 
gives rise to the real-analytic embedding $\mathfrak{T}_{g}\hookrightarrow\D(\Ga)$. In this case,
every group $\Gamma^{\mu}$ is a Fuchsian group. The choice of a
subspace of $\mathcal{B}^{-1,1}(\Gamma)$ consisting of $\mu$
that are identically $0$ on the lower half-plane $\lo$, gives rise to 
the complex-analytic embedding $\mathfrak{T}_{g}\hookrightarrow\D(\Ga)$. In this case
$\Gamma^{\mu}$ is, in general, a quasi-Fuchsian group. Its domain of
discontinuity has two components $\Omega^{\mu}_1=f^{\mu}(\U)$ and
$\Omega^{\mu}_2=f^{\mu}(\lo)$, covering Riemann surfaces $X^{\mu}\simeq\Ga^{\mu}\bk\Omega^{\mu}_{1}$ and
$\bar{X}$ respectively. The Weil-Petersson metric on $\mathfrak{D}(\Gamma)$ restricts to the
Weil-Petersson metric of the Teichm\"uller space $\mathfrak{T}_{g}$. It is given by the same formula \eqref{pairing}, where now $F$ is a fundamental domain for $\Ga$ in $\Omega_{1}$.
We continue to denote by  $d=\pa + \bar\pa$ decomposition of de Rham differential $d$ on $\mathfrak{T}_{g}$ into $(1,0)$ and $(0,1)$ components. The Teichm\"uller space $\mathfrak{T}_{g}$ is the universal covering space for the moduli space $\mathfrak{M}_{g}$ of compact Riemann surfaces of genus $g>1$. 

\subsection{Formal geometry on deformation spaces} \label{formal-geo}
A formal function on a deformation space $\D(\Ga)$ is an element of $C^{\infty}(\D(\Ga))((\hbar))$ --- a formal power series in $\hbar$ with coefficients in $C^{\infty}(\D(\Ga))$. Correspondingly, a formal $1$-form on $\D(\Ga)$ is a formal power series in $\hbar$ with coefficients being $1$-forms on $\D(\Ga)$. For a formal function $\mathcal{F}$ on $\D(\Ga)$, $\pa\mathcal{F}$ and
$\bar{\pa}\mathcal{F}$ are formal $(1,0)$ and $(0,1)$ forms on $\D(\Ga)$. For every $t\in\D(\Ga)$ there is an associated Riemann surface $X_{t}\simeq \Gamma_{t}\backslash\Omega_{t}$, and  $\pa\mathcal{F}(t)$ and $\bar{\pa}\mathcal{F}(t)$ are represented by formal holomorphic and anti-holomorphic quadratic differentials for $\Gamma_{t}$. The Riemann surfaces $X_{t}$ form a holomorphic family parameterized by $\D(\Ga)$.

It follows from Definition \ref{free-energy} 
that the free energy $\mathcal{F}_{X}$  gives rise to a formal function $\mathcal{F}$ 
on the Schottky space $\mathfrak{S}_{g}$, or on the Teichm\"uller space $\mathfrak{T}_{g}$,  depending on the choice of a global coordinate on $X$. Namely, to every point $t\in \mathfrak{S}_{g}$ (or $t\in \mathfrak{T}_{g}$) there is an associated Riemann surface $X_{t}$ with Schottky (or quasi-Fuchsian) global coordinate, and $\mathcal{F}(t)=\mathcal{F}_{X_{t}}\in \frac{1}{\hbar}\CC[[\hbar]]$. 
As it was pointed out in Remark \ref{canonical}, actually
$\mathcal{F}-\frac{1}{2\pi\hbar}S_{cl}$ is a formal function on
the moduli space $\mathfrak{M}_{g}$.

It also follows from Definition \ref{multi-point} that for every $t\in\mathfrak{S}_{g}$ (or $t\in\mathfrak{S}_{g}$) one-point correlation functions $\la\la T(z)X_{t}\ra\ra$ and $\la\la \bar{T}(\z)X_{t}\ra\ra$ are formal quadratic differentials for $\Ga_{t}$ in $z$ and $\z$. We will show in Section \ref{VA-OPCF} that they are holomorphic and anti-holomorphic formal quadratic differentials that represent (up to an additional one-loop term) formal $(1,0)$ and $(0,1)$-forms $\pa\mathcal{F}$ and $\bar{\pa}\mathcal{F}$, where $\mathcal{F}$ is a formal function associated with free energy. Multi-point correlation functions admit similar interpretation. In Sections \ref{TPCF-TT} and \ref{TPCF-TTbar} we present all details for cases $\la\la T(z)T(w)X_{t}\ra\ra$ and $\la\la T(z)\bar{T}(\w)X_{t}\ra\ra$.

\subsection{Variational formulas}\label{TT-VF-VF}
Here we collect variational formulas needed in the next section. For $\mu\in\mathcal{A}^{-1,1}(\Ga)$ quasiconformal mappings $f^{\vep\mu}$ depend holomorphically on $\vep$ in some neighborhood of $0\in\CC$. Setting
\begin{equation*}
\dot{f} = \left.\frac{d}{d\vep} \right|_{\vep=0} f^{\vep\mu},
\end{equation*}
we obtain
\begin{equation} \label{qc-var}
\dot{f}(z)=-\frac{1}{\pi}\iint\limits_{\CC}\frac{z(z-1)\mu(w)}{(w-z)w(w-1)}\,
d^2w\quad\text{and}\quad\dot{f}_{\z}=\mu.
\end{equation}

A tensor of type $(l,m)$ for $\Ga$, where $l$ and $m$ are integers, is a
$C^\infty$-function $\theta$ on $\Omega$ satisfying
\begin{equation*}
\theta\circ \gamma\,(\gamma')^l
(\ov{\gamma^\prime})^m=\theta,\quad \gamma\in\Ga.
\end{equation*}
Let $\theta^{\vep}$ be a smooth family of tensors of type $(l,m)$
for the holomorphic family $\Ga^{\vep\mu}$, where it is always assumed that $\mu \in \Omega^{-1,1} (\Ga)$, and $\vep
\in \C$ is sufficiently small. Set
\begin{equation*}
(f^{\vep\mu})^* (\theta^{\vep}) = \theta^{\vep} \circ f^{\vep\mu}\,
(f_z^{\vep\mu})^l (\bar{f}_{\z}^{\vep\mu})^m,
\end{equation*}
which is a tensor of type $(l,m)$ for $\Ga$ --- a pull-back of the
tensor $\theta^\vep$ for $\Ga^{\vep\mu}$ by $f^{\vep\mu}$. The Lie derivatives of the
family $\theta^{\vep}$ along the vector fields $\frac{\pa}{\pa
t_{\mu}}$ and $\frac{\pa}{\pa \bar{t}_{\mu}}$ are defined in the
standard way,
\begin{equation*}
L_{\mu} \theta = \left.\frac{\pa}{\pa \vep} \right|_{\vep=0}
(f^{\vep\mu})^* (\theta^{\vep})\quad\text{and}\quad L_{\bar{\mu}} \theta =
\left.\frac{\pa}{\pa \bar{\vep}} \right|_{\vep=0}(f^{\vep\mu})^*
(\theta^{\vep}).
\end{equation*}
When $\theta$ is a function on $\D(\Ga)$ --- a tensor of type
$(0,0)$, Lie derivatives reduce to directional derivatives
$L_\mu\theta =
(\pa\theta)(\mu)$ and $L_{\bar\mu}\theta=(\bar\pa\theta)(\bar\mu)$.

When $\theta$ is a $(1,0)$-form
on $\D(\Ga)$, i.e., a family $\theta^{\vep}$ of holomorphic quadratic differentials for $\Ga^{\vep\mu}$, we have 
\begin{equation*}
\pa\theta=\sum_{i=1}^{d}dt_{i}\wedge L_{\mu_{i}}\theta \quad\text{and}\quad \bar{\pa}\theta=\sum_{i=1}^{d}d\bar{t}_{i}\wedge L_{\bar{\mu}_{i}}\theta,
\end{equation*}
where $dt_{1},\dots,dt_{d}$ is the basis for $\Omega^{2,0}(\Ga)$ dual to the basis $\mu_{1},\dots,\mu_{d}$ for $\Omega^{-1,1}(\Ga)$. 

Let $\mathcal{H}^{2,0}(\Ga)$ be the Hilbert space of quadratic differentials for $\Ga$ with the inner product
\begin{equation} \label{inner-quadratic}
(q_{1},q_{2})=\iint\limits_{F}q_{1}(z)\ov{q_{2}(z)}\rho(z)^{-1}d^{2}z,
\end{equation}
where $F$ is a fundamental domain for $\Ga$ in $\Omega$,
and let $P: \mathcal{H}^{2,0}(\Ga)\rightarrow \Omega^{2,0}(\Ga)$ be the orthogonal projection onto the subspace of holomorphic quadratic differentials. It immediately follows from Stokes' theorem that the quadratic differential $\mathcal{D}_{z}h$, where $h$ is a smooth $\Ga$-automorphic function on $\Omega$, is orthogonal to $\Omega^{2,0}(\Ga)$.
 
Now suppose that for a $(1,0)$-form $\theta$ on $\D(\Ga)$,
\begin{equation*}
L_{\mu}\theta(z)=\iint\limits_{F}Q(z,w)\mu(w)d^{2}w,
\end{equation*}
where $Q(z,w)$ is a smooth quadratic differential for $\Ga$ in $z$ and $w$.
Identification of holomorphic tangent and cotangent spaces to $\D(\Ga)$ with $\Omega^{-1,1}(\Ga)$ and
$\Omega^{2,0}(\Ga)$ in Section \ref{TT-VF-TS}, allows us to identify the $(2,0)$-form $\pa\theta$ on $\D(\Ga)$ at the point $\Ga$ with 
$P_{w}P_{z}Q(w,z)-P_{z}P_{w}Q(z,w)$ --- the holomorphic quadratic differential for $\Ga$ in $z$ and $w$,
where $P_{z}$ and $P_{w}$ are corresponding projection operators with respect to $z$
and $w$. Explicitly, 
$$\pa\theta\left(\frac{\pa}{\pa t_{i}},\frac{\pa}{\pa t_{j}}\right)=
-\iint\limits_{F}\iint\limits_{F}Q(z,w)(\mu_{i}(z)\mu_{j}(w)-\mu_{j}(z)\mu_{i}(w))d^{2}zd^{2}w.$$
The symmetric part
$\frac{1}{2}(P_{z}P_{w}Q(z,w)+P_{w}P_{z}Q(w,z))$
corresponds to the following $2$-tensor 
\begin{equation} \label{d-sym}
\pa_{s}\theta = \sum_{i=1}^{d}L_{\mu_{i}}\theta\otimes_{s}dt_{i},
\end{equation}
defined in a coordinate chart of the origin in $\D(\Ga)$, where $\otimes_{s}$ stands for the symmetrized tensor product, $dt_{i}\otimes_{s} dt_{j}=\frac{1}{2}(dt_{i}\otimes dt_{j}+dt_{j}\otimes dt_{i})$.
Explicitly, 
$$\pa_{s}\theta\left(\frac{\pa}{\pa t_{i}},\frac{\pa}{\pa t_{j}}\right)=\frac{1}{2}
\iint\limits_{F}\iint\limits_{F}Q(z,w)(\mu_{i}(z)\mu_{j}(w)+\mu_{j}(z)\mu_{i}(w))d^{2}zd^{2}w.$$
If $S$ is a function on $\D(\Ga)$, then, by definition, $\pa_{s}S=\pa S$, and we have
$$\pa_{s}(\pa_{s} S)=\sum_{i,j=1}^{d}\frac{\pa^{2}S}{\pa t_{i}\pa t_{j}}dt_{i}\otimes_{s}dt_{j},$$
while $\pa(\pa S)=0$.
In general, $\pa_{s}$ can be extended to a linear operator mapping
$(n,0)$-symmetric tensors on a coordinate chart of the origin in $\D(\Ga)$ to symmetric $(n+1,0)$-tensors.

For the Lie derivatives of vector fields $\nu^{\ep \mu} = D^{\ep
\mu} \nu$ we get \cite{Wol}
\begin{equation} \label{var-nu}
L_\mu\nu=0 \quad\text{and}\quad L_{\bar\mu}\nu=-\bar{\pa}\rho^{-1}\bar{\pa} (\Delta_0+\tfrac{1}{2})^{-1}
(\mu\bar{\nu}).
\end{equation}

For every $\Ga^\mu\in \D(\Ga)$, the density $\rho^\mu$ of the
hyperbolic metric on $\Omega^\mu$ is a $(1,1)$-tensor for
$\Ga^\mu$. Lie derivatives of the smooth family of $(1,1)$-tensors
$\rho$ parameterized by $\D(\Ga)$ are given by the following formulas: 
\begin{align}
L_\mu\rho & =L_{\bar\mu}\rho=0, \label{Ahlfors-metric}\\
L_\mu
L_{\bar{\nu}}\rho & =\tfrac{1}{2}\rho(\Delta_0+\tfrac{1}{2})^{-1}(\mu\bar{\nu}), \label{Wolpert-metric}
\end{align}
belonging, correspondingly, to Ahlfors \cite{Ahl} and Wolpert \cite{Wol}.
Since $f^{\vep\mu}$ depends holomorphically on $\vep$, we get from
\eqref{Ahlfors-metric}
\begin{equation} \label{Ahlfors1}
\left.\frac{\pa}{\pa \vep} \right|_{\vep=0} \left(
\varphi^{\vep\mu}_{cl} \circ f^{\vep\mu} \right) = - \dot{f}_z.
\end{equation}

For every $\Ga^{\mu}\in\D(\Ga)$ let $\Omega_1^{\mu}$ be the component of domain of discontinuity of $\Gamma^{\mu}$ such that $X^{\mu}\simeq\Ga^{\mu}\bk\Omega_{1}^{\mu}$,
and let $J_{\mu} : \U\rightarrow
\Omega_1^{\mu}$ be the corresponding covering map. The assignment $\vartheta(\mu)=\mathcal{S}(J_{\mu}^{-1})\in\Omega^{2,0}(\Ga^{\mu})$ defines a $(1,0)$-form on
$\D(\Ga)$. It was proved in \cite{ZT2} for the case $\D(\Ga)=\mathfrak{S}_{g}$, and in \cite{LTT} for the case $\D(\Ga)=\mathfrak{T}_{g}$, that
\begin{equation} \label{var-classical-action}
\pa S_{cl}=2\vartheta,
\end{equation} 
as well as 
\begin{align}
L_{\mu}
\vartheta(z)&=\frac{6}{\pi}
\iint\limits_{\CC} \frac{\mu(w)}{(z-w)^4} \;\; d^2w \label{LS},\\
L_{\bar{\mu}}
\vartheta(z)&=-\frac{1}{2}\rho\,\ov{\mu(z)}. \label{BLS}
\end{align}
The integral in \eqref{LS} is understood in the principal value sense, i.e., as a limit
$$\lim_{\vep\rightarrow 0}\iint\limits_{|w-z|\geq\vep}\frac{\mu(w)}{(z-w)^4} \;\; d^2w,$$
which exists for harmonic $\mu$.
Formula \eqref{BLS} is equivalent to 
\begin{equation} \label{wp-classical}
\bar\pa\pa S_{cl}=-2i\omega_{\wpm},
\end{equation}
whereas interpretation of \eqref{LS} in terms of $\pa_{s}^{2}S_{cl}$ will be given in Section \ref{TPCF-TT-CL}.
\begin{remark} \label{vanishing}
When $\Ga$ is a Fuchsian group and $\left.\mu\,\right\vert_{\lo}=0$, then according to \cite[Lemma 2.12]{TT-I}
$L_{\mu}\vartheta(z)=0$. 
\end{remark}
Lie derivatives of a family of linear operators $\mathcal{O}^{\vep}$ 
mapping tensors of type $(l,m)$ to tensors of type $(l',m')$ are defined by the formulas
\begin{align*}
L_{\mu} \mathcal{O} & = \left.\frac{\pa}{\pa \vep} \right|_{\vep=0}
(f^{\vep\mu})^*\mathcal{O}^{\vep}((f^{\vep\mu})^*)^{-1}, \\
L_{\bar{\mu}} \mathcal{O} & =
\left.\frac{\pa}{\pa \bar{\vep}} \right|_{\vep=0}(f^{\vep\mu})^*
\mathcal{O}^{\vep}((f^{\vep\mu})^*)^{-1},
\end{align*}
and satisfy
\begin{align*}
L_{\mu}(\mathcal{O}(\theta))=L_{\mu}\mathcal{O}(\theta)+\mathcal{O}(L_{\mu}\theta),\quad L_{\bar\mu}(\mathcal{O}(\theta))=L_{\bar\mu}\mathcal{O}(\theta)+\mathcal{O}(L_{\bar\mu}\theta).
\end{align*}
For the families of $\bar{\pa}$-operators mapping $(n,0)$-tensors
to $(n,1)$-tensors, and of $\pa$-operators mapping $(0,1)$-tensors to
$(1,1)$-tensors, we have the following formulas,
\begin{alignat}{3}
L_{\mu}\bar{\pa}& =-\mu\rho^{n}\pa\rho^{-n},&\qquad L_{\bar{\mu}}\bar{\pa}& =0 \label{L-dbar}\\
L_{\mu}\pa&=0,&\qquad L_{\bar{\mu}}\pa&=-\bar\pa\bar\mu. \label{L-d}
\end{alignat}
Hence for the operator $\Delta_0+\tfrac{1}{2}=-\rho^{-1}\pa\bar{\pa}+\tfrac{1}{2}$ we get
\begin{align} \label{var-Laplace}
L_{\mu}(\Delta_0+\tfrac{1}{2})=\rho^{-1}\pa\mu\pa,\qquad L_{\bar{\mu}}
(\Delta_0+\tfrac{1}{2})=\rho^{-1}\bar{\pa}\bar{\mu}\bar{\pa}.
\end{align}
The Lie derivatives of operators $\pa$ and
$\mathcal{D}_z=\rho\pa\rho^{-1}\pa$ as operators from
functions to $(1,0)$ and $(2,0)$ tensors respectively, are given by
\begin{alignat}{3}
L_{\mu}\pa&=0, &\qquad L_{\mu}\mathcal{D}_z&=0, \label{var-del}\\
 L_{\bar{\mu}}\pa & =
-\bar{\mu}\bar{\pa},&\qquad
L_{\bar{\mu}}\mathcal{D}_z & =-\bar{\mu}\bar{\pa}\pa-\rho\pa\rho^{-1}\bar{\mu}\bar{\pa}. \label{var-D}
\end{alignat}

It follows from \eqref{L-dbar} that for a family $\theta^{\vep}(z)$ of holomorphic quadratic differentials $L_{\bar{\mu}}\theta(z)$ is holomorphic in $z$, whereas $L_{\mu}\theta(z)$, in general,  is not.
Thus if 
\begin{equation*}
L_{\bar{\mu}}\theta(z)=\iint\limits_{F}\tilde{Q}(z,w)\ov{\mu(w)}d^{2}w,
\end{equation*}
where $\tilde{Q}(z,w)$ is a smooth quadratic differential for $\Ga$ in $z$ and $\w$, then
the $(1,1)$-form $\bar{\pa}\theta$ on $\D(\Ga)$ at the point $\Ga$ is identified with the quadratic differential $-P_{\bar{w}}\tilde{Q}(z,w)$ for $\Ga$, which is holomorphic in $z$ and anti-holomorphic in $w$. 
Explicitly, 
$$\bar{\pa}\theta\left(\frac{\pa}{\pa t_{i}},\frac{\pa}{\pa \bar{t}_{j}}\right)=
-\iint\limits_{F}\iint\limits_{F}\tilde{Q}(z,w)\mu_{i}(z)\ov{\mu_{j}(w)}d^{2}zd^{2}w.$$

\section{The propagator}\label{PPV}
 The propagator $G(P,Q)$ of quantum Liouville theory --- the integral kernel of the operator  $G=\tfrac{1}{2}(\Delta_{0}+\tfrac{1}{2})^{-1}$, is uniquely characterized by the following properties.
\begin{itemize}
\item[\textbf{P1.}] $G$ is a smooth function on $X\times X\setminus \mathrm{D}$, where $\mathrm{D}$ is the diagonal $P=Q$ in $X\times X$.
\item[\textbf{P2.}] $G$ is symmetric, $G(P,Q) =G(Q,P)$ for $P,Q\in X$.
\item[\textbf{P3.}] For fixed $Q\in X$, $G(P,Q)$ as a function of $P\in X\setminus \{Q\}$ satisfies
\begin{align*}
(\Delta_{0}+\tfrac{1}{2})G =0.
\end{align*}
\item[\textbf{P4.}] For fixed $Q\in X$, the function
\begin{align*}
G(P,Q) + \frac{1}{2\pi}\log|z(P)-z(Q)|^2
\end{align*}
is continuous in some neighborhood $U$ of $Q$, where $z$ is a local coordinate in $U$.
\end{itemize}
It follows from these properties that for every $g\in C^{\infty}(X)$ the function 
$$h(P)=\iint\limits_{X}G(P,Q)g(Q)dQ$$
satisfies the equation
\begin{equation} \label{green-function}
(\Delta_{0}+\tfrac{1}{2})h =\tfrac{1}{2}g.
\end{equation}
In particular,
\begin{equation} \label{integration-G}
\iint\limits_{X}G(P,Q)dQ=1.
\end{equation}

On the upper half-plane $\U$, the kernel for the integral operator $\tfrac{1}{2}(\Delta_0+\tfrac{1}{2})^{-1}$ is given by (see, e.g., \cite{LT-Varenna}) 
\begin{equation} \label{free-propagator}
\mathcal{G}(z,w)=\frac{1}{2\pi}\int_{0}^{1}\frac{t(1-t)}{(t+u)^{2}}dt =\frac{2u+1}{2\pi}\log\frac{u+1}{u}-\frac{1}{\pi},
\end{equation}
where
\begin{align}\label{dist}
u(z,w) =\frac{|z-w|^2}{4\im z\im w}.
\end{align}
The function $\mathcal{G}$ has the property
\begin{equation} \label{invariant}
\mathcal{G}(\sigma z,\sigma w)=\mathcal{G}(z,w),\,\quad\sigma\in\mathrm{PSL}(2,\R).
\end{equation}
It terms of the Fuchsian global coordinate $z$ with the covering map
$J_{F}:\U\rightarrow X$, the propagator $G_{F}(z,w)=G(J_{F}(z), J_{F}(w))$ is given by the method of images,
\begin{equation} \label{method-images}
G_{F}(z,w)= \sum_{\gamma\in \Gamma} \mathcal{G}(z, \gamma
w),
\end{equation}
and it follows from \eqref{invariant} that $G_{F}(z,w)$ is $\Ga$-automorphic in $z$ and $w$,
\begin{equation} \label{G-invariant}
G_{F}(\gamma_{1}z,\gamma_{2}w)=G_{F}(z,w),\,\quad\gamma_{1}, \gamma_{2}\in\Ga.
\end{equation}
If $z_{K}$ is another global coordinate on $X$ with the covering map
$J_{K}:\Omega_{K}\rightarrow X$,
then $G_{K}(z_{K},w_{K})=
G(J_{K}(z_{K}), J_{K}(w_{K}))$
satisfies 
\begin{align} \label{propagator-covers}
G_{K}(z_{K},w_{K})=G_{F}(J^{-1}(z_{K}), J^{-1}(w_{K})),
\end{align}
where $J=J_{K}^{-1}\circ J_{F}$.  
\subsection{Behavior near diagonal and explicit formulas} \label{diagonal}
Here we present basic properties of the propagator of quantum Liouville theory. It is convenient to use the Fuchsian global coordinate  
$z$ on $X$ and to write $G=G_{F}$.

It follows from \eqref{free-propagator}--\eqref{dist} that
\begin{alignat*}{4}
\mathcal{G}(z, w) &=-\frac{1}{2\pi} \log\frac{|z-w|^2}{\im z\im
w} -\frac{1-\log 2}{\pi}+o(1) &
\quad\text{as}&\quad w &\rightarrow z, \\
\intertext{and, therefore,}
G(z,w) &=-\frac{1}{2\pi}
\log|z-w|^2+O(1)&\quad\text{as}&\quad w &\rightarrow z.
\end{alignat*}
Similarly, as $w \rightarrow z$,
\begin{align}
\pa_z\mathcal{G}(z,w) & =-\frac{1}{2\pi}
\frac{w-\z}{(z-w)(z-\z)} 
+o(1), \label{derivative-G-free} \\ 
\pa_z\pa_{w}\mathcal{G}(z,w) &=-\frac{1}{2\pi}
\frac{1}{(z-w)^2}-\frac{1}{\pi}\frac{\z-\w}{z-w}\frac{1}{(z-\z)^{2}}
+o(|z-w|), \label{2-derivative-G-free} \\ 
\pa_z\pa_{\bar{w}}\mathcal{G}(z,w)
&=-\frac{1}{\pi}\,\frac{1}{(z-\z)^2}\log |z-w|^2
+ O(1),\label{z-barz-derivative-G-free}
\end{align}
so that
\begin{align}
\pa_zG(z,w)& =- \frac{1}{2\pi(z-w)}+O(1),  \label{derivative-G}\\ 
\pa_z\pa_{w}G(z,w) &=-\frac{1}{2\pi(z-w)^2} -\frac{1}{\pi}
\frac{\z-\w}{z-w}\frac{1}{(z-\z)^{2}}+O(1), \label{2-derivative-G}\\ 
\pa_z\pa_{\bar{w}}G(z,w) &=-\frac{1}{\pi(z-\z)^2}\log |z-w|^2+O(1).\label{z-barz-derivative-G}
\end{align}

In terms of the Fuchsian global coordinate, the regularization  \eqref{G-reg} at the coincident points is given by
\begin{align} \label{G-z=z}
G(z,z)= \sum_{\substack{
\gamma\in \Gamma\\
\gamma\neq \id }} \mathcal{G}(z, \gamma z)-\frac{1-\log 2}{\pi}
\end{align}
(see \cite{LT1}).
It follows from \eqref{invariant} that $G(z,z)$ is $\Ga$-automorphic,
$$G(\gamma z,\gamma z) = G(z,z),\quad\gamma\in\Ga.$$
Similarly, the regularization \eqref{tadpole} of the tadpole graph $H(z)=\pa_z\pa_zG(z,z)$ is given by
\begin{align} \label{tadpole-F}
H(z)=\sum_{\substack{
\gamma\in \Gamma\\
\gamma\neq \id }} \pa_z\pa_{w}\mathcal{G}(z, \gamma
w)\Bigr\vert_{w=z}
\end{align}
(see \cite{LT2, LT-Varenna}).
It follows from \eqref{invariant} that $H$ is a quadratic differential for $\Ga$, and
\begin{equation} \label{z-derivative-H}
\pa_{z}H(z)=2\sum_{\substack{
\gamma\in \Gamma\\
\gamma\neq \id }} \pa^{2}_z\pa_{w}\mathcal{G}(z, \gamma
w)\Bigr\vert_{w=z}.
\end{equation}
Using property \textbf{P3}, we also obtain
\begin{equation} \label{z-bar-derivative-H}
\pa_{\z}H(z)=\frac{1}{2}\rho(z)\pa_{z}G(z,z).
\end{equation}

It follows from \eqref{propagator-covers} that 
$G(z,z)$ gives rise to a well-defined smooth function $G(P,P)$ on $X$. However,  it follows from \eqref{tadpole} and the classical formula
\begin{gather}  \label{classical-formula} 
\frac{f'(z)f'(w)}{(f(z)-f(w))^{2}}  =\frac{1}{(z-w)^{2}}+ \frac{1}{6}\mathcal{S}(f)(z) +\frac{1}{12}\mathcal{S}(f)'(z)(w-z) + \dots 
\end{gather}
as $w\rightarrow z$, where dots stand for $O(|z-w|^{2})$ term, that
\begin{equation}\label{trans2}
H_{K}(z_{K})=H(J^{-1}(z_{K}))(J^{-1})^{\prime}(z_{K})^2-\frac{1}{12\pi} \mathcal{S}(J^{-1})(z_{K}).
\end{equation}
Hence $H(z)$,  in accordance with \cite{LT2, LT-Varenna}, behaves like ``$-1/12\pi$ of a projective connection'' under changes of global coordinates. 

The following explicit formulas
\begin{equation} \label{DG}
\mathcal{D}_z\mathcal{G}(z,w)
=\frac{1}{2\pi}
\frac{(w-\bar{w})^2}{(z-w)^2(z-\bar{w})^2},
\end{equation}
and
\begin{alignat}{3}
\mathcal{D}_z\pa_{w}\mathcal{G}(z,w)
&=\frac{1}{\pi}\frac{w-\bar{w}}{(z-w)^3(z-\bar{w})},&\qquad
\mathcal{D}_z\mathcal{D}_{w}\mathcal{G}(z,w)
&=\frac{3}{\pi}\frac{1}{(z-w)^4}, \label{DG-cal} \\
\mathcal{D}_z\pa_{\bar{w}}\mathcal{G}(z,w)
&=\frac{1}{\pi}\frac{\bar{w}-w}{(z-w)(z-\bar{w})^3},&\qquad
\mathcal{D}_z\mathcal{D}_{\bar{w}}\mathcal{G}(z,w)
&=\frac{3}{\pi}\frac{1}{(z-\bar{w})^4}, \label{DD-G-cal}
\end{alignat}
give the following asymptotic formulas as $w\rightarrow z$:
\begin{align*}
\mathcal{D}_zG(z,w)
&=\frac{1}{2\pi}\frac{(w-\bar{w})^2}{(z-w)^2(z-\bar{w})^2}+O(1),\\
\mathcal{D}_z\pa_{w}G(z,w)
&=\frac{1}{\pi}\frac{w-\bar{w}}{(z-w)^3(z-\bar{w})}+O(1), \\
\mathcal{D}_z\pa_{\bar{w}}G(z,w)
&=\frac{1}{\pi}\frac{\bar{w}-w}{(z-w)(z-\bar{w})^3}
+O(1),\\
\mathcal{D}_z\mathcal{D}_{w}G(z,w)&=\frac{3}{\pi}\frac{1}{(z-w)^4}
+O(1),
\end{align*}
and explicit formulas
\begin{align} 
\mathcal{D}_z\mathcal{D}_{w}G(z,w) & =\frac{3}{\pi}\sum_{\gamma\in\Ga}\frac{\gamma'(w)^{2}}{(z-\gamma w)^4}, \label{ahlfors-sing}  \\
\mathcal{D}_z\mathcal{D}_{\bar{w}}G(z,w) & =\frac{3}{\pi}\sum_{\gamma\in\Ga}\frac{\gamma'(\w)^{2}}{(z-\gamma\bar{w})^4}. \label{ahlfors}
\end{align}

Since 
\begin{equation} \label{unfolding}
\up=\bigcup_{\gamma\in\Ga}\gamma F,
\end{equation}
where $F$ is a fundamental domain for $\Ga$ in $\up$, we obtain  from \eqref{ahlfors} and the Ahlfors'
classical reproducing formula in \cite[Ch. VI.D., Lemma 2]{A2},
\begin{equation} \label{P-kernel}
(Pq)(z)=4\iint\limits_{F}\mathcal{D}_{z}\mathcal{D}_{\bar{w}}G(z,w)q(w)\rho(w)^{-1}d^{2}w,\quad q\in\mathcal{H}^{2,0}(\Gamma),
\end{equation}
so that $P(z,w)=4\mathcal{D}_z\mathcal{D}_{\bar{w}}G(z,w)$ is an integral kernel of the projection operator $P:\mathcal{H}^{2,0}(\Gamma)\rightarrow\Omega^{2,0}(\Gamma)$.
Representation \eqref{P-kernel} was used  in \cite{LT2, LT-Varenna}, and plays a fundamental role in this paper.

In general,
$\mathcal{D}_{z}\mathcal{D}_{\bar{w}}G(z,w)$, where $z$ and $w$ are  local coordinates on $X$, is a holomorphic quadratic differential on $X$ with respect to the first variable,  and is an anti-holomorphic quadratic differential on $X$ with respect to the second variable. The expression $P(z,w)=4\mathcal{D}_{z}\mathcal{D}_{\bar{w}}G(z,w)$ is an integral kernel of the orthogonal projection operator $P:\mathcal{H}^{2,0}(X)\rightarrow\Omega^{2,0}(X)$, where the  
inner product in $\mathcal{H}^{2,0}(X)$ is defined by using the hyperbolic metric on $X$. 

Similarly, 
$\mathcal{D}_{z}\mathcal{D}_{w}G(z,w)$ is a meromorphic quadratic
differential on $X$ in variables $z$ and $w$, with the fourth order pole at $z=w$. It behaves like a quadratic differential in $z$ and $w$ under a change of global coordinates.
Let $z$ be a global coordinate on $X$. Using \eqref{classical-formula} we get
as $w\rightarrow z$,
\begin{equation} \label{D-short-term}
\mathcal{D}_{z}\mathcal{D}_{w}G(z,w)=
\frac{3}{\pi}\left(\frac{1}{(z-w)^{4}}+\frac{\mathcal{S}(J^{-1})(z)}
{3(z-w)^{2}}-\frac{\mathcal{S}(J^{-1})'(z)}
{6(z-w)}\right) + O(1).
\end{equation}

Finally, 
\begin{equation} \label{R-definition}
R(z,w)=4\rho(z)^{-1}\pa_{\z}\mathcal{D}_{w}G(z,w)
\end{equation}
is a meromorphic quadratic differential on $X$ in $w$ 
with a single simple pole at $w=z$, and is a $(-1,0)$-tensor with respect to $z$. We have 
\begin{equation} \label{asymptotics-R}
R(z,w)=-\frac{1}{\pi(z-w)} + O(1) \quad\text{as}\quad w\rightarrow z,
\end{equation} 
and this expansion does not depend on the choice of local coordinates $z$ and $w$ in the neighborhood of the diagonal in $X\times X$. 
It follows from property \textbf{P3} that for any choice of global coordinate $z$ on $X$, the kernel $R(z,w)$ satisfies the equation
\begin{equation} \label{equation-R}
\pa_{z}R(z,w)+(\pa_{z}\varphi_{cl})(z)R(z,w)=2\mathcal{D}_{w}G(z,w),
\end{equation}
which  implies
\begin{gather} \label{R-zzz}
\pa_{z}^{3}R(z,w)  
=2\mathcal{D}_{z}\mathcal{D}_{w}G(z,w) -(2\pa_{z}R(z,w)+R(z,w)\pa_{z})\mathcal{S}(J^{-1})(z)
\end{gather}
and
\begin{gather}
\mathcal{D}_{z}(R(z,w)\pa_{z}G(z,v)) \label{R-D}\\ 
=2\mathcal{D}_{w}\pa_{z}G(z,w)\pa_{z}G(z,v)+(2\pa_{z}R(z,w) + R(z,w)\pa_{z})\mathcal{D}_{z}G(z,v).\nonumber
\end{gather}
\begin{remark}
The kernel $R(z,w)$ is a Green's function of
$\bar{\pa}$-operator acting on $(-1,0)$-tensors on $X$ --- an integral kernel of the inverse operator  $\bar{\pa}^{-1}$ on the space of trivial Beltrami differentials on $X$, which is an orthogonal complement in $\mathcal{H}^{-1,1}(X)$ to the subspace of harmonic Beltrami differentials. It is also used in the formulation of conformal Ward identities on Riemann surfaces in \cite{EO}. 
\end{remark}
\begin{remark} \label{singular}
Operator $P$ can be also defined on the vector space of quadratic differentials for $\Ga$ which are smooth everywhere on $F$ except at $z=w$, where they have the following asymptotic behavior: 
$$q(z)=\frac{a_{1}}{(z-w)^{4}} +\frac{a_{2}}{(z-w)^{2}} +\frac{a_{3}}{z-w} + O(1)\quad\text{and}\quad \pa_{\z}q(z)=O(1)$$
as $z\rightarrow w$. The integral in \eqref{P-kernel} is understood in the principal value sense, and it follows from the Stokes' theorem, \eqref{D-short-term}, and \eqref{asymptotics-R}-\eqref{R-zzz} that $P(q)\in\Omega^{2,0}(\Ga)$.
\end{remark}
\begin{remark} \label{Lie} For a family of holomorphic quadratic
differentials $\theta^{\vep\mu}(z)$ for $\Gamma^{\vep\mu}$ we have
\begin{align*}
(L_{\mu}\theta)(z)& =\left.\frac{\pa}{\pa\vep}\right |_{\vep=0}
\left(\theta^{\vep\mu}\circ
f^{\vep\mu}(f^{\vep\mu}_z)^2\right)(z)=\left.\frac{\pa}{\pa\vep}\right|_{\vep=0}\theta^{\vep\mu}(z)
\\
&\quad +(\pa_z\theta)(z)\left.\frac{\pa}{\pa\vep}\right|_{\vep=0}f^{\vep\mu}(z)+2\theta(z)
\left.\frac{\pa}{\pa\vep}\right |_{\vep=0}f^{\vep\mu}_{z}(z).
\end{align*}
Suppose that
\begin{align*}
\left.\frac{\pa}{\pa\vep}\right|_{\vep=0}\theta^{\vep\mu}(z)=\iint\limits_F
Q_1(z,w)\mu(w)d^2w,
\end{align*}
where $Q_{1}(z,w)$ is a quadratic differential in $w$.
Then by \eqref{qc-var}
\begin{align*}
(L_{\mu}\theta)(z)=\iint\limits_F Q(z,w)\mu(w)d^2w,
\end{align*}
where
\begin{align*}
Q(z,w)=Q_1(z,w)+(2\pa_z\mathcal{R}(z,w)+\mathcal{R}(z,w)\pa_z)\theta(z),
\end{align*}
and
\begin{align*}
\mathcal{R}(z,w)=-\frac{1}{\pi}\sum_{\gamma\in
\Gamma}\frac{z(z-1)\gamma'(w)^2}{(\gamma w-z)\gamma w (\gamma
w-1)}
\end{align*}
is a meromorphic quadratic differential for $\Ga$ in $w$.  
By the Stokes' theorem, it is easy to prove that
\begin{align*}
P_w\mathcal{R}(z,w)=\mathcal{R}(z,w)+R(z,w).
\end{align*}
Thus we obtain
\begin{gather}
P_w(Q_1)(z,w)+(2\pa_z\mathcal{R}(z,w)+\mathcal{R}(z,w)\pa_z)\theta(z)  \label{Q-Q-1} \\=P_{w}(Q)(z,w) -(2\pa_zR(z,w)+R(z,w)\pa_z)\theta(z), \nonumber
\end{gather}
where $P_{w}(Q_{1})(z,w)$ is holomorphic in $z$. 
\end{remark}
\subsection{Variational formulas} \label{VF-propagator}
Here we collect variational formulas for the propagator $G$, which are
necessary for Sections \ref{VA-OPCF}--\ref{TPCF-TTbar}. 
\begin{lemma} \label{var-propagator} Let $z$ be a global coordinate on $X\simeq \Ga\bk\Omega$, and
$\mu\in\Omega^{-1,1}(\Ga)$. We have the following formulas,
where $F$ is a fundamental domain for $\Ga$ in $\Omega$.
\begin{itemize}
\item[(i)] For $z\neq w$, 
\begin{align*}
L_{\mu}G(z,w)&=2\iint\limits_F \pa_vG(z,v) \pa_v G(v,w)\mu(v)d^2v,\\
L_{\bar{\mu}}G(z,w)&=2\iint\limits_F \pa_{\bar{v}}G(z,v) \pa_{\bar{v}} G(v,w)\ov{\mu(v)}d^2v.
\end{align*}
\item[(ii)] 
\begin{align*}
L_{\mu}G(z,z)&=2\iint\limits_F (\pa_vG(z,v))^{2}\mu(v)d^2v,\\
L_{\bar{\mu}}G(z,z)&=2\iint\limits_F (\pa_{\bar{v}}G(z,v))^{2}\ov{\mu(v)}d^2v.
\end{align*}
\item[(iii)] For $z\neq w$,
\begin{align*}
L_{\mu}\pa_zG(z,w)&=2\iint\limits_F
\pa_z\pa_vG(z,v) \pa_v G(v,w)\mu(v)d^2v,\\
L_{\bar{\mu}}\pa_zG(z,w)&=2\iint\limits_F \pa_z\pa_{\bar{v}}G(z,v) \pa_{\bar{v}}G(v,w)\ov{\mu(v)}d^2v.
\end{align*}
\item[(iv)] For $z\neq w$,
\begin{align*}
L_{\mu}\mathcal{D}_zG(z,w)&=2\iint\limits_F
\mathcal{D}_z\pa_vG(z,v) \pa_v G(v,w)\mu(v)d^2v,\\
L_{\bar{\mu}}\mathcal{D}_zG(z,w)&=2\iint\limits_F
\mathcal{D}_z\pa_{\bar{v}}G(z,v) \pa_{\bar{v}}G(v,w)\ov{\mu(v)}d^2v \\
&\;\;\;\;-\frac{1}{2}\rho(z)\ov{\mu(z)}G(z,w).
\end{align*}
\item[(v)] For $z\neq w$,
\begin{align*}
L_{\mu}\pa_z\pa_{w}G(z,w)&=2\iint\limits_F \pa_z\pa_vG(z,v)
\pa_{w}\pa_v G(v,w)\mu(v)d^2v,\\
L_{\bar{\mu}}\pa_z\pa_{w}G(z,w)&=2\iint\limits_F \pa_z\pa_{\bar{v}}G(z,v) \pa_{w}\pa_{\bar{v}}
G(v,w)\ov{\mu(v)}d^2v.
\end{align*}
\end{itemize}
Both integrals in (ii) and the first integrals in (iii)-(v) are understood in the principal value sense.
\end{lemma}
\begin{proof} 
From the definition of the propagator $G(z,w)$ we obtain
\begin{align*}
\left(L_{\mu}(\Delta_0+\tfrac{1}{2})\right)G(z,w)+(\Delta_0+\tfrac{1}{2})\left(L_{\mu}G(z,w)\right)=0,
\end{align*}
so that using \eqref{var-Laplace} we get for $z\neq w$,
\begin{align*}
L_{\mu}G(z,w)& =-((\Delta_0+\tfrac{1}{2})^{-1} \rho^{-1}\pa\mu\pa)
G(z,w)\\
& =-2\iint\limits_F G(z,v) \pa_{v}\mu(v)\pa_v G(v,w)d^2v\\
&=\;\;2\iint\limits_F \pa_vG(z,v) \pa_v G(v,w)\mu(v)d^2v,
\end{align*}
which proves (i).
Here in the last line we used the Stokes' theorem and elementary fact
\begin{equation} \label{formula}
\oint\limits_{|z|=\vep}\frac{d\bar{z}}{z} =0.
\end{equation}

To prove (ii), it is convenient to use the Fuchsian global coordinate on $X\simeq\Ga\bk\up$.
Using \eqref{Ahlfors1}, we get 
\begin{gather*}
L_{\mu}G(z,z)=\lim_{w\rightarrow z} L_{\mu}\left(G(z,w)
+\frac{1}{2\pi} \left(\log|z-w|^2+\phi_{cl}(z)\right)\right)\\
=\lim_{w\rightarrow z}\left(2\iint\limits_F \pa_vG(z,v) \pa_v
G(v,w)\mu(v)d^2v+\frac{1}{2\pi}\left(\frac{\dot{f}(z)-\dot{f}(w)}{z-w}-\dot{f}_z(z)
\right)\right),
\end{gather*}
where it is easy to justify the interchange of the Lie derivative $L_{\mu}$ and the limit $w\rightarrow z$. Using \eqref{method-images} and \eqref{unfolding} we have
\begin{gather*}
2\iint\limits_F \pa_vG(z,v) \pa_v
G(v,w)\mu(v)d^2v=2\iint\limits_{\U} \pa_v\mathcal{G}(z,v) \pa_v
G(v,w)\mu(v)d^2v\\
=2\iint\limits_{\U} \left(-\frac{1}{2\pi}\frac{1}{v-z}+
h_1(v,z)\right)\left(-\frac{1}{2\pi}\frac{1}{v-w}+
h_2(v,w)\right)\mu(v)d^2v\\
=\frac{1}{2\pi^{2}}\iint\limits_{\U}\Bigl(
\frac{1}{(v-z)(v-w)}-\frac{2\pi}{v-z}\,h_2(v,w)
-\frac{2\pi}{v-w}h_1(v,z)\\
+ 4\pi^{2}h_1(v,z)h_2(v,w)\Bigr)\mu(v)d^2v,
\end{gather*}
where $h_1, h_2$ are bounded functions on $\U\times \U$. The last
three terms are continuous as $w\rightarrow z$. On the other
hand, it follows from \eqref{qc-var} that
\begin{align*}
\frac{\dot{f}(z)-\dot{f}(w)}{z-w}-\dot{f}_z(z)
=-\frac{1}{\pi}\iint\limits_{\U}
\mu(v)\left(\frac{1}{(v-z)(v-w)}-\frac{1}{(v-z)^2}\right)d^2v,
\end{align*}
where the integral is understood in the principal value sense.
Combining this with the previous formula gives (ii).

To prove (iii), we use \eqref{var-del}--\eqref{var-D} to get for $z\neq w$,
\begin{align*}
L_{\mu}\pa_zG(z,w)& =\pa_zL_{\mu}G(z,w) =2\iint\limits_F
\pa_z\pa_vG(z,v) \pa_v G(v,w)\mu(v)d^2v,\\
\intertext{and}
L_{\bar{\mu}}\pa_zG(z,w)&=-\ov{\mu(z)}\,\pa_{\bar{z}}G(z,w)+2\pa_z\iint\limits_F
\pa_{\bar{v}}G(z,v) \pa_{\bar{v}} G(v,w)\ov{\mu(v)}d^2v\\
&=2\iint\limits_F \pa_z\pa_{\bar{v}}G(z,v) \pa_{\bar{v}}
G(v,w)\ov{\mu(v)}d^2v.
\end{align*}
Here we used \eqref{derivative-G} and the elementary formula
\begin{align*}
\frac{\pa}{\pa z}\iint\limits_{|v-z|\leq\vep} \frac{f(z,v)}{\bar{v}-\z}d^2v =
\iint\limits_{|v-z|\leq\vep} \frac{f_{z}(z,v)}{\bar{v}-\z}d^2v-\pi f(z,z) +O(\vep),
\end{align*}
where $f(z,v)$ is smooth at $z=v$, which readily follows from the Stokes' theorem. 

Parts (iv) and (v) are proved similarly. In particular,
using \eqref{var-D} we obtain for $z\neq w$,
\begin{align*}
L_{\bar{\mu}}\mathcal{D}_{z}G(z,w) & =-\ov{\mu(z)}\pa_{z}\pa_{\z}G(z,w)-
\rho(z)\pa_{z}\rho(z)^{-1}\ov{\mu(z)}\pa_{\z}G(z,w) \\
&\quad + 2\rho(z)\pa_{z}\rho(z)^{-1}\pa_{z}\iint\limits_{F}\pa_{\bar{u}}G(z,u)\pa_{\bar{u}}G(u,w)\ov{\mu(u)}d^{2}u \\
& = -\ov{\mu(z)}\pa_{z}\pa_{\z}G(z,w)-
\rho(z)\pa_{z}\rho(z)^{-1}\ov{\mu(z)}\pa_{\z}G(z,w) \\
&\quad + 2\rho(z)\pa_{z}\rho(z)^{-1}\iint\limits_{F}\pa_{\bar{u}}\pa_{z}G(z,u)\pa_{\bar{u}}G(u,w)\ov{\mu(u)}d^{2}u\\
&\quad +\rho(z)\pa_{z}\rho(z)^{-1}\ov{\mu(z)}\pa_{\z}G(z,w)\\
&=  -\frac{1}{2}\rho(z)\ov{\mu(z)}G(z,w) +2\iint\limits_{F}\pa_{\bar{u}}\mathcal{D}_{z}G(z,u)\pa_{\bar{u}}G(u,w)\ov{\mu(u)}d^{2}u.
\end{align*}
\end{proof}
\begin{corollary} \label{var-LD} Let $h^{\vep}$ be a smooth family of $\Ga^{\vep\mu}$--automorphic functions on $\Omega^{\vep\mu}$. Then
\begin{itemize}
\item[(i)]
\begin{align*}
L_{\mu}\iint\limits_{F}\mathcal{D}_{z}G(z,u)h(u)\rho(u)d^{2}u &=
\iint\limits_{F}(L_{\mu}\mathcal{D}_{z}G(z,u)h(u)+
\mathcal{D}_{z}G(z,u)L_{\mu}h(u))\rho(u)d^{2}u, 
\end{align*}
\item[(ii)]
\begin{align*}
L_{\bar{\mu}}\iint\limits_{F}\mathcal{D}_{z}G(z,u)h(u)\rho(u)d^{2}u &=
\iint\limits_{F}(L_{\bar\mu}\mathcal{D}_{z}G(z,u)h(u)+
\mathcal{D}_{z}G(z,u)L_{\bar\mu}h(u))\rho(u)d^{2}u \\
&\quad 
+\frac{1}{2}\rho(z)\ov{\mu(z)}h(z),
\end{align*}
\end{itemize} 
where integrals are understood in the principal value sense.
\end{corollary} 
\begin{proof} Follows from part (iv) of Lemma \eqref{var-propagator} and definition of the principal value integral.
\end{proof}
For a global coordinate $z$ on $X\simeq\Ga\bk\Omega$, set
\begin{align} \label{kernel-K}
K_{\Gamma}(z,w)=\sum_{\gamma \in \Gamma} \frac{\gamma'(w)^2}{(z-\gamma
w)^4}.
\end{align}
The kernel $K_{\Ga}$ is a meromorphic quadratic differential for $\Ga$ in $z$ and $w$, with the fourth order pole at $z=w$. When $z$ is the Fuchsian global coordinate, it follows from  \eqref{ahlfors-sing} that
\begin{equation} \label{K=D}
K_{\Ga}(z,w)=\frac{\pi}{3}\mathcal{D}_{z}\mathcal{D}_{w}G(z,w).
\end{equation}
\begin{lemma} \label{variation-H} 
Let $z$ be a global coordinate $z$ on $X\simeq\Ga\bk\Omega$
and $\mu\in\Omega^{-1,1}(\Ga)$. For $H(z)=\pa_{z}\pa_{z}G(z,z)$ 
we have
\begin{align*}
L_{\mu}H(z) &= 2\iint\limits_{F}
\left(\left(\pa_z\pa_wG(z,w)\right)^2-\frac{1}{4\pi^2}K_{\Ga}(z,w)\right) \mu(w) d^2w,\\
L_{\bar{\mu}}H(z)&=2\iint\limits_{F}
\left(\pa_z\pa_{\bar{w}}G(z,w)\right)^2 \ov{\mu(w)}
d^2w-\frac{1}{4\pi}\rho(z)\ov{\mu(z)}.
\end{align*}
\end{lemma}
\begin{proof}
Let $z$ be  the Fuchsian global coordinate on $X\simeq\Gamma\bk\U$ and $\left.\mu\right|_{\lo}=0$, so that corresponding  $\Gamma^{\vep\mu}=f^{\vep\mu}\circ\Gamma\circ(f^{\vep\mu})^{-1}$ are quasi-Fuchsian groups for $\vep\neq 0$. In order to use representation  \eqref{tadpole-F}, we need to change a quasi-Fuchsian global 
coordinate on $X^{\vep\mu}\simeq\Ga^{\vep\mu}\bk\Omega_{1}^{\vep\mu}$, where $\Omega_{1}^{\vep\mu}=f^{\vep\mu}(\up)$, to the Fuchsian global coordinate, given the covering $J_{\vep\mu}: \up\rightarrow\Omega_{1}^{\vep\mu}$. It follows from \eqref{trans2} that
$$L_{\mu}H(z) =\lim_{z'\rightarrow z} L_{\mu}\left( \pa_z\pa_{z'} G(z,z')
-\pa_z\pa_{z'}\mathcal{G}(z,z')-\frac{1}{12\pi}S(J^{-1})(z)\right).$$
Using Remark \ref{vanishing}, the formula 
$$L_{\mu}\mathcal{G}(z,z')=2\iint\limits_{\U} \pa_w\mathcal{G}(z,w) \pa_w \mathcal{G}(w,z')\mu(w)d^2w,$$
and \eqref{method-images} and \eqref{unfolding}, we obtain
\begin{align*}
L_{\mu}H(z) &=2\lim_{z'\rightarrow z}\iint\limits_{\U}
\pa_z\pa_{w}\mathcal{G}(z,w) \pa_{z'}\pa_{w}(
G(w,z')-\mathcal{G}(w,z'))\mu(w)d^2w\\
&=2\sum_{\substack{
\gamma\in \Gamma\\
\gamma\neq \id }}\iint\limits_{\U}\pa_z\pa_{w}\mathcal{G}(z,w)
\pa_z\pa_{w}\mathcal{G}(\gamma w,z)\mu(w)d^{2}w\\
&=2\iint\limits_{F}\sum_{\substack{
\gamma_{1},\gamma_{2}\in \Gamma\\
\gamma_{1}\neq \gamma_{2} }}\pa_z\pa_{w}\mathcal{G}(z,\gamma_{1}w)
\pa_z\pa_{w}\mathcal{G}(\gamma_{2}w,z)\mu(w)d^{2}w,\\
\end{align*}
where integrals are understood in the principal value sense.
Below we will show that for all $z\in\U$,
\begin{equation} \label{formula-2}
\lim_{\vep\rightarrow 0}\iint\limits_{\U_{\vep}(z)}(
\pa_z\pa_{w}\mathcal{G}(z,w))^2\mu(w)d^2w=0,
\end{equation}
where $\U_{\vep}(z)=\{w\in\U : |w-z|\geq\vep\}$.
Thus we get
$$L_{\mu}H(z)=2\iint\limits_{F} (\pa_z\pa_{w}G(z,w) )^2\mu(w)d^2w,$$
where the integral is understood in the principal value sense.
For a global coordinate given by $X=\Ga\bk\Omega$ we obtain, using \eqref{trans2} and \eqref{LS},
\begin{align*}
L_{\mu}H(z) &= 2\iint\limits_{F}
\left(\pa_z\pa_wG(z,w)\right)^2 \mu(w) d^2w
-\frac{1}{2\pi^2}\iint\limits_{\CC} \frac{\mu(w)}{(z-w)^4} \;\;
d^2w\\
& = 2\iint\limits_{F}
\left(\left(\pa_z\pa_wG(z,w)\right)^2-\frac{1}{4\pi^2}K_{\Ga}(z,w)\right) \mu(w) d^2w.
\end{align*}

To prove \eqref{formula-2}, it is convenient to use the
unit disc $\mathbb{D}=\{z\in\CC : \vert z\vert <1\}$. The kernel $\mathcal{G}(z,w)$ is given by the same formula \eqref{free-propagator}, where now
\begin{align*}
u(z,w)=\frac{|z-w|^2}{(1-|z|^2)(1-|w|^2)}.
\end{align*}
Using the Ahlfors' formula
\begin{align}\label{projection}
\mu(w)=\frac{3(1-|w|^2)^2}{\pi}\iint\limits_{\mathbb{D}}
\frac{\mu(\zeta)}{(1-\zeta\bar{w})^4}d^2\zeta,
\end{align}
we obtain
\begin{gather*}
\lim_{\vep\rightarrow 0}\iint\limits_{\U_{\vep}(z)}(
\pa_z\pa_{w}\mathcal{G}(z,w))^2\mu(w)d^2w  =
\lim_{\vep\rightarrow 0}\iint\limits_{\mathbb{D}_{\vep}(z)}(
\pa_z\pa_{w}\mathcal{G}(z,w))^2\mu(w)d^2w \\
 =\lim_{\vep\rightarrow 0}\frac{3}{\pi}
\iint\limits_{\mathbb{D}_{\vep}(z)}
\iint\limits_{\mathbb{D}}
(\pa_z\pa_{w}\mathcal{G}(z,w))^2\frac{(1-|w|^2)^2}{(1-\zeta\bar{w})^4}
\mu(\zeta) d^2\zeta d^2w\\
=\frac{3}{\pi}\iint\limits_{\mathbb{D}}A(z,\zeta)\mu(\zeta)d^{2}\zeta,
\end{gather*}
where $\mathbb{D}_{\vep}(z)=\{w\in\mathbb{D} : |w-z|\geq\vep\}$ and
\begin{align*}
A(z,\zeta)=\lim_{\vep\rightarrow 0}\iint\limits_{\mathbb{D}_{\vep}(z)}(
\pa_z\pa_{w}\mathcal{G}(z,w))^2\frac{(1-|w|^2)^2}{(1-\zeta\bar{w})^4}d^2w.
\end{align*}
Using \eqref{invariant}, we get
$A(\sigma z, \sigma
\zeta)\sigma'(z)^2\sigma'(\zeta)^2=A(z,\zeta)$
for all $\sigma\in \PSU(1,1)$, and by explicit computation,
\begin{align*}
\pa_z\pa_{w}\mathcal{G}(0,w)=&-\frac{1}{2\pi}\frac{\bar{w}^2}
{(1-|w|^2)}\left(\frac{1-3|w|^2}{|w|^4}-\frac{2}{1-|w|^2}\log|w|^2\right).
\end{align*}
Using polar coordinates, we immediately obtain that 
$A(0, \zeta)=0$ and, therefore, $A(z,\zeta)=0$ for all $z,\zeta\in\mathbb{D}$.

Finally, using \eqref{tadpole} and \eqref{qc-var}, we have for a global coordinate $z$ on $X\simeq\Ga\bk\Omega$,
\begin{align*}
L_{\bar{\mu}}H(z) 
& =\lim_{z'\rightarrow z} L_{\bar{\mu}}\Biggl( \pa_z\pa_{z'}
G(z,z')
+\frac{1}{2\pi}\left(\frac{1}{(z-z')^2}-\frac{1}{2}\rho(z)\frac{\z-\bar{z}'}{z-z'}\right)
\Biggr)\\
& =\lim_{z'\rightarrow z}\Biggl(2\iint\limits_{F}
\pa_z\pa_{\bar{w}}G(z,w) \pa_{z'}\pa_{\bar{w}}
G(w,z')\ov{\mu(w)}d^2w\\
& -\frac{1}{4\pi}\frac{\rho(z)}{z-z'}\left(\ov{\dot{f}(z)}-
\ov{\dot{f}(z')}-(\z-\z')\ov{\dot{f}_z(z)}\right)\Biggr)\\
& =2\iint\limits_{F} (\pa_z\pa_{\bar{w}}G(z,w)
)^2\,\ov{\mu(w)}d^2w -\frac{1}{4\pi}\rho(z)\,\ov{\mu(z)}.
\end{align*}
On the other hand, using \eqref{BLS}, we have for the Fuchsian global coordinate on $X\simeq \Ga\bk\up$,
\begin{align*}
L_{\bar{\mu}}H(z)& =2\iint\limits_{F}
(\pa_z\pa_{\bar{w}}G(z,w))^2\ov{\mu(w)}d^2w 
-2\iint\limits_{\U}(\pa_z\pa_{\bar{w}}\mathcal{G}(z,w)
)^2\,\ov{\mu(w)}d^2w\\&+\frac{1}{24\pi} \rho(z)\ov{\mu(z)}.
\end{align*}
Hence,
\begin{equation} \label{z-bar-w}
\iint\limits_{\U}(\pa_z\pa_{\bar{w}}\mathcal{G}(z,w))^2\ov{\mu(w)}d^2w=\frac{7}{48\pi}\rho(z)\ov{\mu(z)},
\end{equation}
which can be also verified directly. 
\end{proof}
\section{One-point correlation functions}\label{VA-OPCF}
Here we compute one-point correlation functions $\la\la T(z)X\ra\ra$ and $\la\la \bar{T}(\z)X\ra\ra$ in all orders of the perturbation theory.
\begin{theorem} \label{one-point} Let $\mathcal{F}$ be a formal function on the Schottky space $\mathfrak{S}_{g}$, associated with the free energy $\mathcal{F}_{X}$ and defined in Section \ref{TT-VF-TS}. For every $t\in\Sch_{g}$ let $X_{t}\simeq \Gamma_{t}\bk\Omega_{t}$, where $\Gamma_{t}$ is the corresponding Schottky group, and let $J_{t}=J_{S}^{-1}\circ J_{F}$, where $J_{S}$ and $J_{F}$ are covering maps corresponding to the Schottky and Fuchsian uniformizations of $X_{t}$. Then for every $t\in\mathfrak{S}_{g}$ correlation functions $\la\la T(z)X_{t}\ra\ra$ and $\la\la \bar{T}(\z)X_{t}\ra\ra$ are holomorphic and anti-holomorphic quadratic differentials for $\Ga_{t}$,  and
\begin{align}
(\pa\mathcal{F})(t) & = \frac{1}{\pi} \left(\la\la
T(z)X_{t}\ra\ra-\frac{1}{12}\mathcal{S}(J_{t}^{-1})(z)\right), \label{T-ward} \\
(\bar{\pa}\mathcal{F})(t) & = \frac{1}{\pi} \left(\la\la
\bar{T}(\z)X_{t}\ra\ra-\frac{1}{12}\ov{\mathcal{S}(J_{t}^{-1})}(\z)\right), \label{bar-T-ward}
\end{align}
which are understood as equalities in $\frac{1}{\hbar}\Omega^{2,0}(\Ga_{t})[[\hbar]]$ and $\frac{1}{\hbar}\Omega^{0,2}(\Ga_{t})[[\hbar]]$ respectively. The same statement holds for the Teichm\"{u}ller space $\mathfrak{T}_{g}$.
\end{theorem}
\begin{remark} Slightly abusing notations, we will write \eqref{T-ward}--\eqref{bar-T-ward} as
\begin{align}
\pa\log\la X\ra & = -\frac{1}{\pi} \left(\la\la
T(z)X\ra\ra-\frac{1}{12}\mathcal{S}(J^{-1})(z)\right),\label{T-ward-1}\\
 \bar\pa\log\la X\ra & = -\frac{1}{\pi} \left(\la\la
\bar{T}(\z)X\ra\ra-\frac{1}{12}\ov{\mathcal{S}(J^{-1})}(\z)\right).\label{barT-ward-1}
\end{align}
These equations are conformal Ward identities with single insertion of the stress-energy tensor for quantum Liouville theory on compact Riemann surfaces. In particular, it follows from \eqref{T-ward-1} that $\la\la
T(z)X\ra\ra$ is a formal holomorphic quadratic differential on $\Omega_{t}$, i.e., every term in its perturbative expansion is a holomorphic quadratic differential for $\Gamma_{t}$.
\end{remark}
\begin{proof}
Since $\mathcal{F}$ is real-valued, equation \eqref{bar-T-ward} follows from \eqref{T-ward}. We prove \eqref{T-ward} in all orders of the perturbation theory by verifying it at the classical,  one-loop, and higher loops levels. For $t\in\Sch_{g}$ we will abbreviate $\Gamma=\Gamma_{t}$, $X=X_{t}$, $J=J_{t}$, etc.

\subsection{Classical contribution} \label{VA-OPCF-CL}
Formula \eqref{var-classical-action} gives 
\eqref{T-ward} at the classical level.
\subsection{One-loop contribution} \label{VA-OPCF-OLL} 
According to \eqref{ste-1-point}--\eqref{tadpole},
\begin{align*}
\la\la T(z)X\ra\ra_{1-\text{loop}} =-\pi\left( H(z) +\iint\limits_{F}
\mathcal{D}_zG(z,w)G(w,w) \rho(w) d^2w\right),
\end{align*}
where $H(z)=\pa_z\pa_zG(z,z)$ and $F$ is a fundamental domain for $\Gamma$ in $\Omega$, and is given by the following graphs (see Fig. 2): 
\begin{center}
\epsfig{file=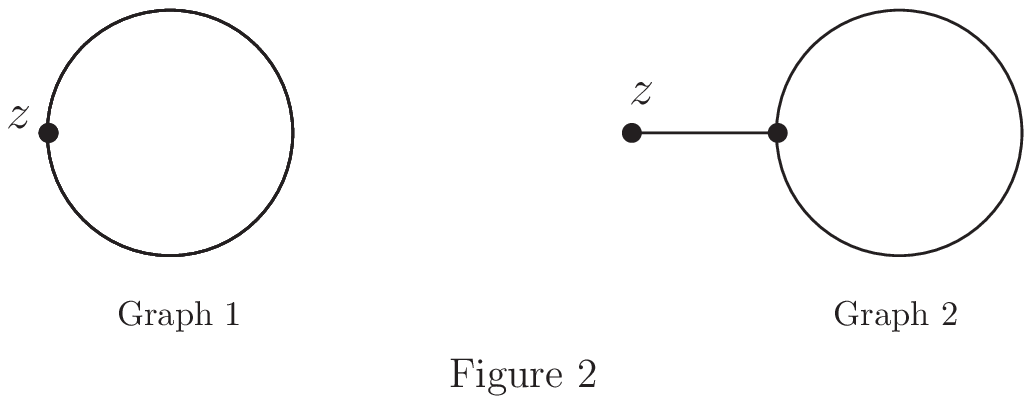, %
height=4cm}
\end{center}

 \noindent
On the other hand, it follows from Definition \ref{free-energy} that at the one-loop level
$\mathcal{F}_{X}=\frac{1}{2}\log Z(2)$.  
It was proved in \cite[Section 3]{TZ}, using the Fuchsian global coordinate on $X\simeq \Gamma\bk\up$ that
\begin{align} \label{partial-Z(2)} 
\pa \log Z(2)=-2P(H),
\end{align}
where $P$ is an orthogonal projection operator onto the space of holomorphic quadratic differentials.
\begin{remark} \label{pa-Z(2)}
For any choice of a global coordinate on $X$, 
$$\pa\log Z(2) =-2P\left(H+ \frac{1}{12\pi}\mathcal{S}(J^{-1})\right).$$
\end{remark}
Using \eqref{partial-Z(2)}, representation \eqref{P-kernel} and the Stokes' theorem,
we get
\begin{align*}
\pa\log Z(2) & = -8\iint\limits_{F}
\mathcal{D}_z\mathcal{D}_{\bar{w}}
G(z,w)H(w) \rho(w)^{-1}d^2w\\
& =-2\lim_{\vep\rightarrow 0}\iint\limits_{F_{\vep}(z)} \pa_{\bar{w}}R(w,z)H(w) d^2w\\
&= 2\iint\limits_{F}R(w,z)\pa_{\w}H(w)d^2w
+i\int\limits_{\pa F_{\vep}(z)} R(w,z)H(w)dw.\\
\end{align*}
Here $R(z,w)$
is given by \eqref{R-definition} and $F_{\vep}(z)=F\setminus \{|w-z| \leq \vep\}$.
Since $R(w,z)H(w)$ is a $(1,0)$ tensor for $\Ga$, 
the line integral over $\pa F$ vanishes. 
Using \eqref{asymptotics-R}, 
we get for the remaining line integral, where $C_{\vep}(z)=\{|w-z|=\vep\}$\footnote{Here and in what follows all contours like
$C_{\vep}(z)$ are oriented counter-clockwise.},
\begin{align*}
-i\lim_{\vep\rightarrow
0}\oint_{C_{\vep}(z)}R(w,z)H(w)dw=-2H(z).
\end{align*}
Now using \eqref{z-bar-derivative-H}, \eqref{equation-R}, \eqref{formula}  and the Stokes' theorem, we obtain
\begin{gather*}
2\iint\limits_{F} R(w,z)\pa_{\w}H(w)d^2w  = \iint\limits_{F}
R(w,z)\pa_w(G(w,w))\rho(w)d^2w\\
=\lim_{\vep\rightarrow 0}\left(-\iint\limits_{F_{\vep}(z)}\pa_w (\rho(w)R(w,z))G(w,w)d^2w -\frac{i}{2}\int_{C_{\vep}(z)}
R(w,z)G(w,w) d\bar{w}\right)\\
= -2\iint\limits_{F}\mathcal{D}_z G(z,w)G(w,w)\rho(w) d^2w.
\end{gather*}
Thus in terms of the the Fuchsian global coordinate on $X$ we have
\begin{align}\label{1loopF}
\pa \log \la X \ra_{1-\text{loop}}= -\frac{1}{\pi} \la\la
T(z)X\ra\ra_{1-\text{loop}}.
\end{align}
Now using \eqref{trans2} and Remark \ref{pa-Z(2)}, we get for the Schottky global coordinate on $X$,
\begin{align}\label{1loop}
\pa \log\la X\ra_{1-\text{loop}}=-\frac{1}{\pi}\left( \la\la
T(z)X\ra\ra_{1-\text{loop}}-\frac{1}{12}
\mathcal{S}(J^{-1})(z)\right),
\end{align}
where $J=J^{-1}_{S}\circ J_{F}$.
\subsection{Higher loops}\label{VA-OPCF-HLL} 
The higher loop terms in $\mathcal{F}_{X}$ do not depend on the 
choice of a global coordinate on $X$, and for convenience we will
be using the Fuchsian global coordinate.

Define the ``forgetful map'' $p_{1}: \mathcal{G}^{(c)}_{\{z\}}\rightarrow\mathcal{G}^{(c)}_{\geq 3}$ by eliminating the labeled vertex of valency $1$ or $2$. Namely, if $\Upsilon$ is a graph with a labeled vertex $v_{1}$ of valency $1$ and $e$ is the edge connecting it to a vertex $v_{2}$ of valency larger than $3$, then $p_{1}(\Upsilon)$ is obtained by removing the vertex $v_{1}$ and the edge $e$. If the valency of $v_{2}$ is $3$, we also remove the vertex $v_{2}$ and replace two remaining edges at $v_{2}$ by a single edge. If $\Upsilon$ is a graph with a labeled vertex $v_1$ of valency $2$ with edges $e_1$ and $e_2$, then $p_1(\Upsilon)$ is obtained by removing the vertex $v_1$ and replacing the edges $e_1$, $e_2$ by a single edge. Clearly, $\chi(\Up)=\chi(p_{1}(\Up))$ and $\mathrm{Aut}(\Up)=\mathrm{Aut}(p_{1}(\Up))$. Conversely, if $\Upsilon'\in p_1^{-1}(\Upsilon)$,
then $\Upsilon'$ is obtained from $\Upsilon$ by one of the
following ways.
\begin{itemize}
\item[(a)] Attach an extra edge $e$ to the
midpoint of an edge of $\Upsilon$, so that one of its endpoints becomes a vertex of valency $3$, and the other becomes a labeled vertex of valency $1$. 
\item[(b)] Insert a labeled vertex $v$ of valency $2$ at the midpoint of an edge of $\Upsilon$.
\item[(c)] Attach an extra edge $e$ to a vertex $v$ of
valency $n$ of $\Upsilon$, so that $v$ becomes a vertex of valency
$n+1$, and the other endpoint of $e$ becomes  a labeled vertex of
valency $1$.
\end{itemize}
We have that in case (a) $V'=V+2$ and $\vep_{1}(\Up')=1$, in case (b) $V'=V+1$ and $\vep_{1}(\Up')=0$, and in case (c) $V'=V+1$ and $\vep_{1}(\Up')=1$.

We will show that at the higher loop level equation \eqref{T-ward} is valid graph by graph, i.e., for every $\Upsilon\in \mathcal{G}^{(c)}_{\geq 3}$ with more than one loop,
\begin{align}\label{onepoint}
\pa W_{\Up}(X)=
 -2\sum_{\Upsilon' \in p_1^{-1}(\Upsilon)}
  (-1)^{|V(\Up)|+|V(\Up')|+\vep_{1}(\Up')}W_{\Upsilon'}(X; z).
\end{align}

Using \eqref{W-integral}--\eqref{W-integrand} and the Leibniz rule, we get
\begin{gather} \label{var-higher-1}
L_{\mu}W_{\Up}(X)=\sum_{e\in E(\Up)}\iint\limits_{X}\iint\limits_{X}
L_{\mu}G(P_{v_{0}(e)},P_{v_{1}(e)})W^{e}_{\Up}(X)(P_{v_{0}(e)},P_{v_{1}(e)})\\
dP_{v_{0}(e)}dP_{v_{1}(e)}, \nonumber
\end{gather}
where
\begin{gather*} 
W^{e}_{\Up}(X)(P_{v_{0}(e)},P_{v_{1}(e)})
=\idotsint\limits_{X^{V'}}\prod_{e'\in E(\Up)\setminus\{e\}}G(P_{v_{0}(e')},P_{v_{1}(e')})\prod_{k=1}^{V'}
dP_{k}.
\end{gather*}
Here $V'=V-2$ 
unless $v_{0}(e)=v_{1}(e)$, in which case $V'=V-1$ and there is a single integration over $P_{v_{0}(e)}$ in \eqref{var-higher-1}. 

First we consider the case $v_{0}(e)\neq v_{1}(e)$. Using part (i) of Lemma \ref{var-propagator}, we get  that the contribution of an edge $e$ into $L_{\mu}W_{\Up}(X)$ is
\begin{align*}
2\iint\limits_F\iint\limits_F  W_{\Up}^{e}(z_{1},z_{2})\left(\iint\limits_F
\pa_zG(z_1,z) \pa_zG(z_2, z)\mu(z)d^{2}z\right) \rho(z_1)\rho(z_2)d^2z_1
d^2z_2.
\end{align*}
Using \eqref{P-kernel}, we get that the contribution of the edge $e$
to $\pa W_{\Up}(X)$ is
\begin{gather*}
\iint\limits_{F}\iint\limits_{F}
W_{\Up}^{e}(z_{1},z_{2})I(z_{1},z_{2})\rho(z_1)\rho(z_2)d^2z_1 d^2z_2,
\end{gather*}
where
\begin{equation*}
I(z_{1},z_{2})=8\iint\limits_{F}\mathcal{D}_z\mathcal{D}_{\bar{w}}G(z,w) \pa_wG(z_1,w) \pa_wG(z_2,
w)\rho(w)^{-1}d^2w,
\end{equation*}
and the change of the order of integrations is easily justified.
Using the Stokes' theorem and setting $F_{\vep}(z,z_{1},z_{2})=F_{\vep}(z)\setminus \left\{\{|w-z_{1}|\leq\vep\}\cup\{|w-z_{2}|\leq\vep\}\right\}$, 
$C_{\vep}(z_{1},z_{2})=\{|w-z_{1}|=\vep\}\cup\{|w-z_{2}|=\vep\}$, we get
\begin{align*}
I & = 
 -2\lim_{\vep\rightarrow 0}\iint\limits_{F_{\vep}(z,z_{1},z_{2})}R(w,z)\pa_{\bar{w}}
(\pa_wG(z_1,w) \pa_wG(z_2, w)) d^2w\\
&+i\lim_{\vep\rightarrow
0}\oint_{C_{\vep}(z)}R(w,z)\pa_wG(z_1,w)
\pa_wG(z_2, w)dw \\
& + i\lim_{\vep\rightarrow
0}
\oint_{C_{\vep}(z_{1},z_{2})}R(w,z)\pa_wG(z_1,w)
\pa_wG(z_2, w)dw\\
 &=I_1 +I_2+I_3.
\end{align*}
As in the one-loop case, using property \textbf{P3},
\eqref{derivative-G} and \eqref{equation-R},
we obtain for $z\neq z_1, z_{2}$,
\begin{align*}
I_1 & = -\iint\limits_{F}\rho(w)R(w,z)
\pa_w(G(z_1,w) G(z_2, w)) d^2w \\
& =2\iint\limits_{F} \mathcal{D}_zG(z,w) G(z_1,w) G(z_2, w)\rho(w)d^2w, \\
 I_2 & =2\pa_zG(z_1,z) \pa_zG(z_2, z), \\
I_3
& =(R(z_{1},z)\pa_{z_{1}}+R(z_{2},z)\pa_{z_{2}})G(z_{1},z_{2}).
\end{align*}

Now it follows from \eqref{W-integral-ste-1}--\eqref{W-integrand-ste-1} that the terms $I_1$ and $I_{2}$ correspond, respectively, to the contribution into \eqref{onepoint} of graphs of type (a) and (b) such that the corresponding edge $e$ is not a loop. Assuming that there are no self-loops starting at $v_{0}(e)$ and $v_{1}(e)$, we can collect terms $I_{3}$ corresponding to all edges having $v_{0}(e)$ or $v_{1}(e)$ as their endpoints. This gives
\begin{gather*}
\iint\limits_{F}  W_{\Upsilon}^{z_1} \rho(z_{1})R(z_1,z)\pa_{z_1}\prod_{k=1}^{n_{1}} G(z_1,
u_k)d^2z_1\\
+\iint\limits_{F}  W_{\Upsilon}^{z_2} \rho(z_{2})R(z_2,z)\pa_{z_2}\prod_{l=1}^{n_{2}} G(z_2,
v_{l})d^2z_2\\
=-2\iint\limits_{F} W_{\Upsilon}^{z_1} \mathcal{D}_{z}
G(z, z_1)\prod_{k=1}^{n_{1}} G(z_1, u_k)\rho(z_1)d^2z_1 \\
-2\iint\limits_{F} W_{\Upsilon}^{z_2} \mathcal{D}_{z}
G(z, z_2)\prod_{l=1}^{n_{2}} G(z_2, v_l)\rho(z_2)d^2z_2,
\end{gather*}
where $u_{1},\dots,u_{n_{1}}$ and $v_{1},\dots,v_{n_{2}}$, respectively, parameterize all vertices in the stars of $v_{1}(e)$ and $v_{2}(e)$. These terms correspond to the contribution into \eqref{onepoint} from the graphs of type (c) such that there are no self-loops starting at $v_{0}(e)$ and $v_{1}(e)$. 

For the remaining case when $v_{0}(e)=v_{1}(e)$, or when there are self-loops starting at $v_{0}(e)$ or $v_{1}(e)$,  we consider the principal value integral
\begin{align*}
\tilde{I}& =8\iint\limits_{F}\mathcal{D}_z\mathcal{D}_{\bar{w}}G(z,w)
(\pa_wG(z_1,w))^{2}\rho(w)^{-1} d^2w \\
& =\tilde{I}_{1}+\tilde{I}_{2}+\tilde{I}_{3},
\end{align*}
where as in the previous case,
\begin{align*}
\tilde{I}_{1} & = 2\iint\limits_{F} \mathcal{D}_zG(z,w) G(z_1,w)^{2}\rho(w)d^2w, \\\
\tilde{I}_{2} & = 2(\pa_zG(z_1,z))^{2}, \\
\tilde{I}_{3} & = i\lim_{\vep\rightarrow 0}\oint_{C_{\vep}(z_{1})}R(w,z)(\pa_wG(z_1,w))^{2}dw.
\end{align*}

It follows from \eqref{W-integral-ste-1}--\eqref{W-integrand-ste-1} that the terms $\tilde{I}_1$ and $\tilde{I}_{2}$ correspond, respectively, to the contribution into \eqref{onepoint} of the graphs of type (a) and (b) such that the corresponding edge $e$ is a loop. To evaluate $\tilde{I}_{3}$, we use
\begin{align*}
(\pa_wG(z_1,w))^{2} & =\sum_{\gamma_1, \gamma_2\in
\Gamma}\pa_w\mathcal{G}(\gamma_1z_1,w) \pa_w\mathcal{G}(\gamma_2z_1, w)\\
& =\sum_{\gamma_1\neq \gamma_2\in
\Gamma}\pa_w\mathcal{G}(\gamma_1z_1,w)
\pa_w\mathcal{G}(\gamma_2z_1, w)+\sum_{\gamma\in
\Gamma}(\pa_w\mathcal{G}(\gamma z_1,w))^{2},
\end{align*}
and write $\tilde{I}_{3}=\tilde{I}_{3,1}+\tilde{I}_{3,2}$. Using \eqref{derivative-G-free}
and \eqref{invariant} we obtain
\begin{align*}
\tilde{I}_{3,1} & =i\sum_{\gamma_1\neq \gamma_2\in
\Gamma}\lim_{\vep\rightarrow 0}\oint_{C_{\vep}(z_{1})}R(w,z)\pa_w\mathcal{G}(\gamma_1z_1,w)
\pa_w\mathcal{G}(\gamma_2z_1, w)dw\\
& =2\sum_{\gamma\neq\mathrm{id}\in \Gamma} \left.R(
z_1,z) \pa_{w}\mathcal{G}(\gamma z_1,
w)\right\vert_{w=z_1} =R(z_1,z)\pa_{z_1}G(z_1,z_1),
\end{align*}
and using property \eqref{equation-R} we get
\begin{align*}
\tilde{I}_{3,2}& = i\sum\limits_{\gamma\in\Gamma}\lim_{\vep\rightarrow 0}\oint_{C_{\vep}(z_{1})}R(w,z)(\pa_w\mathcal{G}(\gamma
z_1,w))^{2} dw
= -\frac{1}{\pi}\mathcal{D}_zG(z, z_1).
\end{align*}

Now collecting all terms $\tilde{I}_{3,1}$ corresponding to edges having the vertex $v_{0}(e)=v_{1}(e)$ as an endpoint, we get
\begin{gather*}
\iint\limits_{F}  W_{\Upsilon}^{z_1} \rho(z_{1})R(z_1,z)\pa_{z_1}\prod_{k=1}^{n_{1}} G(z_1,
u_k)d^2z_1 \\
=-2\iint\limits_{F} W_{\Upsilon}^{z_1} \mathcal{D}_{z}
G(z, z_1)\prod_{k=1}^{n_{1}} G(z_1, u_k)\rho(z_1)d^2z_1, \\
\end{gather*}
and collecting all terms $\tilde{I}_{3,2}$ we obtain
$$-\frac{m_{1}}{\pi}\iint\limits_{F}W_{\Upsilon}^{z_1}\mathcal{D}_zG(z,
z_1)\prod_{k=2}^{n_{1}} G(z_1, u_k)\rho(z_1)d^2z_1,$$
where $m_{1}$ is the number of self-loops at the vertex $v_{0}(e)=v_{1}(e)$. Thus in accordance with the Feynman rules in Section \ref{EMT-OCF}, terms $\tilde{I}_{3}$ correspond to the contribution into \eqref{onepoint} of the remaining graphs of type (c). 
\end{proof}
\begin{remark}
It is elementary to show, using \eqref{method-images} and \eqref{DG},
that
\begin{align*}
\iint\limits_{F} \mathcal{D}_zG(z,w)\rho(w)d^2w=0,
\end{align*}
so that
\begin{align*}
\iint\limits_{F} \mathcal{D}_zG(z,w)
G(w,w)\rho(w)d^2w=\iint\limits_{F} \mathcal{D}_zG(z,w)
\left(G(w,w)+\tfrac{1}{2\pi}\right)\rho(w)d^2w.
\end{align*}
Thus the Feynman rule for vertices with self-loops is  consistent with the regularization at the one loop level.
\end{remark}
\section{Two-point correlation function --- $TT$ equation}\label{TPCF-TT} 
Here we compute two point correlation functions 
$\la\la T(z)T(w)X\ra\ra$ and $\la\la \bar{T}(\bar{z})\bar{T}(\bar{w})X\ra\ra$
in all orders of perturbation theory. Namely, we express them through one point correlation functions
$\la\la T(z)X\ra\ra$ and  $\la\la \bar{T}(\bar{z})X\ra\ra$,
which according to Theorem \ref{one-point}  
can be considered as formal $(1,0)$ and $(0,1)$-forms on the Schottky space $\mathfrak{S}_{g}$. Using notations in Section \ref{TT-VF-VF}, we have the following result.
\begin{theorem} \label{TT-equation}  For every $t\in\mathfrak{S}_{g}$
correlation functions $\la\la T(z)T(w)X_{t}\ra\ra$ and $\la\la \bar{T}(\bar{z})\bar{T}(\bar{w})X_{t}\ra\ra$, where $X_{t}\simeq\Ga_{t}\bk\Omega_{t}$, are meromorphic and anti-meromorphic quadratic differentials for $\Ga_{t}$ respectively,
having only fourth order poles at $z=w$ and $\z=\w$. For $t\in U_{0}\subset\mathfrak{S}_{g}$ --- a coordinate chart of the origin of the Schottky space $\mathfrak{S}_{g}$ --- we have
\begin{align}
\pa_{s}\la\la T(z)X_{t}\ra\ra & =-\frac{1}{\pi}\left(\la\la T(z)T(w)X_{t}\ra\ra -\frac{c}{2}K_{\Ga_{t}}(z,w)\right) \label{ward-TT} \\
&\quad +(2\pa_{z}R(z,w)+R(z,w)\pa_{z})\la\la T(z)X_{t}\ra\ra +\cT(z,w), \nonumber\\
\bar{\pa}_{s}\la\la \bar{T}(\bar{z})X_{t}\ra\ra & =-\frac{1}{\pi}\left(\la\la \bar{T}(\bar{z})\bar{T}(\bar{w})X_{t}\ra\ra -\frac{c}{2}K_{\Ga_{t}}(\bar{z},\bar{w})\right) \label{ward-barTbarT}\\
&\quad +(2\pa_{\z}R(\z,\w)+R(\z,\w)\pa_{\z})\la\la \bar{T}(\z)X\ra\ra
+\ov{\cT(z,w)}.\nonumber
\end{align}
Here $\displaystyle{c=\frac{12}{\hbar}+1}$ and 
$\cT(z,w)=\sum_{n=-1}^{\infty}\hbar^{n}T_{n}(z,w)$ with
$T_{n}(z,w)$ are smooth quadratic differentials for $\Ga_{t}$ in 
$z$ and $w$ which are holomorphic in $w$. Kernels $R(z,w)$ and $K_{\Ga}(z,w)$ are given explicitly by \eqref{R-definition} and \eqref{kernel-K} respectively, and 
$R(\z,\w)=\ov{R(z,w)}$, $K_{\Ga}(\z,\w)=\ov{K_{\Ga}(z,w)}$. 
The same
statement holds for the Teichm\"{u}ller space $\mathfrak{T}_{g}$.
\end{theorem}

Equations \eqref{ward-TT}--\eqref{ward-barTbarT} are conformal Ward identities with two insertions of same type components of the stress-energy tensor for quantum Liouville theory on compact Riemann surfaces.

\begin{proof} For $t\in\mathfrak{S}_{g}$ we abbreviate $\Ga=\Ga_{t}$,
$X=X_{t}$, $J=J_{t}$, $K=K_{\Ga_{t}}$, etc. Equation \eqref{ward-barTbarT} follows from \eqref{ward-TT}, and we will prove \eqref{ward-TT} by computing $\pa_{s}\la\la T(z)X\ra\ra$ in all orders of perturbation theory.
\subsection{Classical contribution}\label{TPCF-TT-CL}  
As it follows from \eqref{LS}, 
\begin{align*}
L_{\nu}T_{cl}
(z)=\frac{6}{\pi\hbar}\iint\limits_{\C}\frac{\nu(w)}{(z-w)^4}d^2w
=\frac{6}{\pi\hbar}\iint\limits_{F}K(z,w)\nu(w)d^2w,
\end{align*}
where $F$ is a fundamental domain for $\Ga$ in $\Omega$.
Using Stokes' theorem, \eqref{asymptotics-R} and  \eqref{R-zzz}, we get 
\begin{gather} \label{projection-K}
P_{w}(K)(z,w)  = 4 \iint\limits_{\Ga\bk\Omega}
\mathcal{D}_w\mathcal{D}_{\bar{u}}G(w,u)K(z,u)\rho(u)^{-1}d^2u\\
=\frac{i}{2}\lim_{\vep\rightarrow 0}\left(\oint_{C_{\vep}(w)}R(u,w)
K(z,u)du+ \oint_{C_{\vep}(z)}R(u,w)K(z,u)du\right) \nonumber \\
=K(z,w)-\frac{\pi}{3}\mathcal{D}_{z}\mathcal{D}_{w}G(z,w) +\frac{\pi\hbar}{6}\left(
2\pa_{z}R(z,w) +R(z,w)\pa_{z}\right)T_{cl}(z). \nonumber
\end{gather}
It follows from Remark \ref{singular} that $P_{w}(K)(z,w)$, in agreement with \eqref{D-short-term}, is smooth quadratic differential in $z$ and $w$ which is holomorphic in $w$, and such that $(P_{w}(K)(z,\,\cdot\,),\nu)=(K(z,\,\cdot\,),\nu)$ for all $\nu\in\Omega^{-1,1}(\Gamma)$.
In particular,
\begin{equation} \label{D-zero}
\iint\limits_{F}\big(2\mathcal{D}_{z}\mathcal{D}_{w}G(z,w) -\hbar(2\pa_{z}R(z,w) +R(z,w)\pa_{z})T_{cl}(z)\big)\nu(w)d^{2}w=0.
\end{equation}

On the other hand,
\begin{align*}\la\la T(z) T(w)X\ra\ra_{cl}
=\frac{2\pi}{\hbar} \mathcal{D}_z\mathcal{D}_w G(z,w)
\end{align*}
is a meromorphic quadratic differential in variables $z$ and $w$ which corresponds to the single tree graph (see Fig. 3):

\begin{center}
\epsfig{file=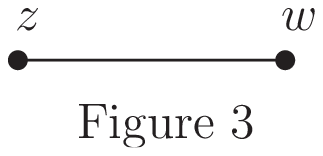, %
height=2cm}
\end{center}

\noindent
Thus we obtain
\begin{gather*}
\pa_{s}T_{cl}(z) =\frac{6}{\pi\hbar}P_{z}(P_{w}(K))(z,w)=
-\frac{1}{\pi}\left(\la\la T(z) T(w)X\ra\ra_{cl}-\frac{6}{\hbar}K(z,w) \right)\\ + 2T_{cl}(z)\pa_{z}R(z,w)+\pa_{z}T_{cl}(z)R(z,w) +\frac{6}{\pi\hbar}T(z,w),
\end{gather*}
where $T(z,w)=P_{z}(P_{w}(K))(z,w)-P_{w}(K)(z,w)$ is a smooth quadratic differential in variables $z$ and $w$, holomorphic in $w$. Note that since $\pa T_{cl}(z)=0$, the kernel $P_{z}(P_{w}(K))(z,w)$ is symmetric. 

\subsection{One-loop contribution}\label{TPCF-TT-OLL} 
At the one loop level,
\begin{align*}
\la\la T(z)X\ra\ra_{1-\text{loop}}= -\pi \left( H(z) +\iint\limits_{F}
\mathcal{D}_zG(z,u)G(u,u) \rho(u) d^2u\right),\\
\end{align*}
corresponding to two graphs in Fig.~2.
 On the other hand, it is easy to see that there are eight graphs
at the one loop level that have two vertices with labels
$z$ and $w$, which contribute to $\la\la T(z)T(w)X\ra\ra_{1-\text{loop}}$. They are given by Fig.~4:

\begin{center}
\epsfig{file=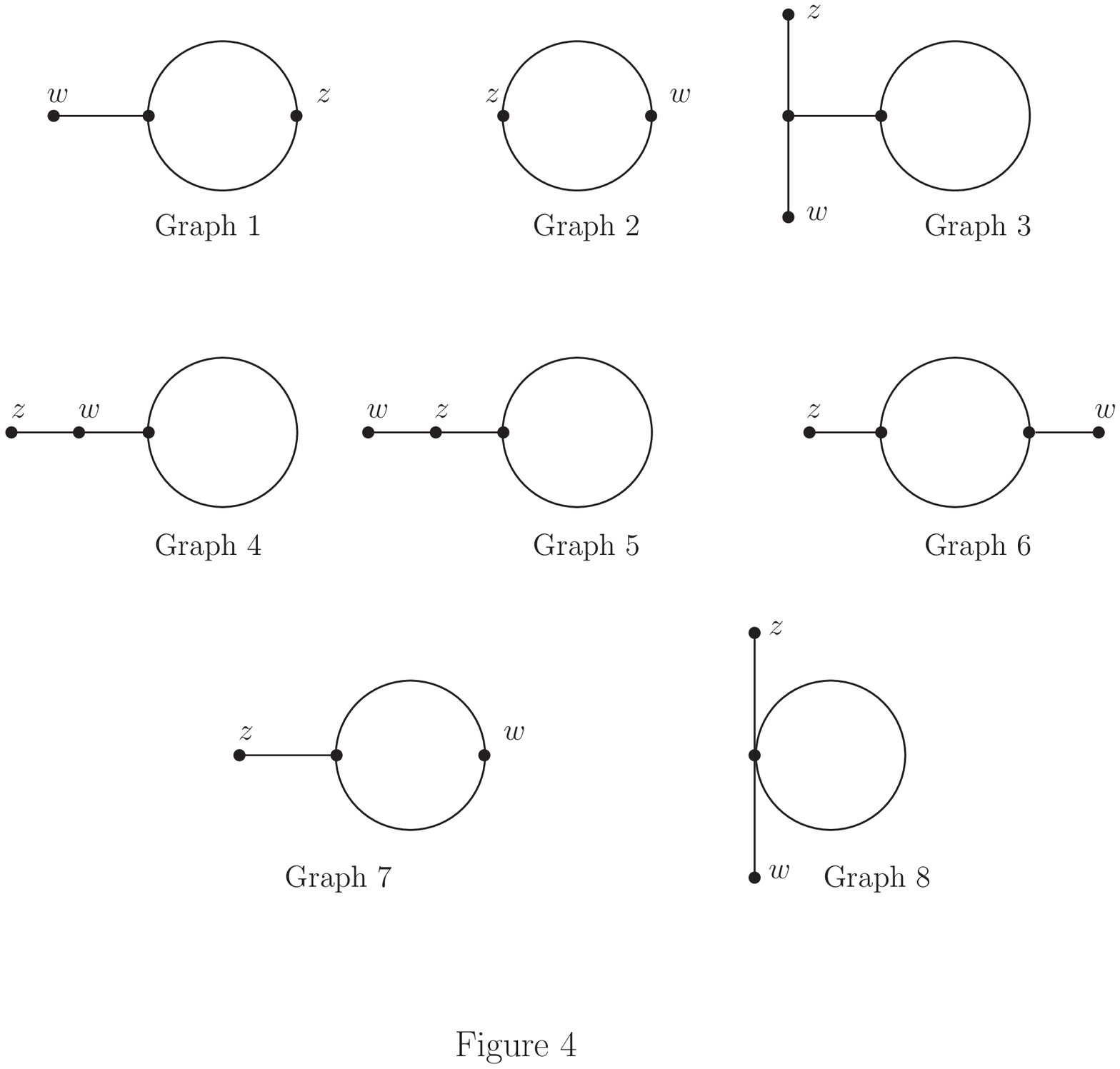, %
height=12.75cm}
\end{center}

\noindent
Using parts (ii) and (iv) of Lemma \ref{var-propagator}, Corollary \ref{var-LD} and Lemma \ref{variation-H}, we get
\begin{align*}
L_{\nu}\la\la T(z)X\ra\ra_{1-\text{loop}}& =-\pi\Bigl(
L_{\nu}H(z)
+\iint\limits_{F} L_{\nu}\mathcal{D}_zG(z,u)G(u,u)\rho(u)d^2u\\
&\;\quad+\iint\limits_{F}
 \mathcal{D}_zG(z,u)L_{\nu}G(u,u)\rho(u)d^2u\Bigr)\\
&=-\pi\iint\limits_{F} E(z,w)\nu(w) d^2w,
\end{align*}
where $E(z,w)=E_1(z,w)+E_2(z,w)+E_3(z,w)$ and
\begin{align*}
E_{1}(z,w) & = 2(\pa_z\pa_wG(z,w))^2-\frac{1}{2\pi^2}K(z,w), \\
E_{2}(z,w) & = 2\iint\limits_{F}\mathcal{D}_{z}\pa_{w}G(z,w)\pa_{w}G(w,u)G(u,u)\rho(u)d^{2}u, \\
E_{3}(z,w) & =2\iint\limits_{F}\mathcal{D}_{z}G(z,u)(\pa_{w}G(w,u))^{2}\rho(u)d^{2}u.
\end{align*}
According to Remark \ref{singular}, $P_{w}(E_{i})(z,w)$, $i=1,2,3$, are holomorphic quadratic differentials in $w$.
We compute them 
by using the Fuchsian global coordinate on $X\simeq\Ga\bk\up$, so $K(z,w)$ now stands for the kernel \eqref{kernel-K} for the Fuchsian group $\Ga$. From the Stokes' theorem and property \textbf{P3} it follows that
\begin{align*}
P(E_1) & =4 \iint\limits_{F}
\mathcal{D}_w\mathcal{D}_{\bar{u}}G(w,u)
\left(2(\pa_z\pa_uG(z,u))^2-\frac{1}{2\pi^{2}}K(z,u)\right)\rho(u)^{-1}d^2u\\
& =-\iint\limits_{F} \rho(u)R(u,w)
\pa_{u}(\pa_zG(z,u))^2 d^2u\\
&\;\;\;\;+ i\lim_{\vep\rightarrow 0} \oint_{C_{\vep}(w)}
R(u,w)
\left((\pa_z\pa_uG(z,u))^2-\frac{1}{4\pi^{2}}K(z,u)\right)du\\
&\;\;\;\;+ i\lim_{\vep\rightarrow 0} \oint_{C_{\vep}(z)}
R(u,w)
\left((\pa_z\pa_uG(z,u))^2-\frac{1}{4\pi^{2}}K(z,u)\right)du \\
& = T_{1} +T_{2}+T_{3}.
\end{align*}
Using again the Stokes' theorem, \eqref{equation-R}, and observing that
$$\lim_{\vep\rightarrow 0}\oint_{C_{\vep}(z)}\rho(u)R(u,w)(\pa_{z}G(z,u))^{2}d\bar{u}=0,$$
we obtain
\begin{equation*}
T_{1}=
2\iint\limits_{F}\mathcal{D}_wG(u,w)(\pa_zG(z,u))^2\rho(u)d^2u.
\end{equation*}
The term $T_{1}$ corresponds to the contribution of graph 1 into $\la\la T(z)T(w)X\ra\ra_{1-\text{loop}}$. Using \eqref{asymptotics-R}, we get
\begin{align*}
T_{2}=2(\pa_z\pa_wG(z,w))^{2} -\frac{1}{2\pi^{2}}K(z,w),
\end{align*}
where the first term corresponds to the graph 2. Since
\begin{equation*}
(\pa_z\pa_uG(z,u))^2-\frac{1}{4\pi^{2}}K(z,u)=2\pa_z\pa_u\mathcal{G}(
z,u)\sum_{\gamma\neq\mathrm{id}\in
\Gamma}\pa_z\pa_u\mathcal{G}(\gamma
z,u)\gamma'(z) + O(1)
\end{equation*}
as $u\rightarrow z$, using \eqref{z-derivative-H} we obtain
\begin{equation} \label{T-3}
T_{3}=(2\pa_{z}R(z,w)  +R(z,w)\pa_{z})H(z).
\end{equation}

To compute $P_{w}(E_{2})$ we observe that by the Stokes' theorem and property \textbf{P3},
\begin{gather*}
8\iint\limits_{F}
\mathcal{D}_w\mathcal{D}_{\bar{u}}G(w,u)\mathcal{D}_z\pa_u
G(z,u)\pa_uG(u,v)\rho(u)^{-1}d^2u \\
=-\iint\limits_{F} \rho(u)R(u,w)\pa_{u}(\mathcal{D}_z
G(z,u)G(u,v))d^2u \\
+i\lim_{\vep\rightarrow 0}\oint_{C_{\vep}}
R(u,w)\mathcal{D}_z\pa_u G(z,u)\pa_uG(u,v)du \\
=2\iint\limits_{F}\mathcal{D}_w G(w,u)\mathcal{D}_z
G(z,u)G(u,v)\rho(u)d^2u 
+2\mathcal{D}_z \pa_{w}G(z,w)\pa_{w}G(w,v)  \\
+ R(v,w)\mathcal{D}_{z}\pa_{v}G(z,v) 
+2\mathcal{D}_{w}\pa_{z}G(z,w)\pa_{z}G(z,v)  \\
 +(2\pa_{z}R(z,w) + R(z,w)\pa_{z})\mathcal{D}_{z}G(z,v),
\end{gather*}
where $C_{\vep}=C_{\vep}(v)\cup C_{\vep}(w)\cup C_{\vep}(z)$,
and in the last line we used \eqref{asymptotics-R} and \eqref{R-D}.
Thus we obtain 
\begin{align*}
P_{w}(E_2)(z,w)& =8\iint\limits_{F}\iint\limits_{F}
\mathcal{D}_w\mathcal{D}_{\bar{u}}G(w,u)\mathcal{D}_z\pa_u
G(z,u)\pa_uG(u,v)G(v,v)\rho(u)^{-1}\rho(v)d^2ud^2v\\
&=T_4+T_5+T_6+T_7+T_{8},
\end{align*}
where
\begin{align*}
T_4 & =2\iint\limits_{F}\iint\limits_{F}
\mathcal{D}_wG(w,u)\mathcal{D}_zG(z,u) G(u,v)G(v,v)
\rho(u)\rho(v)d^{2}ud^2v\\
\intertext{and}
T_5 & =2\iint\limits_F\mathcal{D}_z\pa_w
G(z,w)\pa_wG(w,v)G(v,v)\rho(v)d^2v
\end{align*}
correspond to the graphs 3 and 4, 
\begin{align*}
T_6 &=\iint\limits_F R(v,w)\pa_v\mathcal{D}_zG(z,v)G(v,v)\rho(v)
d^2v,\\
\intertext{while}
T_{7} & = 2\iint\limits_F \mathcal{D}_w\pa_zG(z,w)\pa_zG(z,v)
G(v,v)\rho(v)d^2v
\end{align*}
corresponds to the graph 5, and
\begin{align*}
T_{8} =(2\pa_{z}R(z,w)+ R(z,w)\pa_{z})\iint\limits_{F}\mathcal{D}_{z}G(z,v)G(v,v)\rho(v)d^{2}v.
\end{align*}
Finally, the computation in Section \ref{VA-OPCF-HLL} gives
\begin{align*}
P_{w}(E_3)(z,w)& =4\iint\limits_F
\mathcal{D}_w\mathcal{D}_{\bar{u}}G(w,u)E_3(z,u)
\rho(u)^{-1}d^2u\\
& =8\iint\limits_F\iint\limits_{F} \mathcal{D}_zG(z,v)
\mathcal{D}_w\mathcal{D}_{\bar{u}}G(w,u)
(\pa_uG(u,v))^{2}\rho(u)^{-1}\rho(v)d^{2}u d^2v\\
&=2\iint\limits_F
\iint\limits_{F}\mathcal{D}_zG(z,v)\mathcal{D}_wG(w,u)
G(u,v)^{2}\rho(u)\rho(v)d^2u d^2v\\
&+2\iint\limits_F
\mathcal{D}_zG(z,v)(\pa_wG(w,v))^{2}\rho(v)d^2v\\
&+\iint\limits_F
\mathcal{D}_zG(z,v)R(v,w)
\pa_vG(v,v) \rho(v)d^2v\\
&-\frac{1}{\pi}\iint\limits_F
\mathcal{D}_zG(z,v)\mathcal{D}_wG(w,v) \rho(v) d^2v\\
& =T_9+T_{10}+T_{11}+T_{12}.
\end{align*}
The first two terms $T_{9}$ and $T_{10}$ correspond to the graphs
6 and 7. Using \eqref{equation-R} we see that the terms $T_{6}$, $T_{11}$ and $T_{12}$ correspond to the remaining graph 8. Note that it is the term $T_{12}$ which is responsible for the regularization of the self-loop in Section \ref{EMT-OCF}.

Since $\pa^{2}=0$, we get from Theorem \ref{one-point} that the kernel $P_{z}(P_{w}E))(z,w)$ is symmetric. Thus for the Fuchsian global coordinate on $X$,
\begin{gather} \label{Fuchs-1-loop}
 \pa_{s}\la\la T(z)X\ra\ra_{1-\text{loop}}=-\pi P_{z}(P_{w}(E))(z,w) \\
=-\frac{1}{\pi}\left(\la\la
T(z)T(w) X\ra\ra_{1-\text{loop}}-\frac{1}{2} K(z,w)\right) \nonumber \\ +(2\pa_zR(z,w) +R(z,w) \pa_z)\la\la
T(z)X\ra\ra_{1-\text{loop}} + T_{1}(z,w), \nonumber
\end{gather}
where $T_{1}(z,w)=\pi P_{w}(E)(z,w)-\pi P_{z}(P_{w}(E))(z,w)$ is a smooth quadratic differential in $z$ and $w$ which is holomorphic in $w$. It follows from the symmetry of $\la\la T(z)T(w) X\ra\ra_{1-\text{loop}}$ that it is a meromorphic quadratic differential in $z$ and $w$ with a fourth order pole at $z=w$. 

To find $\pa_{s}\la\la T(z)X\ra\ra_{1-\text{loop}}$ for the Schottky global coordinate we observe that, according to \eqref{trans2}, $\la\la T(z)X\ra\ra_{1-\text{loop}} -\frac{1}{12}\mathcal{S}(J^{-1})(z)$ behaves as a quadratic differential under the change of global coordinates. Using formulas \eqref{K=D} and $\mathcal{S}(J^{-1})(z)=0$, which are valid for the Fuchsian global coordinate on $X$,
we obtain from \eqref{Fuchs-1-loop} that for the Schottky global coordinate,
\begin{gather*}
 \pa_{s}\left(\la\la T(z)X\ra\ra_{1-\text{loop}}-\frac{1}{12}\mathcal{S}(J^{-1})(z)\right) \\=-\frac{1}{\pi}\left(\la\la
T(z)T(w) X\ra\ra_{1-\text{loop}}-\frac{\pi}{6}\mathcal{D}_{z}\mathcal{D}_{w}G(z,w)\right) \\ +(2\pa_zR(z,w) +R(z,w) \pa_z)\left(\la\la
T(z)X\ra\ra_{1-\text{loop}}-\frac{1}{12}\mathcal{S}(J^{-1})(z)\right)  + T_{1}(z,w),
\end{gather*}
Combining this formula with
our computation of $\pa_{s}T_{cl}(z)$ in Section \ref{TPCF-TT-CL}, we finally obtain 
\begin{gather}
 \pa_{s}\la\la T(z)X\ra\ra_{1-\text{loop}}=-\frac{1}{\pi}\left(\la\la
T(z)T(w) X\ra\ra_{1-\text{loop}}-\frac{1}{2} K(z,w)\right) \label{1-loop-TT}  \\ +\,(2\pa_zR(z,w) +R(z,w) \pa_z)\la\la
T(z)X\ra\ra_{1-\text{loop}} + \tilde{T}_{1}(z,w), \nonumber
\end{gather}
where $K(z,w)$ is again the kernel \eqref{kernel-K} for the Schottky group, and  $\tilde{T}_{1}(z,w)$ is a smooth quadratic differential in $z$ and $w$ which is holomorphic in $w$. Using \eqref{1loop} we also get
\begin{gather}
 (\pa_{s} -2\pa_zR(z,w) -R(z,w) \pa_z)\log\la X\ra_{1-\text{loop}} 
  \label{1-loop-dd} \\
  =\frac{1}{\pi^{2}}\left(\la\la
T(z)T(w) X\ra\ra_{1-\text{loop}}-\frac{\pi}{6}\mathcal{D}_{z}\mathcal{D}_{w}G(z,w)\right) + T_{1}(z,w).\nonumber
\end{gather}
\subsection{Higher loops}\label{TPCF-TT-HLL}
Similar to Section \ref{VA-OPCF-HLL}, define the map $p_2: \mathcal{G}^{(c)}_{z,w}\rightarrow \mathcal{G}^{(c)}_{z}$
by eliminating the vertex with label $w$ of valency $1$ or $2$. 
We claim that for every $\Upsilon \in \mathcal{G}^{(c)}_{\geq 3}$ with more than one loop,
\begin{gather*}
\sum_{\Upsilon'\in p_1^{-1}(\Upsilon)}(-1)^{|V(\Upsilon')| +\vep_{1}(\Upsilon')}(\pa_{s} -2\pa_{z}R(z,w)-R(z,w)\pa_{z})W_{\Upsilon'}(X;z) \\
= -2\sum_{\Upsilon^{\prime\prime} \in (p_2\circ
p_1)^{-1}(\Upsilon)}(-1)^{|V(\Upsilon'')| +\vep_{1}(\Upsilon'')}W_{\Upsilon''}(X;z,w). 
\end{gather*}
This readily follows from the arguments in Sections  \ref{VA-OPCF-HLL} and \ref{TPCF-TT-OLL}. Namely, when using the Leibniz rule we concentrate on the edge $e$ of $\Upsilon'$ with neither of its endpoints being a labeled vertex of valency $1$ or $2$, the result follows as in Section \ref{VA-OPCF-HLL}. When one of the endpoints of $e$ is a vertex with label $z$ of valency $1$ or $2$, repeating arguments in Section \ref{TPCF-TT-OLL} yields an extra contribution 
\begin{align*}
(2\pa_z R(z,w)+R(z,w)\pa_{z})W_{\Upsilon'}(X;z).
\end{align*}
Putting everything together proves the theorem.
\end{proof}
\begin{remark} Equation \eqref{ward-TT} can be stated as
$$ (P_{w}Q)(z,w) -
(2\pa_zR(z,w) + R(z,w)\pa_{z})\la\la T(z)X\ra\ra=-\frac{1}{\pi}\left(\la\la
T(z)T(w)X\ra\ra-\frac{c}{2}K(z,w)\right),$$
with the kernel $Q(z,w)$ given by
$$L_{\nu}\la\la T(z)X\ra\ra =\iint\limits_{F}Q(z,w)\nu(w)d^{2}w.$$
Using Remark \ref{Lie}, we can also state Theorem \ref{TT-equation} as
$$P_w(Q_1)(z,w)+(2\pa_z\mathcal{R}(z,w)+\mathcal{R}(z,w)\pa_z)\la\la T(z)X\ra\ra=-\frac{1}{\pi}\left(\la\la
T(z)T(w)X\ra\ra-\frac{c}{2}K(z,w)\right),$$
which clearly shows that  $\la\la T(z)T(w)X\ra\ra$ is meromorphic in $z$ and $w$.
\end{remark}
\begin{corollary} \label{LL}
For $\mu,\nu\in\Omega^{-1,1}(\Ga)$,
\begin{align*}
L_{\nu}\la\la T(z)X\ra\ra
& = -
\frac{1}{\pi}\iint\limits_{F}
\left(\la\la
T(z)T(w) X\ra\ra -\frac{c}{2}K(z,w)
\right)\nu(w)d^{2}w, \\
L_{\bar{\nu}}\la\la \bar{T}(\z)X\ra\ra &=-\frac{1}{\pi}\iint\limits_{F}
\left(\la\la
\bar{T}(\z)\bar{T}(\w) X\ra\ra -\frac{c}{2}K(\z,\w)
\right)\ov{\nu(w)}d^{2}w,
\end{align*}
where integrals are understood in the principal value sense.
\end{corollary}
\begin{proof}
\begin{align*}
L_{\nu}\la\la T(z)X\ra\ra & =\iint\limits_{F}Q(z,w)\nu(w)d^{2}w\\ &=
\iint\limits_{F}\left(Q_1(z,w)+(2\pa_z\mathcal{R}(z,w)+\mathcal{R}(z,w)\pa_z)\la\la
T(z)X\ra\ra\right)\nu(w)d^{2}w\\
&=\iint\limits_{F}\left(P_wQ_1(z,w)+(2\pa_z\mathcal{R}(z,w)+\mathcal{R}(z,w)\pa_z)\la\la
T(z)X\ra\ra\right)\nu(w)d^{2}w.
\end{align*}
\end{proof}
\begin{remark}
It follows from Corollary \ref{LL} that for $\nu\in\Omega^{-1,1}(\Ga)$
\begin{gather} \label{T-zero}
\iint\limits_{F}(2\pa_{z}R(z,w)+R(z,w)\pa_{z})\la\la T(z)X\ra\ra)\nu(w)d^{2}w=0. 
\end{gather}
In fact, using \eqref{equation-R} and orthogonality of $\mathcal{D}_{w}G$ to harmonic Beltrami differentials,  we get for $q\in \Omega^{2,0}(\Gamma)$ and $\nu\in\Omega^{-1,1}(\Gamma)$
$$\iint\limits_F (2\pa_zR(z,w)+R(z,w)\pa_z)q(z)\nu(w)d^2w=0.$$
We also have, in agreement with \eqref{D-zero} and \eqref{T-zero}, that
$$\iint\limits_{F}\mathcal{D}_{z}\mathcal{D}_{w}G(z,w)\mu(w)d^{2}w=0.$$
\end{remark}
\begin{corollary} \label{LL-X} For $\mu,\nu\in\Omega^{-1,1}(\Ga)$,
\begin{align*}
L_{\mu}L_{\nu}\log\la X\ra & =
\frac{1}{\pi^{2}}\iint\limits_{F}\iint\limits_{F}\Bigl(\la\la
T(z)T(w) X\ra\ra -\frac{6}{\hbar}K(z,w)
\Bigr)\mu(z)\nu(w)d^{2}zd^{2}w, \\
L_{\bar{\mu}}L_{\bar{\nu}}\log\la X\ra & =
\frac{1}{\pi^{2}}\iint\limits_{F}\iint\limits_{F}\Bigl(\la\la
\bar{T}(\z)\bar{T}(\w) X\ra\ra -\frac{6}{\hbar}K(\z,\w)
\Bigr)\ov{\mu(z)\nu(w)}d^{2}z d^{2}w,
\end{align*}
where integrals are understood in the principal value sense.
\end{corollary}
\begin{proof} Follows from Theorem \ref{one-point}, equation \eqref{LS} and Corollary \ref{LL}.
\end{proof}
\section{Two-point correlation function --- $T\bar{T}$ equation}\label{TPCF-TTbar}
Here we compute the two point correlation function $\la\la T(z)\bar{T}(\w)X\ra\ra$ in all orders of the perturbation theory. Using notations in Section \ref{TT-VF-VF}, we have the following result.
\begin{theorem} \label{T-barT-equation} 
On the Schottky space $\mathfrak{S}_{g}$,
\begin{equation}
\bar{\pa}\la\la T(z)X\ra\ra =\frac{1}{\pi}\la\la T(z) \bar{T}(\bar{w})X\ra\ra.
\end{equation}
The same statement holds for the Teichm\"{u}ller space $\mathfrak{T}_{g}$.
\end{theorem}
\begin{proof} We follow the proof of Theorem \ref{TT-equation}, using
$L_{\bar\nu}$ instead of $L_{\nu}$.
\subsection{Classical contribution}\label{TPCF-TTbar-CL}
It follows from \eqref{BLS} and \eqref{P-kernel} that
\begin{align*}
L_{\bar{\nu}}T_{cl}(z)
=-\frac{1}{2\hbar}\rho(z)\ov{\nu(z)}=-\frac{2}{\hbar}\iint\limits_{F}
\mathcal{D}_z\mathcal{D}_{\bar{w}}G(z,w)\ov{\nu(w)}d^2w.
\end{align*}
Using identification in Section 4.3, we get
\begin{align*}
\bar{\pa}T_{cl}(z)
=\frac{2}{\hbar}\mathcal{D}_z\mathcal{D}_{\bar{w}}G(z,w)=\frac{1}{\pi}\la\la
T(z)\bar{T}(\bar{w})X\ra\ra_{cl}.
\end{align*}
This equation corresponds to the single tree graph in Fig.~3 with $w$ replaced by $\w$, and  according to \eqref{var-classical-action}, it is equivalent to
\begin{equation} \label{wp-classical-1}
\bar{\pa}\pa\log\la X\ra_{cl} =\frac{i}{\pi\hbar}\omega_{\wpm}.
\end{equation}
\subsection{One-loop contribution} \label{TPCF-TTbar-OLL}
At the one loop level, using \eqref{P-kernel}, parts (ii) and (iv) of Lemma 5.1, Corollary \ref{var-LD} and Lemma \ref{variation-H}, we get 
\begin{align*}
L_{\bar{\nu}}\la\la T(z)X\ra\ra_{1-\text{loop}}& =-\pi \Bigl(
L_{\bar{\nu}}H(z)
 +\iint\limits_F L_{\bar{\nu}}\mathcal{D}_zG(z,u)G(u,u) \rho(u)
d^2u \\
&\;\quad +\iint\limits_F \mathcal{D}_zG(z,u)L_{\bar{\nu}}G(u,u) \rho(u)
d^2u
+\frac{1}{2}\rho(z)\ov{\nu(z)}G(z,z)\Bigr) \\
&=-\pi\iint\limits_F \tilde{E}(z,w)\ov{\nu(w)} d^2w,
\end{align*}
where $\tilde{E}(z,w)=\tilde{E}_1(z,w)+\tilde{E}_2(z,w)+\tilde{E}_3(z,w)+\tilde{E}_4(z,w)+\tilde{E}_5(z,w)$, and
\begin{align*}
\tilde{E}_1(z,w) & = 2(\pa_z\pa_{\bar{w}}G(z,w))^2
-\frac{1}{\pi}\mathcal{D}_z\mathcal{D}_{\bar{w}}G(z,w), \\
\tilde{E}_2(z,w) & = 2\iint\limits_F\mathcal{D}_z\pa_{\bar{w}}G(z,w)
\pa_{\bar{w}}G(w,u)G(u,u) \rho(u) d^2u, \\
\tilde{E}_3(z,w) & =2\iint\limits_{F}\mathcal{D}_{z}(z,u)(\pa_{\w}G(w,u))^{2}\rho(u)d^{2}u,\\
\tilde{E}_{4}(z,w)&=-2 \mathcal{D}_z\mathcal{D}_{\w}G(z,w)\iint\limits_{F}G(z,u)G(u,u)
\rho(u)d^{2}u,\\
\tilde{E}_{5}(z,w)&=2\mathcal{D}_z\mathcal{D}_{\w}G(z,w)G(z,z).
\end{align*}

Here $\tilde{E}_{4}(z,w)$ and $\tilde{E}_{5}(z,w)$ are already 
anti-holomorphic quadratic differentials for $\Ga$ in variable $w$,
and we compute the corresponding orthogonal projections 
of $\tilde{E}_1(z,w)$, $\tilde{E}_2(z,w)$ and $\tilde{E}_3(z,w)$
by using the Fuchsian global coordinate on $X\simeq \Ga\bk\up$. From the Stokes' theorem, property \textbf{P3} and \eqref{equation-R} it follows that
\begin{align*}
P_{\w}(\tilde{E_{1}})& =4\iint\limits_{F}\mathcal{D}_{\bar{w}}\mathcal{D}_uG(w,u)\left(
2(\pa_z\pa_{\bar{u}}G(z,u))^{2}- \frac{1}{\pi}\mathcal{D}_z\mathcal{D}_{\bar{u}}G(z,u)\right)\rho(u)^{-1}d^2u\\
&=-\iint\limits_{F}\rho(u)R(\bar{u},\w)\pa_{\bar{u}}(
\pa_zG(z,u))^2d^2u - \frac{1}{\pi}\mathcal{D}_z\mathcal{D}_{\bar{w}}G(z,w)\\
&\quad -i\lim_{\vep\rightarrow 0}
\oint_{C_{\vep}(w)}R(\bar{u},\w)
(\pa_z\pa_{\bar{u}}G(z,u))^2d\bar{u}\\
&=2\iint\limits_{F}\mathcal{D}_{\bar{w}}G(w,u)(\pa_zG(z,u))^2\rho(u)d^2u
+2(\pa_z\pa_{\bar{w}}G(z,w))^2 \\
&\quad -\frac{1}{\pi}\mathcal{D}_z\mathcal{D}_{\bar{w}}G(z,w)
-\frac{i}{2}\lim_{\vep\rightarrow 0}\oint_{C_{\vep}(z)}R(\bar{u},\w)(
\pa_zG(z,u))^2\rho(u)du\\
&=\tilde{T}_{1}+\tilde{T}_{2}+\tilde{T}_{3} + \tilde{T}_{4}.
\end{align*}
Terms $\tilde{T}_{1}$ and $\tilde{T}_{2}$ correspond to the contributions of graphs 1 and 2, with $w$ replaced by $\w$, into $\la\la T(z)\bar{T}(\w)X\ra\ra_{1-\text{loop}}$, and
$$\tilde{T}_{3}=-\frac{1}{\pi}\mathcal{D}_z\mathcal{D}_{\bar{w}}G(z,w).$$
To compute $\tilde{T}_{4}$, we use \eqref{derivative-G-free}, \eqref{equation-R} and 
\begin{align*}
(\pa_zG(z,u))^2 =&\sum_{\gamma_1\neq \gamma_2\in
\Gamma}\pa_z\mathcal{G}(z,\gamma_1 u)\pa_z\mathcal{G}(z,\gamma_2 u)+\sum_{\gamma\in
\Gamma}(\pa_z\mathcal{G}(z,\gamma u))^2,
\end{align*}
to obtain
\begin{align*}
\tilde{T}_{4} & = 
\rho(z)R(\z,\w)\sum_{\gamma \neq \id\in\Gamma}
\left.\pa_z\mathcal{G}(z,\gamma u)\right\vert_{u=z}
-\frac{i}{2}\lim_{\vep\rightarrow 0}\oint_{C_{\vep}(z)}R(\bar{u},\w)(
\pa_z\mathcal{G}(z,u))^2\rho(u)du \\
& = \frac{1}{2}\rho(z)R(\z,\w)\pa_{z}(G(z,z)) + \frac{1}{\pi}\mathcal{D}_z\mathcal{D}_{\bar{w}}G(z,w).
\end{align*}

To compute $P_{\w}(\tilde{E}_2)$ we observe that by the Stokes' theorem, property \textbf{P3} and \eqref{equation-R},
\begin{gather*}
8\iint\limits_{F}\mathcal{D}_{\bar{w}}\mathcal{D}_uG(w,u) \mathcal{D}_z\pa_{\bar{u}}G(z,u)
\pa_{\bar{u}}G(u,v)\rho(u)^{-1} d^2u\\
=-\iint\limits_{F}R(\bar{u},\w)\pa_{\bar{u}}(\mathcal{D}_zG(z,u)
G(u,v))\rho(u)d^2u\\
-i\lim_{\vep\rightarrow 0}\oint_{C_{\vep}}R(\bar{u},\w)
\mathcal{D}_z\pa_{\bar{u}}G(z,u)
\pa_{\bar{u}}G(u,v)d\bar{u} \\
=2\iint\limits_{F}
\mathcal{D}_{\bar{w}}G(w,u)\mathcal{D}_zG(z,u) G(u,v)
\rho(u)d^{2}u +2\mathcal{D}_{z}\pa_{\w}G(z,w)\pa_{\w}G(v,w)\\
+ 2\pa_{z}\mathcal{D}_{\w}G(z,w)\pa_{z}G(z,v)+R(\bar{v},\w)\mathcal{D}_{z}\pa_{\bar{v}}G(z,v)+2\mathcal{D}_{z}\mathcal{D}_{\w}G(z,w)G(z,v), 
\end{gather*}
where $C_{\vep}=C_{\vep}(v)\cup C_{\vep}(w)\cup C_{\vep}(z)$.
Thus we obtain
\begin{align*}
P_{\w}(\tilde{E}_{2})(z,w)& =8\iint\limits_{F}\iint\limits_{F}\mathcal{D}_{\bar{w}}\mathcal{D}_uG(w,u) \mathcal{D}_z\pa_{\bar{u}}G(z,u)
\pa_{\bar{u}}G(u,v)G(v,v)\rho(u)^{-1}\rho(v) d^2ud^{2}v \\
& =\tilde{T}_{5} + \tilde{T}_{6} +\tilde{T}_{7} +\tilde{T}_{8} +\tilde{T}_{9},
\end{align*}
where 
\begin{align*}
\tilde{T}_{5}&=2\iint\limits_{F}\iint\limits_{F}
\mathcal{D}_{\bar{w}}G(w,u)\mathcal{D}_zG(z,u) G(u,v)G(v,v)
\rho(u)\rho(v)d^{2}ud^{2}v, \\
\tilde{T}_{6}& =2\iint\limits_F\mathcal{D}_z\pa_{\bar{w}}G(z,w)
\pa_{\bar{w}}G(w,v)G(v,v)\rho(v)d^2v, \\
\tilde{T}_{7} & = 2\iint\limits_F\pa_{z}\mathcal{D}_{\w}G(z,w)
\pa_{z}G(z,v)G(v,v)\rho(v)d^2v
\end{align*}
correspond, respectively, to  graphs 3, 4 and 5 with $w$ replaced by $\bar{w}$, while 
\begin{align*}
\tilde{T}_{8} & =\iint\limits_{F}R(\bar{v},\w)\mathcal{D}_{z}\pa_{\bar{v}}G(z,v)G(v,v)\rho(v)d^{2}v
\end{align*}
and
$$\tilde{T}_{9}=-\tilde{E}_{4}(z,w).$$

Finally, as in Section \ref{VA-OPCF-HLL}, we obtain
\begin{align*}
P_{\w}(\tilde{E}_3)(z,w)& =8\iint\limits_F\iint\limits_F \mathcal{D}_zG(z,v)
\mathcal{D}_{\bar{w}}\mathcal{D}_{u}G(w,u)(\pa_{\bar{u}}G(u,v))^{2}
\rho(u)^{-1}\rho(v)d^{2}ud^2v\\
& =2\iint\limits_F\iint\limits_F
\mathcal{D}_zG(z,v)\mathcal{D}_{\bar{w}}G(w,u)
G(u,v)^{2}\rho(u)\rho(v)d^2u d^2v\\
&\quad +2\iint\limits_F
\mathcal{D}_zG(z,v)\pa_{\bar{w}}G(w,v)\pa_{\bar{w}}G(w,v)\rho(v)d^2v\\
&\quad+\iint\limits_F
\mathcal{D}_zG(z,v)R(\bar{v},\w)
\pa_{\bar{v}}G(v,v) \rho(v)d^2v \\
&\quad -\frac{1}{\pi}\iint\limits_F
\mathcal{D}_zG(z,v)\mathcal{D}_{\bar{w}}G(w,v) \rho(v) d^2v\\
=& \tilde{T}_{10}+\tilde{T}_{11}+\tilde{T}_{12}+\tilde{T}_{13}.
\end{align*}
The first two terms, $\tilde{T}_{10}$ and $\tilde{T}_{11}$, correspond to graphs 6 and 7. Using the Stokes' theorem and \eqref{equation-R}, we see that the sum $\tilde{T}_{3}+\tilde{T}_{4}+\tilde{E}_{5}+\tilde{T}_{8}+\tilde{T}_{12}+\tilde{T}_{13}$ corresponds to the remaining graph 8. Thus we have proved
$$\la\la
T(z)\bar{T}(\bar{w})X\ra\ra_{1-\text{loop}} =\pi^{2}P_{\w}(\tilde{E})(z,w),$$
so that $\la\la T(z)\bar{T}(\bar{w})X\ra\ra_{1-\text{loop}}$ is holomorphic in $z$ and anti-holomorphic in $w$.
Hence
\begin{align*}
\bar{\pa}\la\la T(z)X\ra\ra_{1-\text{loop}} =\frac{1}{\pi}\la\la
T(z)\bar{T}(\bar{w})X\ra\ra_{1-\text{loop}} .
\end{align*}
\subsection{Higher loops}\label{TPCF-TTbar-HLL}
As in Section \ref{TPCF-TT-HLL}, we claim that for every $\Upsilon \in \mathcal{G}^{(c)}_{\geq 3}$ with more than one loop,
\begin{gather*}
\sum_{\Upsilon'\in p_1^{-1}(\Upsilon)}(-1)^{|V(\Upsilon')| +\vep_{1}(\Upsilon')}\bar\pa W_{\Upsilon'}(X;z) \\
= -2\sum_{\Upsilon^{\prime\prime} \in (p_2\circ
p_1)^{-1}(\Upsilon)}(-1)^{|V(\Upsilon'')| +\vep_{1}(\Upsilon'')}W_{\Upsilon''}(X;z,\w), 
\end{gather*}
where now $p_2: \mathcal{G}^{(c)}_{z,\w}\rightarrow \mathcal{G}^{(c)}_{z}$
is the map eliminating the labeled vertex $\w$ of valency $1$ or $2$.

This readily follows from the arguments in Sections  \ref{VA-OPCF-HLL}, \ref{TPCF-TT-HLL} and \ref{TPCF-TTbar-OLL}.
Since the only graph in $\mathcal{G}_{z}^{(c)}$ that contains an edge with both end points being a vertex of valency two is a one-loop tadpole graph, the computation is even simpler than is Section \ref{TPCF-TT-HLL}.
\end{proof}
It follows from Theorem  \ref{T-barT-equation} that
\begin{equation} \label{bar-L-T}
L_{\bar{\nu}}\la\la T(z) X\ra\ra =-\frac{1}{\pi}\iint\limits_{F}\la\la T(z)\bar{T}(\w)X\ra\ra\ov{\nu(w)}d^{2}w.
\end{equation}
Combining Theorems \ref{one-point} and \ref{T-barT-equation}, we obtain
\begin{corollary} \label{L-barL-X}
\begin{align*}
\bar{\pa}\pa\log\la X\ra =-\frac{1}{\pi^2}\left(\la\la
T(z)\bar{T}(\bar{w})X\ra\ra-\frac{\pi}{6}\mathcal{D}_z\mathcal{D}_{\bar{w}}G(z,w)\right),
\end{align*}
or, equivalently,
\begin{align*}
L_{\mu}L_{\bar{\nu}}\log\la X\ra =\frac{1}{\pi^2}\iint\limits_{F}\iint\limits_{F}\left(\la\la
T(z)\bar{T}(\bar{w})X\ra\ra-\frac{\pi}{6}\mathcal{D}_z\mathcal{D}_{\bar{w}}G(z,w)\right)\mu(z)\ov{\nu(w)}d^{2}zd^{2}w.
\end{align*}
\end{corollary}
\section{Conformal Ward identities and modular geometry}\label{MG}
According to Belavin, Polyakov and
Zamolodchikov \cite{BPZ}, conformal symmetry of the two-dimensional quantum field theory on the Riemann sphere is expressed by the so-called conformal Ward identities for correlation functions with insertions of the stress-energy tensor. In particular, one-point Ward identities determine conformal dimensions of primary fields, while two-point Ward identities describe the Virasoro algebra symmetry of a theory. BPZ conformal Ward identities were generalized to higher genus Riemann surfaces in \cite{EO}.

As we have already mentioned, equations \eqref{T-ward-1}-\eqref{barT-ward-1} and \eqref{ward-TT}-\eqref{ward-barTbarT}, \eqref{T-barT-equation} are one-point and two-point Ward identities for quantum Liouville theory on the higher genus Riemann surfaces. One-point Ward identities for the punctured Riemann sphere were discussed previously in \cite{LT2,LT-Varenna}\footnote{We plan to address this case in the forthcoming publication.}.  Here we only observe that from  
\eqref{asymptotics-R} we obtain the following asymptotic for two-point correlation functions: 
\begin{align*}
 \la\la T(z)T(w)X \ra\ra & =\frac{c/2}{(z-w)^4} + \frac{2}{(z-w)^2}
\la\la T(z)X\ra\ra - \frac{1}{z-w} \pa_{z} \la\la T(z)X \ra\ra \\ 
&\quad+\;\text{regular terms as}\; w\rightarrow z, \\
 \la\la T(z)\bar{T}(\bar{w})X \ra\ra &=\quad\;\text{regular terms as}\; w\rightarrow z,\\
 \la\la \bar{T}(\bar{z})\bar{T}(\bar{w}) X\ra\ra &= \frac{c/2}{(\z-\bar{w})^4} + \frac{2}
 {(\z-\bar{w})^2}\la\la \bar{T}(\bar{z})X\ra\ra - \frac{1}{\z-\bar{w}} \pa_{\bar{z}} \la\la
 \bar{T}(\bar{z})X
 \ra\ra,\\
&\quad+ \;\text{regular terms as}\; w\rightarrow z,
\end{align*}
where $\displaystyle{c=\frac{12}{\hbar}+1}$. The leading terms in these equations are precisely BPZ Ward identities, where $c$ is
the central charge of the theory. 

As was pointed out in \cite{LT-Varenna}, equations \eqref{T-ward-1}-\eqref{barT-ward-1} and \eqref{ward-TT}-\eqref{ward-barTbarT}, \eqref{T-barT-equation} also admit interpretation in terms of ``modular geometry'' of Friedan and Shenker. Actually, these equations give precise meaning to the discussion in \cite[Section 3]{FS}. Namely, introducing\footnote{It is interesting to interpret this finite one-loop redefinition of the free energy in invariant terms.}
\begin{equation} \label{F-new}
\tilde{\mathcal{F}}=\mathcal{F} +\frac{1}{24\pi}S_{cl},
\end{equation}
and using \eqref{LS}, we can rewrite \eqref{T-ward-1}--\eqref{barT-ward-1} as
\begin{align} 
\pa\tilde{\mathcal{F}}_{X} = \frac{1}{\pi}\la\la T(z)X\ra\ra, \label{1,0} \\
\bar\pa\tilde{\mathcal{F}}_{X} = \frac{1}{\pi}\la\la \bar{T}(\z)X\ra\ra,\label{0,1}
\end{align}
where $\pa$ and $\bar\pa$ are $(1,0)$ and $(0,1)$ components of de Rham differential on $\mathfrak{S}_{g}$. Interpreting $e^{\tilde{\mathcal{F}}}$ as a Hermitian metric\footnote{Here we are tacitly assuming that $\tilde{\mathcal{F}}$ is a smooth function on $\mathfrak{S}_{g}$. Of course, it is only a formal function, so all geometric objects should be interpreted in a formal category.} in a trivial holomorphic line bundle $\mathfrak{S}_{g}\times\CC\rightarrow\mathfrak{S}_{g}$, we see that
$\frac{1}{\pi}\la\la T(z)X\ra\ra$ and $ \frac{1}{\pi}\la\la \bar{T}(\z)X\ra\ra$
are $(1,0)$ and $(0,1)$ components of the corresponding canonical connection\footnote{The connection which is compatible with the Hermitian metric and complex structure in the line bundle.} in the unitary frame. It was proved by Zograf (see \cite[Theorem 3.1]{Zo}) that the Hermitian metric $\exp\left\{\frac{1}{12\pi}S_{cl}\right\}$ in $\mathfrak{S}_{g}\times\CC$ descends to the Hermitian metric
in the Hodge line bundle $\lambda_{H}$ over the moduli space $\mathfrak{M}_{g}$. Since $e^{\tilde{\mathcal{F}}}=e^{\frac{c}{24\pi}S_{cl}}e^{\mathcal{F}_{0}}$, 
where $\mathcal{F}_{0}$ is a (formal) function on $\mathfrak{M}_{g}$, we see that  the trivial holomorphic line bundle $\mathfrak{S}_{g}\times\CC\rightarrow\mathfrak{S}_{g}$ with the Hermitian
metric $e^{\tilde{\mathcal{F}}}$ descends to a ``projective holomorphic line bundle'' $\cE_{c}=\lambda_{H}^{c/2}$ over the moduli space $\mathfrak{M}_{g}$ (see \cite{FS} for the definition of a projective line bundle).

Correspondingly, 
Corollaries \ref{LL} and \ref{L-barL-X}
can be interpreted as curvature computations for $\cE_{c}$. Namely, denote by $\textbf{1}$ the section of $\cE_{c}$ whose pull-back to the trivial bundle over $\mathfrak{S}_{g}$ is a section identically equal to $1$, and
by $\nabla_{\mu}$, $\mu\in\Omega^{-1,1}(\Ga)$ --- covariant derivative 
of the canonical connection. Using Corollary \ref{LL} we have 
\begin{align*}
\nabla_{\mu}\nabla_{\nu} \textbf{1}&= 
\frac{1}{\pi}\left(L_{\mu} + \frac{1}{\pi}\iint\limits_{F}\la\la
 T(w)X\ra\ra\mu(w)d^{2}w\right)\iint\limits_F \la\la
 T(z)X\ra\ra
 \nu(z)d^2z\\
&= -\frac{1}{\pi^{2}}\iint\limits_{F}\iint\limits_{F}\Bigl(\la\la
T(z)T(w) X\ra\ra -\frac{c}{2}K(z,w)
\Bigr)\mu(w)\nu(z)d^{2}wd^{2}z \\
&\quad +\frac{1}{\pi^2}\iint\limits_X \la\la
 T(z)X\ra\ra
 \nu(z)d^2z \iint\limits_X \la\la
 T(w)X\ra\ra
 \mu(w)d^2w,
\end{align*}
which is symmetric in $\nu$ and $\mu$, so that the $(2,0)$
component of the curvature tensor vanishes. 

Similar statements hold for $(0,2)$ components. Finally, it follows from  Corollary \ref{L-barL-X} that
\begin{align*}
\nabla_{\mu} \nabla_{\bar{\nu}} \textbf{1}& =0,\\
 \nabla_{\bar{\nu}}\nabla_{\mu}\textbf{1}& =\frac{1}{\pi}\left(L_{\bar{\nu}}
 \iint\limits_F\la\la T(z)X\ra\ra \mu(z) d^2z\right) \\
 &=-\frac{1}{\pi^2}\iint\limits_F\iint\limits_F
 \la\la T(z)\bar{T}(\bar{w})X\ra\ra\mu(z)\ov{\nu(w)} d^2zd^2w.
\end{align*}
Thus using the identification in Section \ref{TT-VF-VF} we see that the $(1,1)$ component of the curvature tensor is given by
\begin{align*}
\frac{1}{\pi^2}\la\la T(z)\bar{T}(\bar{w})X\ra\ra.
\end{align*}
\begin{remark} \label{Potential} Since the Hodge line bundle $\lambda_{H}$ is positive, the projective line bundle $\cE_{c}$ is also positive for $c>0$. Moreover, assuming that $\tilde{\mathcal{F}}$ is a function on
$\mathfrak{S}_{g}$ given by the actual integral \eqref{X}, the 
curvature form $\frac{1}{\pi^2} \la\la T(z)\bar{T}(\bar{w})X\ra\ra$ of the canonical connection on $\cE_{c}$ is a positive definite $(1,1)$ form
on $\mathfrak{M}_{g}$. Indeed, denoting by $\cD_{L}\varphi=e^{-\frac{1}{2\pi\hbar}S(\varphi)}\cD\varphi$ the corresponding measure on $\cC\cM(X)$ and using that
$$\la X\ra =\underset{\cC\cM(X)}{\pmb{\int}}\cD_{L}\vp,$$
we obtain for $\mu\in\Omega^{-1,1}(\Ga)$,
\begin{gather*}
\la X\ra^{2} \iint\limits_F \iint\limits_F \la\la
T(z)\bar{T}(\bar{w})X\ra\ra \mu(z) \ov{\mu(w)} d^2zd^2w \\
= \la X\ra \underset{\cC\cM(X)}{\pmb{\int}}\left|\iint\limits_{F}T(\vp)(z)\mu(z)d^{2}z\right|^{2}\!\!\cD_{L}\vp 
- \left| \underset{\cC\cM(X)}{\pmb{\int}}\!\iint\limits_{F}T(\vp)(z)\mu(z)d^{2}z\;\cD_{L}\vp\right|^{2},
\end{gather*}
which is non-negative by Cauchy-Bunyakovskii inequality.
In this way we get a K\"ahler metric $\omega$ on $\mathfrak{M}_{g}$, whose pull-back to $\mathfrak{S}_{g}$ has a K\"ahler potential $-\tilde{\mathcal{F}}$. The corresponding symplectic form $\omega$ is given by the following power series in $\hbar$,
\begin{equation} \label{omega}
\omega=\frac{1}{2\pi\hbar}\,\omega_{\wpm}+\sum_{n=0}^{\infty}\hbar^{n}\omega_{(n)}.
\end{equation}
It would be very interesting to give a geometric interpretation of these ``quantum corrections'' to the Weil-Petersson metric, and to understand the series \eqref{omega} non-perturbatively.
\end{remark}
\appendix
\section{Belavin-Knizhnik theorem and the $T\bar{T}$ equation}
 
Here we compare the one-loop $T\bar{T}$ equation in Corollary \ref{L-barL-X} with the special case of Belavin-Knizhnik theorem \cite{BK} --- a local index theorem for families of $\bar{\pa}$-operators on Riemann surfaces ---
a formula for the Chern form of Quillen's metric in the corresponding determinant line bundle over $\mathfrak{M}_{g}$. Using $\log \la X\ra_{1-\text{loop}}=-\frac{1}{2}\log Z(2)$,  we get from Corollary \ref{L-barL-X},
\begin{align} 
L_{\mu}L_{\bar{\nu}}\log Z(2)& =-\frac{2}{\pi^{2}}\iint\limits_{F}\iint\limits_{F}\la\la T(z)\bar{T}(\w)X\ra\ra\mu(z)\ov{\nu}(w)d^{2}zd^{2}w \label{LL-Z(2)} \\
&\quad +\frac{1}{12\pi}(\mu,\nu),\nonumber
\end{align}
where $(\mu,\nu)$ stands for the inner product \eqref{pairing} in $\Omega^{-1,1}(\Ga)$. On the other hand, using D'Hoker-Phong 
formula \cite{dhoker-phong} $\det\Delta_{2}=c_{g}Z(2)$, where $\Delta_{2}$ is the Laplace operator of the hyperbolic metric acting on quadratic differentials on $X$ and $c_{g}$ is a constant depending only on genus, the Belavin-Knizhnik formula for the family of $\bar{\pa}$-operators  acting on quadratic differentials can be written in the form
\begin{align} \label{BK}
L_{\mu}L_{\bar{\nu}} \log Z(2) - L_{\mu}L_{\bar{\nu}}\log \det N =\frac{13}{12\pi}(\mu,\nu).
\end{align}
Here $N$ is a Gram matrix with respect to the inner product \eqref{inner-quadratic} of the bases of holomorphic quadratic differentials on the Riemann surfaces $X_{t}$, which depend holomorphically on $t\in\mathfrak{T}_{g}$ (see \cite{TakZog87:Localindexthm} for details and references).

We show how to obtain the Belavin-Knizhnik formula \eqref{BK} from \eqref{LL-Z(2)}. First, using \eqref{Wolpert-metric}, \cite[Lemma 1]{TakZog87:Localindexthm}, formulas (2.8) in \cite{TakZog87:Localindexthm} and (1.3) in \cite{TZ}, it is elementary to obtain
\begin{align}
L_{\mu}L_{\bar{\nu}}\log \det N &=-\iint\limits_{F}\iint\limits_{F} 
P(z,z)G(z,w)\mu(w)\ov{\nu(w)}\rho(w)\rho(z)^{-1}d^2zd^2w \nonumber  \\
&\quad -\iint\limits_{F}\iint\limits_{F}
P(z,w)G(z,w)\mu(z)\ov{\nu(w)}d^2zd^2w, \label{LL-N-1}
\end{align}
where $P(z,w)=4\mathcal{D}_z\mathcal{D}_{\bar{w}}G(z,w)$. 
\begin{remark} Formula \eqref{LL-N-1} coincides with Wolpert's formula \cite{Wol} for the Ricci tensor of the Weil-Petersson metric on $\mathfrak{T}_{g}$. 
\end{remark}
Now using the Fuchsian global coordinate on $X\simeq\Ga\bk\up$, we 
rewrite the first term in \eqref{LL-N-1} as 
\begin{align*}
&-4\iint\limits_{F}\iint\limits_{F}  \sum_{\gamma\in \Gamma}
\left.\mathcal{D}_{z'}\mathcal{D}_{\bar{z}} \mathcal{G}(z',\gamma
z)
\right\vert_{z'=z}G(z,w)\mu(w)\ov{\nu(w)}\rho(w)\rho(z)^{-1}d^2zd^2w.\\
\end{align*}
Using \eqref{DD-G-cal} and \eqref{integration-G},
we obtain
\begin{gather*}
-4\iint\limits_{F}\iint\limits_{F} \left.\mathcal{D}_{z'}\mathcal{D}_{\bar{z}}
\mathcal{G}(z', z)
\right\vert_{z'=z}G(z,w)\mu(w)\ov{\nu(w)}\rho(w)\rho(z)^{-1}d^2zd^2w
=-\frac{3}{4\pi}(\mu, \nu).
\end{gather*}
Using equations  \eqref{mu-harm}, \eqref{derivative-G}, \eqref{DD-G-cal},  property \textbf{P3} and the Stokes'  theorem, we can rewrite the remaining part of the first term in \eqref{LL-N-1} as
\begin{gather*}
4\iint\limits_{F}\iint\limits_{F} \sum_{\gamma\neq \id \in \Gamma}\left.
\pa_{z'}\mathcal{D}_{\bar{z}} \mathcal{G}(z',\gamma
z)
\right\vert_{z'=z}\pa_zG(z,w)\mu(w)\ov{\nu(w)}\rho(w)\rho(z)^{-1}d^2zd^2w\\
=-2\iint\limits_{F}\iint\limits_{F} \sum_{\gamma\neq \id \in \Gamma}\left.
\pa_{z'}\pa_{\bar{z}} \mathcal{G}(z',\gamma z)\right\vert_{z'=z}G(z,w)\mu(w)\ov{\nu(w)}\rho(w)d^2zd^2w\\
-2\iint\limits_{F}\iint\limits_{F} \sum_{\gamma\neq \id \in \Gamma}\left.
\pa_{\bar{z}} \mathcal{G}(z',\gamma z)
\right\vert_{z'=z}\pa_zG(z,w)\mu(w)\ov{\nu(w)}\rho(w)d^2zd^2w\\
+2i\lim_{\vep\rightarrow 0}\iint\limits_F
\oint_{C_{\vep}(w)}\sum_{\gamma\neq \id \in \Gamma}\left.
\pa_{z'}\pa_{\bar{z}} \mathcal{G}(z',\gamma z)
\right\vert_{z'=z}\pa_zG(z,w)\mu(w)\ov{\nu(w)}\rho(w)\rho(z)^{-1}dzd^2w\\
=\iint\limits_{F}\iint\limits_{F} \sum_{\gamma\neq \id \in \Gamma}\left.
\mathcal{G}(z',\gamma z)
\right\vert_{z'=z}G(z,w)\mu(w)\ov{\nu(w)}\rho(z)\rho(w)d^2zd^2w\\
+2\iint\limits_F \sum_{\gamma\neq \id \in \Gamma}\left.
\pa_{z'}\pa_{\bar{z}} \mathcal{G}(z',\gamma z)
\right\vert_{z'=z}\mu(z)\ov{\nu(z)}d^2z=J_1+J_2.
\end{gather*}

Similarly, the second term in \eqref{LL-N-1} can be rewritten as
\begin{align*}
-4 \iint\limits_{\U}\iint\limits_{F}
\mathcal{D}_z\mathcal{D}_{\bar{w}}\mathcal{G}(z,w)\sum_{\gamma\in
\Gamma}\mathcal{G}(z,\gamma w)\mu(z)\ov{\nu(w)}d^2zd^2w.
\end{align*}
To compute the contribution from $\gamma=\mathrm{id}\in\Ga$, we use the identity
\begin{align*}
\iint\limits_{\up}\mathcal{D}_z\mathcal{D}_{\bar{u}}\mathcal{G}(z,u)
\mathcal{G}(z,
u)\mathcal{D}_u\mathcal{D}_{\bar{w}}\mathcal{G}(u,w)\rho(u)^{-1}d^2u =\frac{1}{12\pi}\mathcal{D}_z\mathcal{D}_{\bar{w}}\mathcal{G}(z,w).
\end{align*}
Indeed, denoting the integral by $B(z,w)$, we get from \eqref{invariant} that  
$$B(\sigma z,\sigma w)\sigma'(z)^{2}\ov{\sigma'(w)}^{2}=B(z,w)$$
for all $\sigma\in\mathrm{PSL}(2,\RR)$, so it is sufficient to compute it at a fixed $z$. Using the unit disk $\mathbb{D}$ as a model for the hyperbolic plane (cf. with the proof of Lemma \ref{variation-H}), it is easy to compute that
$B(0,w)=\frac{1}{4\pi^{2}}$, and the identity follows. 
Therefore,
\begin{gather*}
-4 \iint\limits_{\U}\iint\limits_{F}
\mathcal{D}_z\mathcal{D}_{\bar{w}}\mathcal{G}(z,w)\mathcal{G}(z,
w)\mu(z)\ov{\nu(w)}d^2zd^2w \\ =
-16\iint\limits_F\iint\limits_{\up}\mu(z) B(z,w)\ov{\nu(w)}d^{2}zd^{2}w
=-\frac{1}{3\pi}(\mu, \nu).
\end{gather*}

Similarly, the remaining part of the second term in \eqref{LL-N-1} can be rewritten as
\begin{gather*}
4 \iint\limits_{\U}\iint\limits_{F}
\pa_z\mathcal{D}_{\bar{w}}\mathcal{G}(z,w)\sum_{\gamma\neq \id\in
\Gamma}\pa_z\mathcal{G}(z,\gamma w)\mu(z)\ov{\nu(w)}d^2zd^2w\\
-\iint\limits_F\sum_{\gamma\neq \id\in
\Gamma}\mathcal{G}(z,\gamma
z)\mu(z)\ov{\nu(z)}\rho(z)d^2z\\
=-4\iint\limits_{\U}\iint\limits_{F}
\pa_z\pa_{\bar{w}}\mathcal{G}(z,w)\sum_{\gamma\neq \id\in
\Gamma}\pa_z\pa_{\bar{w}}\mathcal{G}(z,\gamma
w)\mu(z)\ov{\nu(w)}d^2zd^2w\\
+2i\lim_{\vep\rightarrow 0}
\iint\limits_F\oint_{C_{\vep}(z)}\sum_{\gamma\neq \id\in
\Gamma}\pa_z\pa_{\bar{w}}\mathcal{G}(z,\gamma
w)
\pa_z\mathcal{G}(z,
w)\mu(z)\ov{\nu(w)}dwd^2z\\
-\iint\limits_F\sum_{\gamma\neq \id\in
\Gamma}\mathcal{G}(z,\gamma
z)\mu(z)\ov{\nu(z)}\rho(z)d^2z \\
=J_{3}-J_{2}+J_{4}.
\end{gather*}
Thus 
$$L_{\mu}L_{\bar{\nu}}\log\det N=-\frac{13}{12\pi}(\mu,\nu) + J_{1}+J_{3}+J_{4},$$
and using \eqref{G-z=z}, \eqref{z-bar-w} and \eqref{integration-G}, we finally obtain
\begin{align*}
L_{\mu}L_{\bar{\nu}}\log \det N 
& = \iint\limits_{F}\iint\limits_{F}G(z, z)
G(z,w)\mu(w)\ov{\nu(w)}\rho(w)\rho(z)d^2zd^2w -\frac{1}{2\pi}(\mu,\nu) \\
&\quad -4\iint\limits_{F}\iint\limits_{F}
\Bigl(\pa_z\pa_{\bar{w}}G(z,w)\Bigr)^2\mu(z)\ov{\nu(w)}d^2zd^2w \\
&\quad -\iint\limits_FG(z, z)\mu(z)\ov{\nu(z)}\rho(z)d^2z.
\end{align*}

Using this representation for $L_{\mu}L_{\bar{\nu}}\log \det N$, we get the Belavin-Knizhnik theorem \eqref{BK} by carefully analyzing the contribution of each one-loop graph into \eqref{LL-Z(2)}. The corresponding computation is quite tedious and is based on the repeated use of the Stokes' theorem. In a sense, it reverses the computation in  Section \ref{TPCF-TTbar-OLL}. We leave details to the interested reader. Instead, here we present a shortcut which uses \eqref{var-nu} and Remark \ref{pa-Z(2)}. Namely, from \eqref{var-nu} and Lemma \ref{variation-H} we get
\begin{align}
L_{\mu}L_{\bar{\nu}}\log Z(2)
& =-2L_{\bar{\nu}}\iint\limits_F
\left(H(z)+\frac{1}{12\pi}\mathcal{S}(J^{-1})(z)\right) \mu(z)d^2z  \nonumber \\
&=-4\iint\limits_{F}
\iint\limits_F(\pa_z\pa_{\bar{w}}G(z,w))^2\mu(z)\ov{\nu(w)}d^{2}zd^2w
+ \frac{7}{12\pi} (\mu,\nu)\nonumber\\
&\quad +4\iint\limits_F\iint\limits_{F}
H(z)\mathcal{D}_{\z}G(z,w)\mu(w)\ov{\nu(w)}\rho^{-1}(z)\rho(w)d^{2}zd^{2}w.
\nonumber
\end{align}
Using \eqref{z-bar-derivative-H}, Stokes' theorem and \eqref{green-function},  we can rewrite the last term as
\begin{gather*}
-2\iint\limits_F\iint\limits_{F}
\pa_{z}G(z,z)\pa_{\z}G(z,w)\mu(w)\ov{\nu(w)}\rho^{-1}(z)\rho(w)d^{2}zd^{2}w \\
=\iint\limits_{F}\iint\limits_{F}
G(z,z)G(z,w)\mu(w)\ov{\nu(w)}\rho(z)\rho(w)d^{2}zd^{2}w-\iint\limits_F
G(z,z)\mu(z)\ov{\nu(z)}\rho(z)d^2z.
\end{gather*}
Combining this with the obtained expression for $L_{\mu}L_{\bar{\nu}}\log \det N $ gives \eqref{BK}.
\begin{remark} Thus the one-loop term in the $T\bar{T}$ equation can be viewed as another ``packaging'' of the local index theorem for families of $\bar{\pa}$-operators on Riemann surfaces. It would be interesting to find geometric interpretation of higher loop terms.
\end{remark}
\section{The stress-energy tensor and the action functional}
Let $z$ be a Schottky global coordinate on $X\simeq\Ga\bk\Omega$. 
For $\mu\in\mathcal{H}^{-1,1}(\Ga)$ and sufficiently small $\vep\in\CC$, consider the holomorphic family $X^{\vep\mu}\simeq\Ga^{\vep\mu}\bk\Omega^{\vep\mu}$, where
$\Omega^{\vep\mu}=f^{\vep\mu}(\Omega)$ and $\Ga^{\vep\mu}=f^{\vep\mu}\circ\Ga\circ (f^{\vep\mu})^{-1}$.  For given $\varphi\in\cC\cM(X)$, let $\varphi^{\vep\mu}\in\cC\cM(X^{\vep\mu})$ be a smooth family defined by
\begin{equation} \label{family}
\vp^{\vep\mu}\circ f^{\vep\mu} +\log|f_{z}^{\vep\mu}|^{2}=\vp.
\end{equation}
\begin{lemma} \label{ste-action} Let $S:\cC\cM(X)\rightarrow\RR$ be the Liouville action functional defined by \eqref{action-Schottky}, and let $T(\vp)=\vp_{zz}-\frac{1}{2}\vp_{z}^{2}$ be the corresponding $(2,0)$ component of the stress-energy tensor. We have
\begin{equation}
\left.\frac{\pa}{\pa\vep}\right|_{\vep=0}S(\vp^{\vep\mu}) =2\iint\limits_{F}T(\vp)(z)\mu(z)d^{2}z.
\end{equation}
\end{lemma} 
\begin{proof} It repeats verbatim the proof of Theorem 1 in \cite{ZT2}! Namely, condition \eqref{family}, which replaces Ahlfors lemma used in \cite{ZT2}, gives
\begin{align*}
\dot{\vp}_{z}+\vp_{zz}\dot{f}& =-\vp_{z}\dot{f}_{z}-\dot{f}_{zz}, \\
\dot{\vp}_{\z}+\vp_{z\z}\dot{f}& =-\vp_{z}\dot{f}_{\z}-\dot{f}_{z\z}
\end{align*}
where 
$$\dot{\vp}=\left.\frac{\pa}{\pa\vep}\vp^{\vep\mu}\right|_{\vep=0},$$
and the corresponding computation in \cite{ZT2} works line by line. The Gauss-Bonnet theorem, used in \cite{ZT2}, is replaced by the equation 
$$\iint\limits_{F^{\vep\mu}}e^{\vp^{\vep\mu}}d^{2}z=\iint\limits_{F} e^{\vp}(1-|\vep\mu|^{2})d^{2}z,$$
which follows from \eqref{family}.
\end{proof}
Lemma \ref{ste-action} gives a derivation of the stress-energy tensor from the Liouville action functional. We stress that the ``transformation law''  
\eqref{family}, and the form \eqref{action-Schottky} of the action functional, both play a crucial role in this computation. The same statement holds for the Liouville action functional for the quasi-Fuchsian global coordinate, and the proof repeats verbatim the proof of Theorem 4.1 in \cite{LTT}. 

In conclusion, we present a heuristic derivation of the one-point conformal Ward identity, which clarifies corresponding arguments in \cite{BPZ}.
Namely, considering \eqref{family} as a ``change of variables'' in the functional integral
$$\la X^{\vep\mu}\ra =\underset{\cC\cM(X^{\vep\mu})}{\pmb{\int}}e^{-\frac{1}{2\pi\hbar}S(\vp^{\vep\mu})}\cD\vp^{\vep\mu},$$ 
and assuming that $\cD\vp^{\vep\mu}=\cD\vp$, we obtain
\begin{align*}
L_{\mu}\la X\ra & = \left.\frac{\pa}{\pa\vep}\right|_{\vep=0}\la X^{\vep\mu}\ra  = \underset{\cC\cM(X)}{\pmb{\int}}L_{\mu}S(\vp)e^{-\frac{1}{2\pi\hbar}S(\vp)}\cD\vp \\
& = \iint\limits_{F}\la T(z) X\ra\mu(z)d^{2}z.
\end{align*}
Now every infinitesimally trivial Beltrami
differential $\mu$ gives rise to a family $X^{\vep\mu}$ conformally equivalent to $X$, so that $L_{\mu}\la X\ra=0$. This shows that $\la T(z)X\ra$ is a holomorphic quadratic differential for $\Ga$. 

As we have shown, there is a one-loop correction to this naive form of the Ward identity, which is due to the regularization of the divergent tadpole graph. Thus rigorous definition of the ``integration measure'' $\cD\vp$ (which, in particular, would make this and similar arguments work) is a non-trivial problem.
\bibliographystyle{amsalpha}
\bibliography{Liou}
\end{document}